\documentclass[aps,amsmath,twocolumn,superscriptaddress,floatfix]{revtex4-2}
\usepackage{amsmath,amssymb,bm}
\usepackage{braket}
\usepackage{graphicx}
\usepackage{color}
\usepackage{comment}
\PassOptionsToPackage{hyphens}{url}
\usepackage{hyperref}
\hypersetup{setpagesize=false,
bookmarksnumbered=true,
bookmarksopen=true,
colorlinks=true,
linkcolor=blue,
citecolor=red,
}

\usepackage{here}
\usepackage{mathrsfs}
\numberwithin{equation}{section}
\usepackage{physics}
\newcommand{\site}{\bm r}
\newcommand{\sited}{\bm r'}

\newcommand{\siteA}{\site_{\textrm{A}}}
\newcommand{\siteB}{\site_{\textrm{B}}}
\newcommand{\siteAd}{\site_{\textrm{A}}'}
\newcommand{\siteBd}{\site_{\textrm{B}}'}
\newcommand{\wn}{\bm k}

\newcommand{\magnet}{\textrm{AFM}}
\newcommand{\metal}{\textrm{NM}}
\newcommand{\inter}{\textrm{int}}
\newcommand{\interA}{\textrm{int-A}}
\newcommand{\interB}{\textrm{int-B}}

\begin{document}

\title{Microscopic theory of spin Seebeck effect in antiferromagnets}

\author{Keisuke Masuda}
\email{21nd104s@vc.ibaraki.ac.jp}
\affiliation{Department of Physics, Ibaraki University, Mito,
 Ibaraki 310-8512, Japan}

\author{Masahiro Sato}
\email{sato.phys@chiba-u.jp}
\affiliation{Department of Physics, Ibaraki University, Mito,
 Ibaraki 310-8512, Japan}
\affiliation{Department of Physics, Chiba University, Chiba 263-8522, Japan}

\date{\today}

\begin{abstract}
 We develop a microscopic theory for the spin Seebeck effect (SSE) in N\'eel and canted phases of antiferromagnetic insulators.
 We calculate DC spin current tunneling from an antiferromagnet to an attached metal, incorporating the spin-wave theory and the non-equilibrium Green's function approach.
 Our result shows a sign change of the spin current at the spin-flop phase transition between N\'eel and canted phases, which is in agreement with a recent experiment for the SSE on $\rm Cr_2O_3$ in a semi-quantitative level.
 The sign change can be interpreted from the argument based on the density of states of up-spin and down-spin magnons, which is related to polarized-neutron-scattering spectra.
 The theory also demonstrates that the spin current in the N\'eel phase is governed by magnon correlation, whereas that in the canted phase consists of two parts: contributions from not only magnon dynamics but also static transverse magnetization. This result leads to a prediction that
 at sufficiently low temperature, the spin current non-monotonically changes as a function of magnetic field in the canted phase.
 Towards a more unified understanding of the SSE in antiferromagnets, we discuss some missing pieces in SSE theories: interface properties, effects of the transverse spin moment in the canted phase, spin-orbit coupling in metals, etc.
 Finally, we compare the SSE of antiferromagnets with that of different magnetic phases such as ferromagnets, ferrimagnets, a one-dimensional spin liquid, a spin-nematic liquid, and a spin-Peierls (dimerized) phase.
\end{abstract}
\maketitle
\section{Introduction}
\label{Sec:Introduction}
\begin{figure}[t]
 \includegraphics[width=\linewidth]{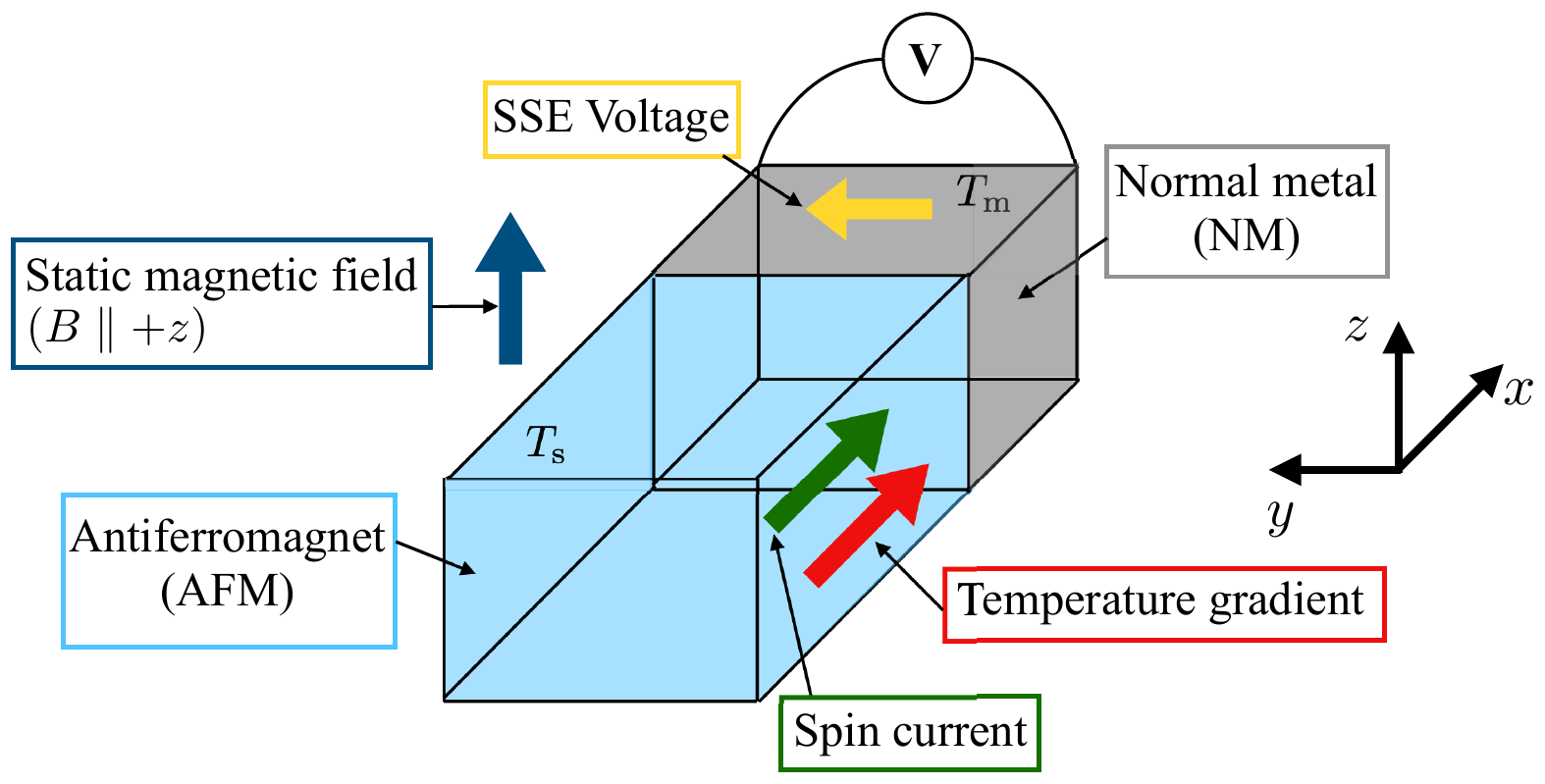}
 \caption{(Color online)
 Schematic setup of the spin Seebeck effect in antiferromagnets.
 A bilayer structure consisting of an antiferromagnetic insulator and a paramagnetic metal is placed in a static magnetic field whose direction is along the $z$-axis.
 The temperature gradient is applied parallel to the $x$ axis.
 A spin current is generated along the same $x$ direction in the antiferromagnet and is injected into the metal via an interfacial interaction.
 The injected spin current is converted into an electromotive force via the inverse spin Hall effect in the metal.
 We denote the averaged temperature of the antiferromagnet as $T_{\textrm{s}}$ and that of the metal as $T_{\textrm{m}}$ ($T_{\textrm{s}}>T_{\textrm{m}}$).
 }
 \label{FIG:SSE-Antiferromagnets}
\end{figure}
The spin Seebeck effect (SSE) \cite{Uchida2008,Uchida2010b,Jaworski2010,Uchida2010} is an established way to generate a DC spin current using a temperature gradient.
Its usual setup is depicted in Fig.~\ref{FIG:SSE-Antiferromagnets}, in which a temperature gradient is applied to the junction system of a magnet (typically a magnetic insulator) and a metal film (typically Pt).
The direction of the gradient is perpendicular to the interface.
The generated spin current runs along the gradient in the magnet and is then injected into the metal.
In the metal, the spin current is converted into electric voltage via the inverse spin Hall effect (ISHE) \cite{Saitoh2006,Valenzuela2006,Kimura2007}, and the electric field of the voltage is perpendicular to both the spin polarization and the spin-current flow.
Measuring the voltage provides clear evidence of the spin-current generation, i.e., the SSE.
Since the spin current is basically a non-conserved quantity and any direct method of observing spin current has never been developed, we usually attach a metal as shown in Fig~\ref{FIG:SSE-Antiferromagnets}.
The attachment of a metal is an essential difference between the SSE and the electronic Seebeck effect.

The SSE in a ferromagnetic insulator was first detected in 2010 \cite{Uchida2010}, and since then, experimental \cite{Jaworski2011,Uchida2012,Qu2013,Kikkawa2013a,Ramos2013,UchidaIEEE,Uchida2014a,Kikkawa2015,Kehlberger2015,Seki2015,Wu2016,Geprags2016,Li2020a,DeLima2023a} and theoretical \cite{Xiao2010,Adachi2010,Adachi2011,Zhang2012,Ohnuma2013,Hoffman2013,Rezende2014,Rezende2016a,Reitz2020,Yamamoto2021a} studies on the SSE in ordered magnets have been developed.
Beyond conventional ordered magnets, the SSE has also been studied in exotic magnetic materials such as the paramagnetic phase of a geometrically frustrated magnet $\rm Gd_3Ga_5O_{12}$~\cite{Wu2015,Liu2018}, a one-dimensional spin liquid in $\rm Sr_2CuO_3$~\cite{Hirobe2017d}, a spin-nematic liquid in $\rm LiCuVO_4$~\cite{Hirobe2019}, a spin-Peierls (dimerized) state in $\rm CuGeO_3$~\cite{Chen2021}, and a magnon-condensation state in $\rm Pb_2V_3O_9$~\cite{Xing2022}.

Antiferromagnetic insulators are one of the most fundamental classes of these magnetic materials.
They have also attracted much attention as a new platform for ultrafast spintronics~\cite{Baltz2018a,Jungwirth2016,Zelezny2018,Sato2016} because their magnetic excitations (magnons) are in the higher-energy THz regime (typically $10^{11-13}$ Hz) compared with the GHz-regime magnons in ferromagnets.
It is therefore important to deeply understand the spin dynamics and the resulting nonequilibrium phenomena in a wide class of antiferromagnets, including the SSE.

N\'eel and canted ordered phases usually appear in most antiferromagnets in the space of the temperature and an external magnetic field.
Recent experiments on antiferromagnets~\cite{Seki2015,Wu2016,Li2020a} have focused on SSEs in both phases.
Reference~\cite{Seki2015} performs an experiment on the SSE in the antiferromagnet $\rm Cr_2O_3$.
The authors detect a large jump of the SSE voltage at the spin-flop phase transition from the N\'eel phase to the canted one, whereas the SSE voltage almost vanishes in the N\'eel phase and the sign of the voltage does not change.
The SSE experiment on another antiferromagnet $\rm MnF_2$ is investigated in Ref.~\cite{Wu2016}, and the authors also observe a voltage jump at the spin-flop transition and no sign change.
However, the authors in Ref.~\cite{Li2020a} observe a sign change of the SSE voltage at the spin-flop transition in a SSE experiment on $\rm Cr_2O_3$.

At present, we have no theory explaining the whole of the above experimental results in a unified way.
However, theoreticians have continuously studied the SSE in antiferromagnets to attain a thorough understanding.
The authors in Ref.~\cite{Rezende2016a} indicates that the sign of the spin current in the SSE of antiferromagnets is opposite to that in the SSE of ferromagnets.
Reference~\cite{Reitz2020} explains the sign change observed in Ref.~\cite{Li2020a} using a low-energy semi-classical theory for antiferromagnets and a phenomenological theory for bulk transport.
The authors in Ref.~\cite{Yamamoto2021a} compute the spin current tunneling from an antiferromagnet to a metal based on the Ginzburg-Landau theory assuming that on the interface, both ferromagnetic and antiferromagnetic spin moments in the antiferromagnet are independently coupled to conducting electrons in the metal.
They then analyze the existence and disappearance of the sign change by tuning the ratio of two couplings.
In Ref.~\cite{Arakawa2022}, Arakawa computes the temperature dependence of thermal spin-current conductivity in bulk antiferromagnets by applying the Green's function method with magnon interactions.

In this paper, we add a new theoretical result for the SSE of ordered antiferromagnetic insulators.
Incorporating the spin-wave theory and the non-equilibrium Green's function approach, we derive the microscopic formula for a tunneling DC spin current flowing from the antiferromagnet to a normal metal \cite{Jauho1994,Adachi2011,Ohnuma2013} under the assumption that the relevant interfacial interaction is a standard exchange coupling between localized spins on the antiferromagnet and conducting electron spins on the metal.
Our theory explains the sign reversal of the SSE voltage, which semi-quantitatively agrees with the experiment of Ref.~\cite{Li2020a}.
The sign change is clearly understood from the argument based on the density of states (DoS) of up-spin and down-spin magnons, which can be observed through a polarized neutron-scattering experiment.
The theory also reveals that the leading term of the spin current stems from the magnon's DoS in the N\'eel phase, whereas there are two main contributions to the spin current in the canted phase: One is the magnon's DoS, and the other is the transverse magnetization term inducing the spin flipping of conducting electrons in the metal.
This microscopic origin of the spin current further leads to a prediction that in the canted phase at low temperature, the spin current non -monotonically varies as a function of applied magnetic field.
This low-temperature behavior could be potentially observed in future experiments.

In addition to the above microscopic calculation, we qualitatively discuss several effects of the interface and the bulk transport that are missing from our theory, and we argue the possibility that these effects explain the different experimental results of Refs.~\cite{Seki2015,Wu2016}.
We also compare the SSE voltages of antiferromagnets with those of several magnetic states such as ferromagnets, ferrimagnets, a 1D spin liquid, a spin-nematic liquid, and a spin-Peierls state.
In particular, we show that the present microscopic approach for tunneling spin current predicts reasonable values of the spin current in both antiferromagnets and ferromagnets.

The remaining part of this paper is organized as follows.
In Sec.~\ref{Sec:Model}, we define an antiferromagnetic Heisenberg model on a cubic lattice and explain its magnetic phase diagram including both N\'eel and canted phases.
Then we derive the magnon (spin-wave) band dispersion in both N\'eel and canted phases.
The magnon representations of spin operators are important for computing the spin current.
A simple model for conduction electrons in a normal metal is also introduced.
In Sec. \ref{Sec:SpinCurrent}, based on the Keldysh Green's function method, we derive the formula of the DC spin current flowing from the antiferromagnet to the normal metal via an interfacial exchange interaction.
The formula demonstrates that in the canted phase, the DC spin current has two main components: a magnon component dependent on temperature and a transverse magnetization component almost independent of temperature.
We explain that the sign of the spin current is mainly determined by the magnons' density of states.
The main results of this paper are in Secs.~\ref{Sec:SSE} and~\ref{Sec:Comparison}.
In Sec.~\ref{Sec:SSE}, we estimate the magnetic-field and temperature dependence of the tunneling DC spin current.
We show that the sign of the SSE voltage reverses at the first-order spin-flop transition owing to the change of the dominant spin-current carrier between spin-up and spin-down magnons.
The computed tunneling spin current semi-quantitatively agrees with a recent experimental result for the SSE in the antiferromagnet $\rm Cr_2O_3$~\cite{Li2020a}.
We also predict that in the low-temperature regime, the spin current of the canted phase exhibits a non-monotonic magnetic-field dependence.
In Sec.~\ref{Sec:Comparison}, we quantitatively compare the SSE in antiferromagnets with that in ferromagnets using the tunneling spin-current formalism.
The result is in good agreement with typical experimental results for SSEs.
In addition, we compare SSE results in various types of magnets.
From the diverse behavior of SSE voltages, we argue that the SSE can be used not only for spintronics function but also as a new probe to detect features of different magnets.
Section~\ref{Sec:Discussion} is devoted to the argument about missing parts of the tunneling spin-current formalism.
From this, we point out some important issues associated with the SSE.
Finally, we simply summarize our results in Sec.~\ref{Sec:Conclusion}.
Details of several calculations in this paper are given in the Appendices.
\section{Model}
\label{Sec:Model}
\begin{figure}[t]
 \includegraphics[width=\linewidth]{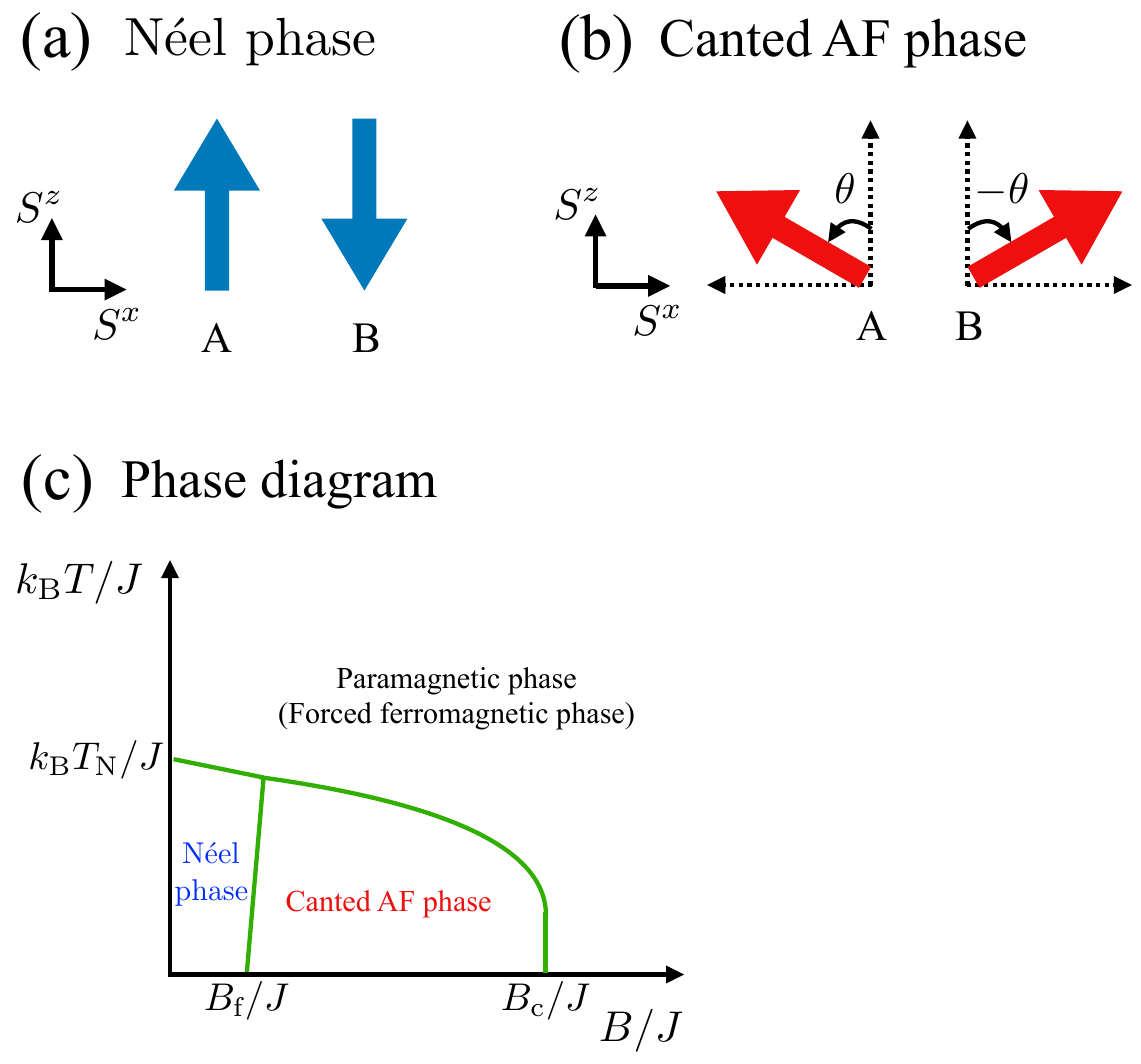}
 \caption{(Color online)
  (a),(b) Schematic views of the spin arrangement of the model Eq. (\ref{ModelHamiltonian-AFM}) in the classical ground state.
  The model Eq. (\ref{ModelHamiltonian-AFM}) has two ordered phases, (a) N\'eel phase along the $S^{z}$-axis and (b) canted AF phase in the $S^{z}$-$S^{x}$ plane.
  The magnetic structure consists of two sublattices A and B.
  Blue and red arrows represent spins localized on the sublattices.
  (c) Schematic view of the phase diagram of the model Eq. (\ref{ModelHamiltonian-AFM}) in the space $(k_{\textrm B}T,B)$.
 }
 \label{FIG:AntiferromagneticPhase}
\end{figure}
In this section, we review the basic physical nature of antiferromagnetic insulators.
In this paper, we focus on a simple Heisenberg antiferromagnet on a cubic lattice (the lattice space $a_{0}$), whose Hamiltonian is given by
\begin{equation}
 \mathscr{H}_{\textrm{AFM}}=
 J\sum_{\langle\site,\sited\rangle}\bm S_{\site}\cdot\bm S_{\sited}
 +D\sum_{\site}\qty(S_{\site}^{z})^{2}
 -B\sum_{\site}S_{\site}^{z},
 \label{ModelHamiltonian-AFM}
\end{equation}
where $\bm S_{\site}$ is the spin-$S$ operator on site $\site$, $J>0$ is the antiferromagnetic exchange coupling constant, $D<0$ is the easy-axis anisotropy, and $B=g\mu_{\textrm{B}}H\geq 0$ is the magnitude of the external magnetic field ($g$, $\mu_{\textrm{B}}$, and $H$ are the $g$ factor, the Bohr magneton, and the magnetic field, respectively).
The thermodynamic properties of this model are well understood.
In Sec.~\ref{subSec:Phases}, we briefly explain the phase diagram of Eq.~(\ref{ModelHamiltonian-AFM}), which includes N\'eel and canted ordered phases.
Then we derive the spin-wave band dispersions in both ordered phases in Secs.~\ref{subSec:Model-Neel} and \ref{subSec:Model-Cant}.
The results will be used to compute the SSE spin current.
Section~\ref{subSec:Model_Metal} defines a simple model for a bulk normal metal in the nonequilibrium steady state of the SSE setup.
\subsection{Magnetic phase diagram}
\label{subSec:Phases}
Here, we briefly summarize the phase diagram of the model (\ref{ModelHamiltonian-AFM}) in the space $(k_{\textrm B}T,B)$, where $T$ is temperature and $k_{\rm B}$ is the Boltzmann constant.
As shown in Fig.~\ref{FIG:AntiferromagneticPhase}(c),
the antiferromagnetic Heisenberg model has two ordered phases, the N\'eel and canted phases.
Let us consider the case of sufficiently low temperature with a finite $D$.
The N\'eel phase is stable against a small magnetic field $B$,
whereas the canted phase becomes stable when $B$ exceeds a certain value.
The first-order phase transition between the N\'eel and canted phases is
called the spin-flop transition.

Some phase transition points can be simply estimated from mean-field theory and the energy of a classical spin configuration on a semi-quantitative level.
The N\'eel temperature $T_{\textrm{N}}$ at zero field $B=0$ is calculated from the mean-field approximation
\begin{equation}
 k_{\textrm{B}}T_{\textrm{N}}
 =\frac{S\qty(S+1)}{3}
 \qty(6J+2|D|)
 \sim 2S\qty(S+1)J,
 \label{NeelTemperature}
\end{equation}
where we assume $|D|$ is much smaller than $J$.
The magnetic field $B_{\textrm{f}}=g\mu_{\textrm{B}} H_{\textrm{f}}$ of
the spin-flop transition at $T=0$ is computed by comparing the ground-state energies in the N\'eel and canted states.
The result is
\begin{equation}
 B_{\textrm{f}}
 =2S\sqrt{|D|\qty(6J-|D|)}
 \sim 2S\sqrt{6J|D|},
 \label{SpinFlopField}
\end{equation}
where we again assume $|D|\ll J$.
Similarly, the critical magnetic field $B_\textrm{c}$ between the canted state and fully polarized states at $T=0$ is given by
\begin{equation}
 B_{\textrm{c}}
 =2S\qty(6J-|D|).
 \label{CriticalField}
\end{equation}

Experiments on the SSE of antiferromagnets have been done on the two materials $\textrm{Cr}_{2}\textrm{O}_3$ and $\textrm{MnF}_{2}$.
Using the above estimation of the phase transitions, we can approximately determine the values of $J$ and $D$  in these compounds (although their crystal structures are not a simple cubic type).
Since the $S=3/2$ antiferromagnet $\textrm{Cr}_{2}\textrm{O}_3$ has $T_{\textrm{N}}\sim300\ \textrm{K}$ and $H_{\textrm{f}}\sim6\ \textrm{T}$,
we obtain $J\sim4\ \textrm{meV}$ and $|D|\sim5\times10^{-3}\ \textrm{meV}$.
Similarly, since the $S=5/2$ antiferromagbnet $\textrm{MnF}_{2}$ has $T_{\textrm{N}}\sim70\ \textrm{K}$ and $H_{\textrm{f}}\sim9\ \textrm{T}$,
$J\sim0.4\ \textrm{meV}$ and $|D|\sim4\times10^{-2}\ \textrm{meV}$ are obtained.
\subsection{Spin-wave dispersion in the N\'eel phase}
\label{subSec:Model-Neel}
\begin{figure}[t]
 \includegraphics[width=\linewidth]{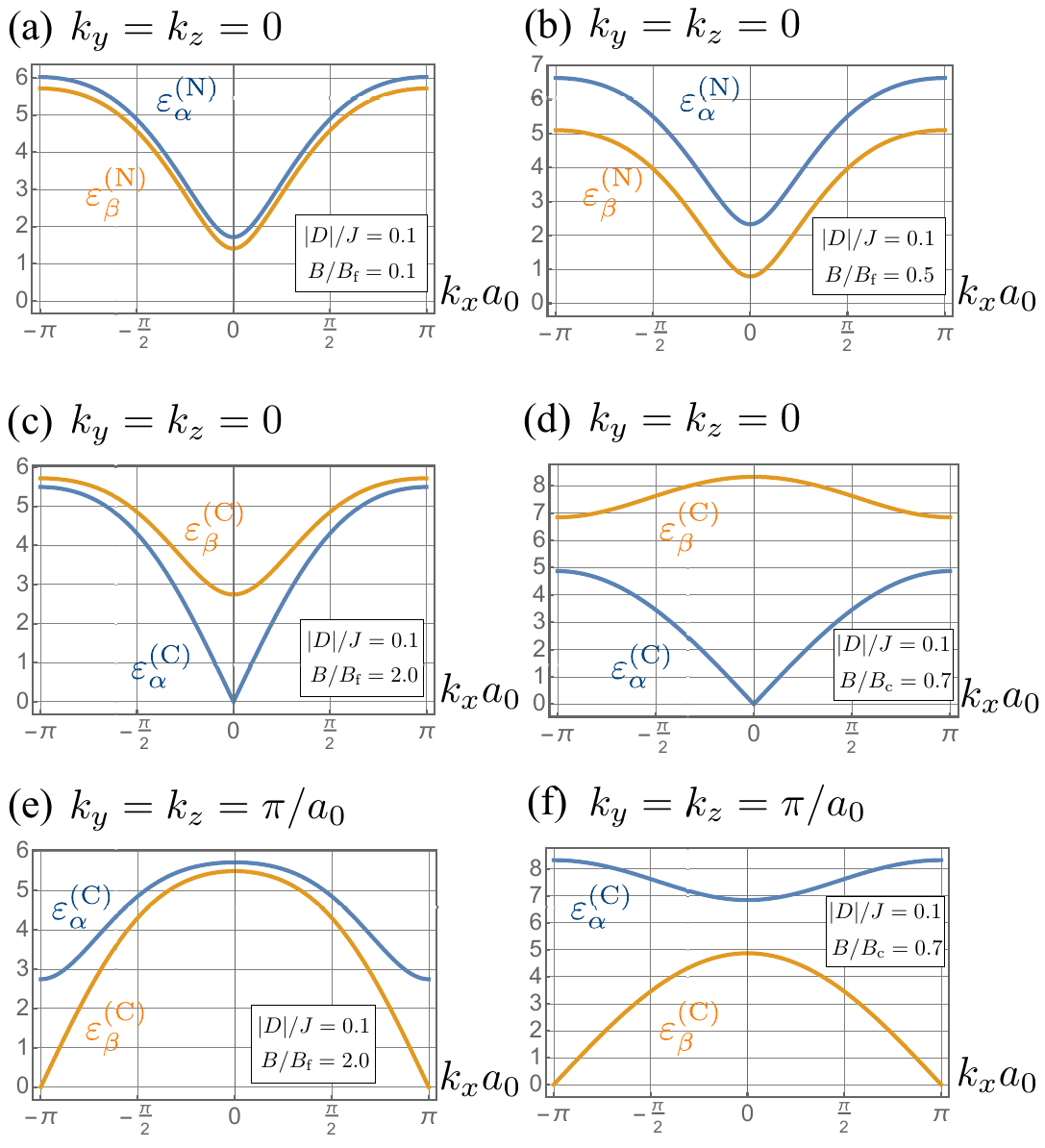}
 \caption{(Color online)
 Panels (a) and (b) are the magnon dispersions in the N\'eel phase near the $\Gamma$ point, $\wn=(0,0,0)$.
 Blue and yellow lines respectively correspond to the $\alpha$ mode $\varepsilon_{\alpha}^{(\textrm{N})}\qty(\wn)/J$ and $\beta$ mode $\varepsilon_{\beta}^{(\textrm{N})}\qty(\wn)/J$.
 Panels (c)-(f) are the magnon dispersions in the canted phase near the $\Gamma$ point, $\wn=(0,0,0)$ [(c) and (d)] and near the Brillouin zone edge, $\wn=(\pi/a_{0},\pi/a_{0},\pi/a_{0})$ [(e) and (f)].
 Blue and yellow lines respectively correspond to $\varepsilon_{\alpha}^{(\textrm{C})}\qty(\wn)/J$ and $\varepsilon_{\beta}^{(\textrm{C})}\qty(\wn)/J$.
 In all the panels, parameters are set to $S=1$, $a_{0}=1$, and $J=1$.
 In the canted phase, we have two gapless points, $\varepsilon_{\alpha}^{(\textrm{C})}\qty(0,0,0)=0$ and $\varepsilon_{\beta}^{(\textrm{C})}\qty(\pi/a_{0},\pi/a_{0},\pi/a_{0})=0$ owing to the spontaneous breaking of the U(1) symmetry around the $S^z$ axis.
 }
 \label{FIG:MagnonDispersion-AFHeisenberg}
\end{figure}
In this subsection, we summarize the spin-wave theory for the N\'eel phase of Eq.~(\ref{ModelHamiltonian-AFM})~\cite{Kubo1952,Yosida}.
The ordered state consists of two sublattices A and B, and the spin moment is parallel to the $z$ axis owing to the anisotropy $D$, as shown in Fig.~\ref{FIG:AntiferromagneticPhase}(a).
Following the standard Holstein-Primakoff (HP) transformation, we can approximate the spin operators on A and B sublattices as
\begin{align}
  & S_{\siteA}^{z}=S-a_{\siteA}^{\dagger}a_{\siteA}, \,\,\,\,\,\,\,\,  
 S_{\siteA}^{+}\simeq\sqrt{2S}a_{\siteA},\,\,\,\,\,\,\,\,
 S_{\siteA}^{-}\simeq\sqrt{2S}a_{\siteA}^{\dagger},
 \nonumber                                                             \\
  & S_{\siteB}^{z}=-S+b_{\siteB}^{\dagger}b_{\siteB}, \,\,\,\,\,\,\,\, 
 S_{\siteB}^{+}\simeq\sqrt{2S}b_{\siteB}^{\dagger},\,\,\,\,\,\,\,\,
 S_{\siteB}^{-}\simeq\sqrt{2S}b_{\siteB},
 \label{LSWA-Neel}
\end{align}
where $a_{\siteA}^{\dagger}$ and $a_{\siteA}$
($b_{\siteB}^{\dagger}$ and $b_{\siteB}$) are, respectively, the creation and annihilation operators of magnons on an A-sublattice site $\siteA$ (B-sublattice site $\siteB$).
The Fourier transformations of magnon operators are defined as
\begin{align}
  & a_{\wn}= 
 \sqrt{\frac{2}{N}}
 \sum_{\siteA}e^{-i\wn\cdot\siteA}a_{\siteA},\,\,\,\,\,\,\,\,
 a_{\wn}^{\dagger}=
 \sqrt{\frac{2}{N}}
 \sum_{\siteA}e^{i\wn\cdot\siteA}a_{\siteA}^{\dagger},
 \nonumber   \\
  & b_{\wn}= 
 \sqrt{\frac{2}{N}}
 \sum_{\siteB}e^{i\wn\cdot\siteB}b_{\siteB},\,\,\,\,\,\,\,\,
 b_{\wn}^{\dagger}=
 \sqrt{\frac{2}{N}}
 \sum_{\siteB}e^{-i\wn\cdot\siteB}b_{\siteB}^{\dagger},
 \label{FourierTransformation-Neel}
\end{align}
where $N$ is the total number of sites in the antiferromagnet, and $\wn=\qty(k_{x},k_{y},k_{z})$ is the wave number in the first Brillouin zone of the sublattice.
Substituting Eqs.~(\ref{LSWA-Neel}) and
(\ref{FourierTransformation-Neel}) into the model~(\ref{ModelHamiltonian-AFM}) and neglecting magnon interactions under the assumption that the magnon density is low enough, we obtain the low-energy
spin-wave Hamiltonian.
Through a suitable Bogoliubov transformation, we finally arrive at the diagonalized spin-wave Hamiltonian
\begin{equation}
 \mathscr{H}_{\textrm{SW}}^{(\textrm{N\'eel})}=
 \sum_{\wn}\varepsilon^{(\textrm{N})}_{\alpha}\qty(\wn)\alpha_{\wn}^{\dagger}\alpha_{\wn}
 +\sum_{\wn}\varepsilon^{(\textrm{N})}_{\beta}\qty(\wn)\beta_{\wn}^{\dagger}\beta_{\wn}
 +(\textrm{const.}),
 \label{SpinWave_Neel_Diagonalization}
\end{equation}
where $\alpha_{\wn}$ and $\beta_{\wn}$ are the magnon annihilation operators, and the corresponding two magnon (spin-wave) dispersions are given by
\begin{align}
  & \varepsilon^{(\textrm{N})}_{\alpha}\qty(\wn)= 
 S\sqrt{\qty(6J+2|D|)^{2}-\qty(2J\gamma_{\wn})^{2}}+B ,
 \label{MagnonDispersion-AlphaMode-Neel}          \\
  & \varepsilon^{(\textrm{N})}_{\beta}\qty(\wn)=  
 S\sqrt{\qty(6J+2|D|)^{2}-\qty(2J\gamma_{\wn})^{2}} -B.
 \label{MagnonDispersion-BetaMode-Neel}
\end{align}
The parameter $\gamma_{\wn}$ is defined as
\begin{equation}
 \gamma_{\wn}  =\cos\qty(k_{x}a_{0})+\cos\qty(k_{y}a_{0})+\cos\qty(k_{z}a_{0}).
 \label{Cosine}
\end{equation}

In the magnon description, the spin operators are represented as
\begin{align}
  & S_{\siteA}^{+}\simeq                          
 \sqrt{2S}
 \sqrt{\frac{1}{N/2}}\sum_{\wn}e^{i\wn\cdot\siteA}\qty(\cosh\phi_{\wn}\alpha_{\wn}+\sinh\phi_{\wn}\beta_{\wn}^{\dagger}),
 \label{MagnonRepresentation-AsublatticeNeelSpin} \\
  & S_{\siteB}^{+}\simeq                          
 \sqrt{2S}
 \sqrt{\frac{1}{N/2}}\sum_{\wn}e^{i\wn\cdot\siteB}\qty(\sinh\phi_{\wn}\alpha_{\wn}+\cosh\phi_{\wn}\beta_{\wn}^{\dagger}),
 \label{MagnonRepresentation-BsublatticeNeelSpin}
\end{align}
where we neglect the terms corresponding to more than two magnons.
The angle $\phi_{\wn}$ is used in the Bogoliubov transformation and is determined by
\begin{equation}
 \tanh\qty(2\phi_{\wn})=\frac{-J\gamma_{\wn}}{3J+|D|}.
 \label{BogoliubovTransformation-NeelPhase-Condition}
\end{equation}
These representations of the spins are very essential to compute the spin current of the SSE.

We show some typical magnon bands of the N\'eel state in Figs.~\ref{FIG:MagnonDispersion-AFHeisenberg}(a) and (b).
In the N\'eel phase, two spin-wave modes have opposite spin polarizations (angular momenta).
Since the $\alpha$-mode magnon has the down-spin polarization and the $\beta$-mode one has the up-spin polarization, as the field increases, $\varepsilon^{(\textrm{N})}_{\alpha}\qty(\wn)$ increases while $\varepsilon^{(\textrm{N})}_{\beta}\qty(\wn)$ decreases.
In the zero field case $B=0$, the two modes are degenerate.
The N\'eel phase spontaneously breaks one-cite translation symmetry, whereas the SU(2) spin-rotation symmetry is also spontaneously broken at $B=D=0$.
In the latter case, the two spin-wave excitations become gapless, which correspond to the Nambu-Goldstone modes.
\subsection{Spin-wave dispersion in the canted AF phase}
\label{subSec:Model-Cant}
Here, we summarize the result of the spin-wave theory
for the canted phase, in which we assume that the spin moment is in the $S^{z}$-$S^{x}$ plane.
This phase also has two sublattices A and B, as shown in Fig.~\ref{FIG:AntiferromagneticPhase}(b).
From the condition of minimizing the ground-state energy, the canted angle $\theta$ defined in Fig.~\ref{FIG:AntiferromagneticPhase}(b) is given by
\begin{equation}
 \cos\theta=
 \frac{B}{2S\qty(6J-|D|)}.
 \label{CantedAngle}
\end{equation}
Based on this classical ground state, we can obtain the spin-wave Hamiltonian by taking into account the quantum fluctuation around the canted moment.
The resultant Hamiltonian is given by
\begin{equation}
 \mathscr{H}_{\textrm{SW}}^{(\textrm{Cant})}=
 \sum_{\wn}\varepsilon^{(\textrm{C})}_{\alpha}\qty(\wn)\alpha_{\wn}^{\dagger}\alpha_{\wn}
 +\sum_{\wn}\varepsilon^{(\textrm{C})}_{\beta}\qty(\wn)\beta_{\wn}^{\dagger}\beta_{\wn}
 +(\textrm{const.}),
 \label{SpinWaveHamiltonian-CantedPhase}
\end{equation}
where the two magnon dispersions are
\begin{widetext}
 \begin{align}
   & \varepsilon^{(\textrm{C})}_{\alpha}\qty(\wn)= 
  \sqrt{
   2JS\qty(3-\gamma_{\wn})
   \qty\Big{
    2JS(3+\gamma_{\wn})
    -2|D|S
    +B\cos\theta
    -4JS(3+\gamma_{\wn})\cos^{2}\theta
    +4|D|S\cos^{2}\theta
   }
  } ,
  \label{MagnonDispersion-AlphaMode-Canted}        \\
   & \varepsilon^{(\textrm{C})}_{\beta}\qty(\wn)=  
  \sqrt{
   2JS\qty(3+\gamma_{\wn})
   \qty\Big{
    2JS(1-2\cos^{2}\theta)(3-\gamma_{\wn})
    -2|D|S(1-2\cos^{2}\theta)
    +B\cos\theta
   }
  } .
  \label{MagnonDispersion-BetaMode-Canted}
 \end{align}
 Using these magnon operators, we can represent the spin operators as
 \begin{multline}
  S_{\siteA}^{+}
  \simeq
  \frac{\sqrt{S}}{2}
  \qty(1+\cos\theta)
  \sqrt{\frac{1}{N/2}}\sum_{\wn}e^{i\wn\cdot\siteA}
  \qty(\cosh\varphi_{\wn}^{\alpha}\alpha_{\wn}
  +\sinh\varphi_{\wn}^{\alpha}\alpha_{-\wn}^{\dagger}
  +\sinh\varphi_{\wn}^{\beta}\beta_{\wn}^{\dagger}
  +\cosh\varphi_{\wn}^{\beta}\beta_{-\wn})
  \\
  -\frac{\sqrt{S}}{2}
  \qty(1-\cos\theta)
  \sqrt{\frac{1}{N/2}}\sum_{\wn}e^{-i\wn\cdot\siteA}
  \qty(\cosh\varphi_{\wn}^{\alpha}\alpha_{\wn}^{\dagger}
  +\sinh\varphi_{\wn}^{\alpha}\alpha_{-\wn}
  +\sinh\varphi_{\wn}^{\beta}\beta_{\wn}
  +\cosh\varphi_{\wn}^{\beta}\beta_{-\wn}^{\dagger})
  -S\sin\theta,
  \label{MagnonRepresentation-AsublatticeCantedAFSpin}
 \end{multline}
 \begin{multline}
  S_{\siteB}^{+}
  \simeq
  -\frac{\sqrt{S}}{2}
  \qty(1+\cos\theta)
  \sqrt{\frac{1}{N/2}}\sum_{\wn}e^{-i\wn\cdot\siteB}
  \qty(\sinh\varphi_{\wn}^{\alpha}\alpha_{\wn}^{\dagger}
  +\cosh\varphi_{\wn}^{\alpha}\alpha_{-\wn}
  -\cosh\varphi_{\wn}^{\beta}\beta_{\wn}
  -\sinh\varphi_{\wn}^{\beta}\beta_{-\wn}^{\dagger})
  \\
  +\frac{\sqrt{S}}{2}
  \qty(1-\cos\theta)
  \sqrt{\frac{1}{N/2}}\sum_{\wn}e^{i\wn\cdot\siteB}
  \qty(\sinh\varphi_{\wn}^{\alpha}\alpha_{\wn}
  +\cosh\varphi_{\wn}^{\alpha}\alpha_{-\wn}^{\dagger}
  -\cosh\varphi_{\wn}^{\beta}\beta_{\wn}^{\dagger}
  -\sinh\varphi_{\wn}^{\beta}\beta_{-\wn})
  +S\sin\theta,
  \label{MagnonRepresentation-BsublatticeCantedAFSpin}
 \end{multline}
\end{widetext}
where we neglect higher-order magnon terms.
The angles $\varphi_{\wn}^{\alpha}$ and $\varphi_{\wn}^{\beta}$ are given by
\begin{align}
  & \tanh\qty(2\varphi_{\wn}^{\alpha})= 
 \frac{-2\textrm{C}_{4}+\textrm{C}_{2}(\wn)}{\textrm{C}_{1}-\textrm{C}_{3}(\wn)},
 \nonumber                              \\
  & \tanh\qty(2\varphi_{\wn}^{\beta})=  
 \frac{-2\textrm{C}_{4}-\textrm{C}_{2}(\wn)}{\textrm{C}_{1}+\textrm{C}_{3}(\wn)},
 \label{BogoliubovTrans-CantedPhase}
\end{align}
where $\textrm{C}_{1-4}$ are defined in Eq.~(\ref{App_ParametersC1_4}) (see App.~\ref{App:BGT_Canted}).

Figures~\ref{FIG:MagnonDispersion-AFHeisenberg}(c)-(f) depict typical magnon bands in the canted state.
In this phase, both the spin-wave modes ($\alpha$ and $\beta$ modes) averagely possess the down-spin polarization (negative angular momentum along the $z$ axis) because an uniform magnetization exists.
The $\alpha$ mode is gapless around $\bm k=(0,0,0)$, whereas the $\beta$ mode is so around $\wn=(\pi/a_{0},\pi/a_{0},\pi/a_{0})$.
This gaplessness originates from the spontaneous breaking of the U(1) spin-rotation symmetry around the $S^z$ axis.
Comparing Figs.~\ref{FIG:MagnonDispersion-AFHeisenberg}(c) and (d) [(e) and (f)], one sees that the band shape, especially the higher-energy band, strongly depends on the value of the magnetic field $B$.
\subsection{Normal metal}
\label{subSec:Model_Metal}
Here, we introduce the model for conduction electrons in the attached metal of Fig.~\ref{FIG:SSE-Antiferromagnets}.
We define a simple free-electron model for the metal, and its Hamiltonian is given by
\begin{equation}
 H_{\textrm{NM}}
 =\sum_{\bm q,\alpha}
 \epsilon_{\bm q}f_{\bm q,\alpha}^{\dagger}f_{\bm q,\alpha},
 \label{Dynamics_Metal}
\end{equation}
where $f_{\bm q,\alpha}$ $\qty(f_{\bm q,\alpha}^{\dagger})$ is the annihilation (creation) operator of a conducting electron with wave vector $\bm q$ and spin $\alpha=\uparrow,\downarrow$, and
$\epsilon_{\bm q}=\qty(\hbar^{2}q^{2})/2m$ is the kinetic energy ($q=|\bm q|$).
In a real setup, the metal possesses a finite spin-orbit (SO) interaction to yield the ISHE, but we assume that the effects of the SO interaction are negligible when we estimate the tunneling spin current.

The quantum dynamics (i.e., time evolution) of conducting electrons obeys the Hamiltonian~(\ref{Dynamics_Metal}) (e.g., $f_{\bm q,\alpha}(t)=e^{iH_{\textrm{NM}}t/\hbar}f_{\bm q,\alpha}e^{-iH_{\textrm{NM}}t/\hbar}$).
However, in a bilayer (or multilayer) system consisting of a normal metal and a magnet, a spin-dependent chemical potential~\cite{Cornelissen2016,Maekawa2017} appears in the vicinity of the interface when an nonequilibrium steady state (NESS) is realized by applying an external force (an electric field, a temperature gradient, etc.) to the system.
Therefore, to describe such NESSs, we also define another Hamiltonian for the metal~\cite{Kato2020}:
\begin{equation}
 \mathcal{H}_{\textrm{NM}}=
 \sum_{\bm q}\qty(\xi_{\bm q}-\frac{\delta\mu_{\textrm{s}}}{2})f_{\bm q,\uparrow}^{\dagger}f_{\bm q,\uparrow}
 +\sum_{\bm q}\qty(\xi_{\bm q}+\frac{\delta\mu_{\textrm{s}}}{2})f_{\bm q,\downarrow}^{\dagger}f_{\bm q,\downarrow},
 \label{ThermalAverage_Metal}
\end{equation}
where $\mu_{\alpha}$ ($\alpha=\uparrow,\downarrow$) is the spin-dependent chemical potential, $\xi_{\bm q}=\epsilon_{\bm q}-\mu$ is the kinetic energy measured from the averaged chemical potential $\mu=\qty(\mu_{\uparrow}+\mu_{\downarrow})/2$, and $\delta\mu_{\textrm{s}}=\mu_{\uparrow}-\mu_{\downarrow}$ is the potential difference on the interface.
In the NESS of the metal, the thermal average should be taken using the second Hamiltonian $\mathcal{H}_{\textrm{NM}}$ as $\langle\cdots\rangle=\textrm{Tr}\qty(e^{-\beta\mathcal{H}_{\textrm{NM}}}\cdots)/\textrm{Tr}\qty(e^{-\beta\mathcal{H}_{\textrm{NM}}})$~\cite{Kato2020}.
We note that in the SSE setup, the conduction electrons relax to the NESS through dissipation process, which cannot be described by simple free-electron models such as Eq.~(\ref{Dynamics_Metal}).

The magnitude of $\delta\mu_{\textrm{s}}$ is known to be much smaller than the other typical energy scales of the bilayer SSE setup such as magnetic exchanges and the Fermi energy (or band width).
For instance, Ref.~\cite{Cornelissen2016} estimates $\delta\mu_{\textrm{s}}\sim\mathcal{O}(10^{1})$ \textmu\textrm{V} at the interface of YIG and Pt.
Therefore, effects of $\delta\mu_{\textrm{s}}$ are often negligible, but (as we will explain later) a finite $\delta\mu_{\textrm{s}}$ will play a crucial role when we consider the SSE in the canted phase.

In the SSE setup, one applies a static magnetic field and hence a Landau-level structure seems to emerge in the attached metal. However, a dirty metal with impurities and a polycrystalline structure are usually used in SSE experiments, and the energy scales of the metal (the Fermi energy, the Coulomb interaction, the impurity potential, etc.) are much larger than those of the applied magnetic field and the exchange interaction of the magnet. Thus, it is probably difficult to observe fine energy scales such as the Landau levels from SSE signals.
\section{Tunneling Spin Current}
\label{Sec:SpinCurrent}
This section explains the generic formula of tunneling spin current from a magnet to a metal, which will be applied to the SSE of antiferromagnets in the next section.
The calculation details are in App.~\ref{App:PerturbativeCalculation}, and here, we mainly discuss the important nature of the formula.
We note that if we apply a temperature gradient in the setup of Fig.~\ref{FIG:SSE-Antiferromagnets} and a long time passes, a NESS is realized. In the present study, we will always focus on such a NESS with a finite tunneling spin current.

As we briefly mentioned in the Introduction, the tunneling spin current is proportional to the measured SSE voltage in the metal~\cite{Saitoh2006,Valenzuela2006,Kimura2007}.
Therefore, one can understand the dependences of several parameters (such as magnetic field and temperature) on the spin current from the SSE voltage.
Additionally, it is enough to compute the tunneling spin current instead of the SSE voltage.
In Secs.~\ref{SubSec:SpinCurrent}- \ref{SubSec:SpinCurrent_Cant}, based on the approach of non-equilibrium Green's function, we derive the formula of the thermally generated DC spin current tunneling from an antiferromagnetic insulator to a paramagnetic metal via the interfacial interaction~\cite{Jauho1994,Adachi2011,Ohnuma2013}.
This type of tunneling-current formula was first developed in mesoscopic physics to analyze the tunneling electric current from left to right leads through a quantum dot~\cite{Jauho1994}.
Adachi, {\it et al}, applied this approach to the SSE in bilayer systems consisting of a ferromagnet and a metal~\cite{Adachi2011}.
The driving force for electric current is a chemical potential gradient (i.e., electric field) in the dot systems, whereas that for spin current is a temperature gradient.
This spin-current formula has been applied and developed to SSEs of different magnets such as a 1D spin liquid~\cite{Hirobe2017d}, a spin-nematic liquid~\cite{Hirobe2019}, and a spin-Peierls state~\cite{Chen2021}.
The formula has succeeded in explaining the magnetic-field dependence of these SSEs on a semi-quantitative level.
We will apply it to the SSE in antiferromagnets.

In Sec.~\ref{Sec:DoS}, we show that the tunneling spin current and its sign can be interpreted from the viewpoint of the magnon's density of states, which can be observed via inelastic-neutron scattering experiments.
This interpretation is essential for understanding the magnetic-field dependence of the SSE in antiferromagnets in the next section.
\subsection{Perturbation calculation of tunneling spin current}
\label{SubSec:SpinCurrent}
We focus on the bilayer system in Fig.~\ref{FIG:SSE-Antiferromagnets}, which consists of an antiferromagnet and a normal metal.
These two materials are weakly interacting through an s-d exchange interaction at the interface.
In a real experiment, the temperature is smoothly changed as a function of the coordinate $x$, but here, we approximate such a varying temperature as two representative values, the averaged temperature of the antiferromagnet, $T_{\textrm{s}}$, and that of the normal metal, $T_{\textrm{m}}$, and we assume $T_{\textrm{s}}>T_{\textrm{m}}$ for simplicity.
We consider a spin current injected from the magnet to the metal in the non-equilibrium steady state with a temperature difference of $T_{\textrm{s}}>T_{\textrm{m}}$.
\begin{figure}[t]
 \includegraphics[width=\linewidth]{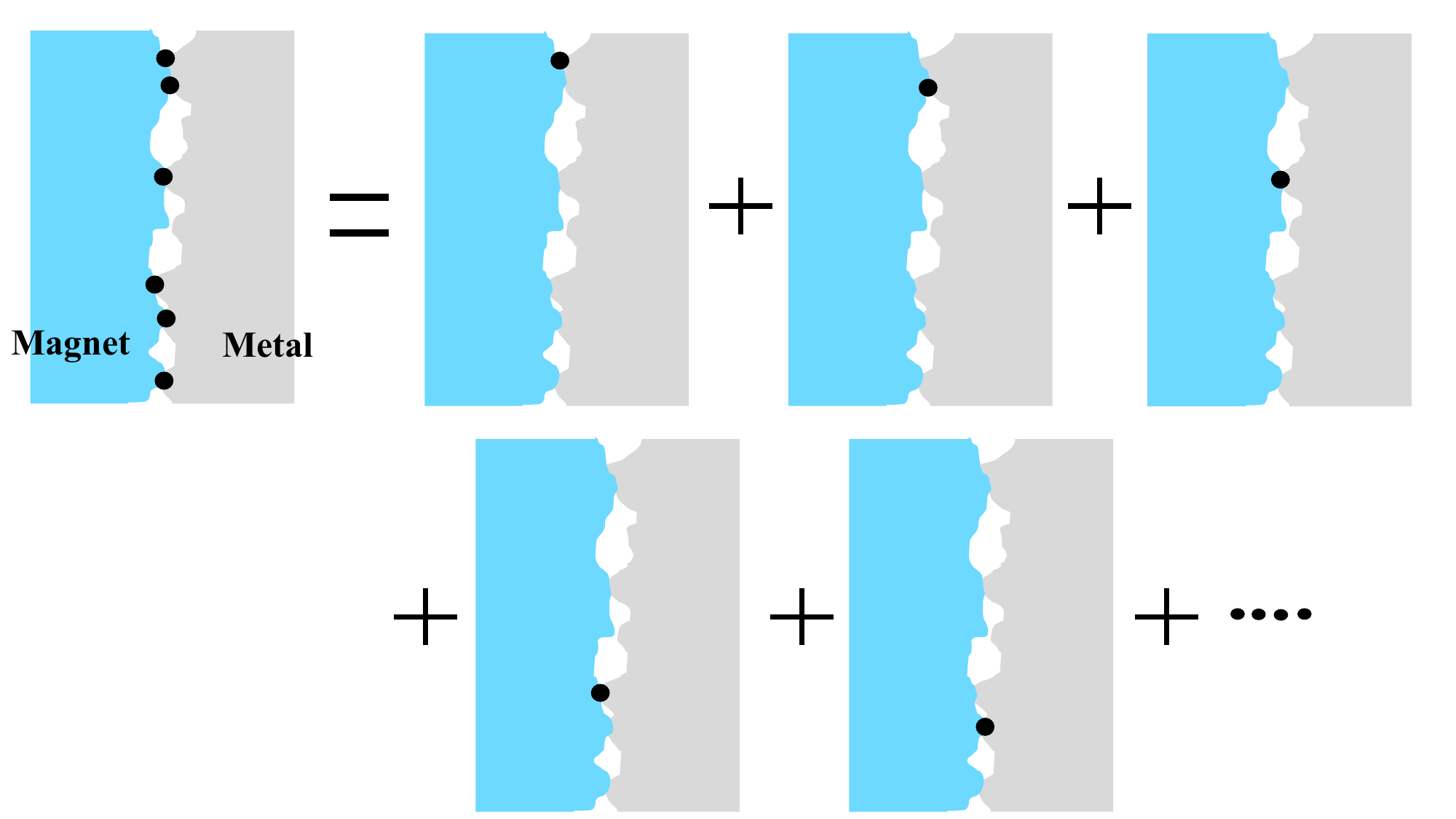}
 \caption{(Color online)
  Schematic image of randomly distributed exchange interactions (black points) on the interface between a magnet and a metal.
  The total number of interface sites is supposed to be much smaller than that of sites in the magnet or the metal near the interface, and thus the correlation between neighboring sites on the interface is expected to be quite weak.
  Under this assumption, the tunneling spin current is described by the sum of the tunneling spin currents through each site on the interface.
 }
 \label{Fig:Interface}
\end{figure}

The total Hamiltonian of the bilayer system consists of the following three parts:
\begin{equation}
 \mathscr{H}=
 \mathscr{H}_{\textrm{AFM}}
 +\mathscr{H}_{\textrm{NM}}
 +\mathscr{H}_{\textrm{int}}.
 \label{System-Hamiltonian}
\end{equation}
The first term $\mathscr{H}_{\textrm{AFM}}$ is the antiferromagnetic Heisenberg model.
The second term $\mathscr{H}_{\textrm{NM}}$ represents the Hamiltonian of conduction electrons in the normal metal.
The third term $\mathscr{H}_{\textrm{int}}$ is the s-d exchange interaction at the interface, and we assume that the spins localized on the A sublattice and B one couple with the conduction electrons with equal weight, respectively:
\begin{equation}
 \mathscr{H}_{\textrm{int}}  =
 \sum_{\siteA\in\interA}
 J_{\textrm{sd}}\qty(\siteA)
 \bm S_{\siteA}\cdot\bm\sigma_{\siteA}
 +\sum_{\siteB\in\interB}
 J_{\textrm{sd}}\qty(\siteB)
 \bm S_{\siteB}\cdot\bm\sigma_{\siteB},
 \label{ExchngeInteraction-Interface}
\end{equation}
where $J_{\textrm{sd}}\qty(\bm r_{\textrm{X}})$ is the s-d exchange coupling constant at a $\bm r_{\textrm{X}}$ site (X=A or B) on the interface, and $\sum_{\bm r \in\textrm{int-X}}$ stands for summation w.r.t. all the X-sublattice sites on the interface.
The symbol $\bm{\sigma_{\site_{\textrm{X}}}}$ denotes the conduction-electron spin of the metal coupled to the localized spin of the magnet $\bm S_{\site_{\textrm{X}}}$ at an X-sublattice site $\bm r_{\textrm{X}}$ on the interface.

We define the tunneling DC spin-current operator as the time derivative of the spin-polarization density of conduction electrons at the interface:
\begin{align}
 I_{\textrm{S}} & =                                                        
 \sum_{\siteA\in\interA}
 \pdv{t}\sigma_{\siteA}^{z}(t)
 +\sum_{\siteB\in\interB}
 \pdv{t}\sigma_{\siteB}^{z}(t)
 \nonumber                                                                 \\
                & =\qty(-\frac{i}{\hbar})\sum_{\siteA\in\interA}           
 \comm{\sigma_{\siteA}^{z}(t)}{\mathscr{H}}
 \nonumber                                                                 \\
                & \quad\quad+\qty(-\frac{i}{\hbar})\sum_{\siteB\in\interB} 
 \comm{\sigma_{\siteB}^{z}(t)}{\mathscr{H}},
 \label{TunnelingDCSpinCurrent-Operator}
\end{align}
where $\sigma_{\site}^{z}(t)=e^{i\mathscr{H}t/\hbar}\sigma_{\site}^{z}e^{-i\mathscr{H}t/\hbar}$, and $\sigma_{\site}^{z}=\frac{1}{2}\qty(f_{\site,\uparrow}^{\dagger}f_{\site,\uparrow}-f_{\site,\downarrow}^{\dagger}f_{\site,\downarrow})$ with $f_{\site,\alpha}$ and $f_{\site,\alpha}^{\dagger}$ being creation and annihilation operators of conducting electrons with spin $\alpha=\uparrow,\downarrow$ in the metal, respectively.
Using the commutation relation $\comm{\sigma_{\site}^{z}}{\sigma_{\sited}^{\pm}}=\pm\delta_{\site,\sited}\sigma_{\site}^{\pm}$, we obtain
\begin{align}
 I_{\textrm{S}}=
  & \sum_{\siteA\in\interA}  
 \frac{J_{\textrm{sd}}\qty(\siteA)}{2\hbar}
 \qty{(-i)S_{\siteA}^{-}(t)\sigma_{\siteA}^{+}(t)+\qty(\textrm{H.c})}
 \nonumber                   \\
  & +\sum_{\siteB\in\interB} 
 \frac{J_{\textrm{sd}}\qty(\siteB)}{2\hbar}
 \qty{(-i)S_{\siteB}^{-}(t)\sigma_{\siteB}^{+}(t)+\qty(\textrm{H.c})}.
 \label{TunnelingDCSpinCurrent-Operator-ver2}
\end{align}
From Eq.~(\ref{TunnelingDCSpinCurrent-Operator-ver2}), the statistical average of $I_{\textrm{S}}$ under the non-equilibrium steady state in the SSE setup is given by
\begin{align}
 \langle I_{\textrm{S}}\rangle
 = & \sum_{\siteA\in\interA} 
 \frac{J_{\textrm{sd}}\qty(\siteA)}{\hbar}
 \textrm{Re}
 \qty[
  (-i)
  \Big\langle
  S_{\siteA}^{-}(t)\sigma_{\siteA}^{+}(t)
  \Big\rangle
 ]
 \nonumber                   \\
   & +                       
 \sum_{\siteB\in\interB}
 \frac{J_{\textrm{sd}}\qty(\siteB)}{\hbar}
 \textrm{Re}
 \qty[
  (-i)
  \Big\langle
  S_{\siteB}^{-}(t)\sigma_{\siteB}^{+}(t)
  \Big\rangle
 ]\nonumber                  \\
   & =                       
 \sum_{\siteA\in\interA}
 \langle I_{\siteA}\rangle
 +
 \sum_{\siteB\in\interB}
 \langle I_{\siteB}\rangle,
 \label{Expectation}
\end{align}
where $\langle\cdots\rangle$ denotes the statistical average for the total Hamiltonian Eq.~(\ref{System-Hamiltonian}) under the non-equilibrium SSE setup, and $I_{\siteA}$ ($I_{\siteB}$) is the spin current tunneling through a sublattice-A (B) site $\siteA$ ($\siteB$) on the interface.

In general, the s-d interaction at the interface is considered to be weaker than the energy scale of magnets and metals.
Hence, taking $\mathscr{H}_{\magnet}+\mathscr{H}_{\metal}$ as the unperturbed Hamiltonian and $\mathscr{H}_{\inter}$ as the perturbation, we may apply the approach of the non-equilibrium Green's function~\cite{HaugJauhoSpringer2008,StefanucciCambridge2013} to Eq.~(\ref{Expectation}).
Note that in this perturbation theory, the average $\langle\cdots\rangle$ of the unperturbed system is equivalent to the thermal average for the decoupled antiferromagnet with $T=T_{\rm s}$ and the metal with $T=T_{\rm m}$.
We here assume that interface sites $\siteA$ and $\siteB$ are randomly distributed and the averaged distance between neighboring interface sites is much longer than the lattice spaces of the antiferromagnet and the metal.
This means the interface is assumed to be very dirty, as shown in Fig.~\ref{Fig:Interface}.
Under this condition, we can neglect the correlation between the local tunneling spin currents, $I_{\site_{\textrm{X}}}$ and $I_{{\site'}_{\textrm{X}'}}$ ($\site_\textrm{X}\neq {\site'}_{\textrm{X}'}$).
Therefore, we can compute the average of each local spin current $\langle I_{\site_{\textrm{X}}}\rangle$ independently, and the total spin current is given by the sum of the independent local currents (see Fig.~\ref{Fig:Interface}).

To proceed the perturbation calculation, we here define some correlation functions.
The local spin current is re-expressed as
\begin{align}
 \langle I_{\siteA}\rangle
  & =                                 
 \frac{J_{\textrm{sd}}\qty(\siteA)}{\hbar}
 \lim_{\delta\to+0}
 \textrm{Re}
 \qty[
  F_{+-}^{<}\qty(\siteA,t;\siteA,t')
 ],
 \label{LocalSpinCurrent-Asublattice} \\
 \langle I_{\siteB}\rangle
  & =                                 
 \frac{J_{\textrm{sd}}\qty(\siteB)}{\hbar}
 \lim_{\delta\to+0}
 \textrm{Re}
 \qty[
  F_{+-}^{<}\qty(\siteB,t;\siteB,t')
 ],
 \label{LocalSpinCurrent-Bsublattice}
\end{align}
where $t'=t+\delta$, and $F_{+-}^{<}\qty(\site,t;\sited,t')$ is the lesser component of the two-point function $F_{+-}\qty(\site,t;\sited,t')=-i\big\langle T_{\textrm{C}}\sigma_{\site}^{+}(t)S_{\sited}^{-}(t')\big\rangle$ with $T_{\textrm{C}}$ being the time-ordered product on the Keldysh contour \cite{HaugJauhoSpringer2008,StefanucciCambridge2013}.
The spin correlation function of the decoupled metal and that of the antiferromagnet are respectively defined as
\begin{align}
 \chi_{+-}\qty(\site,t;\sited,t')
  & =-i\big\langle T_{\textrm{C}}\tilde{\sigma}_{\site}^{+}(t)\tilde{\sigma}_{\sited}^{-}(t')\big\rangle_{0},                 
 \label{SpinCorrelationFunction_Metal}
 \\
 G_{+-}^{(\textrm{X})}\qty(\site_{\textrm{X}},t;\sited_{\textrm{X}},t')
  & =-i\big\langle T_{\textrm{C}}\tilde{S}_{\site_{\textrm{X}}}^{+}(t)\tilde{S}_{\sited_{\textrm{X}}}^{-}(t')\big\rangle_{0}, 
 \label{SpinCorrelationFunction_Magnet}
\end{align}
where $\langle\cdots\rangle_{0}$ stands for the statistical average for the unperturbed Hamiltonian and $\textrm{X}=\textrm{A},\ \textrm{B}$.
The symbol $\tilde{}$ stands for time evolution under the unperturbed Hamiltonian.
Using these correlators and the Langreth rule~\cite{HaugJauhoSpringer2008,StefanucciCambridge2013}, we arrive at the following expression of the local spin current (for more detail, see App.~\ref{App:PerturbativeCalculation}):
\begin{widetext}
 \begin{align}
  \langle I_{\siteA}\rangle
   & =            
  \frac{1}{2}
  \qty(\frac{J_{\textrm{sd}}\qty(\siteA)}{\hbar})^{2}
  \int_{-\infty}^{\infty}\frac{d\omega}{2\pi}
  \textrm{Re}
  \bqty{
  \chi_{+-}^{\textrm{R}}\qty(\siteA,\omega)
  G_{+-}^{(\textrm{A})<}\qty(\siteA,\omega)
  +\chi_{+-}^{<}\qty(\siteA,\omega)
  G_{+-}^{(\textrm{A})\textrm{A}}\qty(\siteA,\omega)
  }\nonumber      \\
   & =\frac{1}{2} 
  \qty(\frac{J_{\textrm{sd}}\qty(\siteA)}{\hbar})^{2}
  \frac{1}{N_{\textrm{m}}\qty(N/2)}
  \sum_{\bm{p},\wn}
  \int_{-\infty}^{\infty}\frac{d\omega}{2\pi}
  \textrm{Re}
  \bqty{
  \chi_{+-}^{\textrm{R}}\qty(\bm{p},\omega)
  G_{+-}^{(\textrm{A})<}\qty(\wn,\omega)
  +\chi_{+-}^{<}\qty(\bm{p},\omega)
  G_{+-}^{(\textrm{A})\textrm{A}}\qty(\wn,\omega)
  }.
  \label{LocalSpinCurrentA}
 \end{align}
 Here, we have defined four types of correlation functions:
 \begin{align}
   & \chi_{+-}^{\textrm{R}}\qty(\bm r,t;\bm r',t')=                                    
  -i\theta\qty(t-t')
  \Big\langle
  \qty[\tilde{\sigma}^{+}_{\bm r}(t),\tilde{\sigma}^{-}_{\bm r'}(t')]
  \Big\rangle_{0},
  \label{RetardedFunction_Metal}
  \\
   & \chi_{+-}^{<}\qty(\bm r,t;\bm r',t')=                                             
  -i\Big\langle
  \tilde{\sigma}^{-}_{\bm r'}(t')
  \tilde{\sigma}^{+}_{\bm r}(t)
  \Big\rangle_{0},
  \label{LesserFunction_Metal}
  \\
   & G_{+-}^{(\textrm{X})\textrm{A}}\qty(\bm r_{\textrm{x}},t;\bm r_{\textrm{x}}',t')= 
  i\theta\qty(t'-t)
  \Big\langle
  \qty[\tilde{S}^{+}_{\bm r_{\textrm{x}}}(t),\tilde{S}^{-}_{\bm r_{\textrm{x}}'}(t')]
  \Big\rangle_{0},
  \label{AdvancedFunction_Magnet}
  \\
   & G_{+-}^{(\textrm{X})<}\qty(\bm r_{\textrm{x}},t;\bm r_{\textrm{x}}',t')=          
  -i
  \Big\langle
  \tilde{S}^{-}_{\bm r_{\textrm{x}}'}(t')
  \tilde{S}^{+}_{\bm r_{\textrm{x}}}(t)
  \Big\rangle_{0}.
  \label{LesserFunction_Magnet}
 \end{align}
 In the first line in Eq.~(\ref{LocalSpinCurrentA}), $\chi_{+-}^{\textrm{R}[<]}\qty(\siteA,\omega)$ is the Fourier component of the retarded [lesser] part of $\chi_{+-}\qty(\site_{\textrm{A}}, t-t')$ in the frequency $\omega$ space, and $G_{+-}^{(\textrm{A})\textrm{A}[<]}\qty(\siteA,\omega)$ is the Fourier component of the advanced [lesser] part of $G_{+-}^{(\textrm{A})}\qty(\site_{\textrm{A}}, t-t')$ in the $\omega$ space:
 $\chi_{+-}^{\textrm{R}[<]}\qty(\siteA,\siteAd,\omega)=
  \int_{-\infty}^{\infty}d(t-t')
  e^{i\omega\qty(t-t')}
  \chi_{+-}^{\textrm{R}[<]}\qty(\siteA,\siteAd,t-t')$ and
 $G_{+-}^{(\textrm{A})\textrm{A}[<]}
  \qty(\siteA,\siteAd,\omega)=
  \int_{-\infty}^{\infty}d(t-t')
  e^{i\omega\qty(t-t')}
  G_{+-}^{(\textrm{A})\textrm{A}[<]}\qty(\siteA,\siteAd,t-t')$.
 In the second line in Eq.~(\ref{LocalSpinCurrentA}), $N_{\textrm{m}}$ is the total number of sites of the metal, and $\chi_{+-}^{\textrm{R}[<]}\qty(\bm{p},\omega)$ and $G_{+-}^{(\textrm{A})\textrm{A}[<]}\qty(\bm{k},\omega)$ are, respectively, the Fourier components of $\chi_{+-}^{\textrm{R}[<]}\qty(\siteA,\omega)$ and $G_{+-}^{(\textrm{A})\textrm{A}[<]}\qty(\siteA,\omega)$:
 $\chi_{+-}^{\textrm{R}[<]}\qty(\bm{p},\omega)=\sum_{\substack{\siteA-\siteAd (\siteA,\siteAd\in\textrm{NM})}}
  e^{-i\bm{p}\cdot\qty(\siteA-\siteAd)}
  \chi_{+-}^{\textrm{R}[<]}\qty(\siteA-\siteAd,\omega)$ and
 $G_{+-}^{(\textrm{A})\textrm{A}[<]}
  \qty(\bm{k},\omega)=\sum_{\substack{\siteA-\siteAd                  (\siteA,\siteAd\in\textrm{AFM})}}
  e^{-i\bm{k}\cdot\qty(\siteA-\siteAd)}
  G_{+-}^{(\textrm{A})\textrm{A}[<]}\qty(\siteA-\siteAd,\omega)$.
 Through a similar algebra, we also obtain the expression of $\langle I_{\siteB}\rangle$. Substituting these results into Eq.(\ref{Expectation}), we obtain
 \begin{multline}
  \langle I_{\textrm{S}}\rangle=
  \frac{1}{2}\qty(\frac{J_{\textrm{sd}}}{\hbar})^{2}
  \frac{N_{\inter}/2}{N_{\textrm{m}}\qty(N/2)}
  \sum_{\bm{p},\wn}
  \int_{-\infty}^{\infty}\frac{d\omega}{2\pi}
  \textrm{Re}
  \bqty{
  \chi_{+-}^{\textrm{R}}\qty(\bm{p},\omega)
  G_{+-}^{(\textrm{A})<}\qty(\wn,\omega)
  +\chi_{+-}^{<}\qty(\bm{p},\omega)
  G_{+-}^{(\textrm{A})\textrm{A}}\qty(\wn,\omega)
  }
  \\
  +\frac{1}{2}\qty(\frac{J_{\textrm{sd}}}{\hbar})^{2}
  \frac{N_{\inter}/2}{N_{\textrm{m}}\qty(N/2)}
  \sum_{\bm{p},\wn}
  \int_{-\infty}^{\infty}\frac{d\omega}{2\pi}
  \textrm{Re}
  \bqty{
  \chi_{+-}^{\textrm{R}}\qty(\bm{p},\omega)
  G_{+-}^{(\textrm{B})<}\qty(\wn,\omega)
  +\chi_{+-}^{<}\qty(\bm{p},\omega)
  G_{+-}^{(\textrm{B})\textrm{A}}\qty(\wn,\omega)
  },
  \label{TunnelingDCSinCurrent-Expectation-PerturbationCalculation}
 \end{multline}
 where we have introduced an averaged s-d exchange $J_{\textrm{sd}}$ via $\sum_{\siteA\in\interA}J_{\textrm{sd}}\qty(\siteA)^2=\sum_{\siteB\in\interB}J_{\textrm{sd}}\qty(\siteB)^2=:N_{\inter} J_{\textrm{sd}}^2/2$ ($N_{\inter}$ is the number of sites of the interface).
\end{widetext}
This is the formula for the DC spin current flowing from the antiferromagnet to the normal metal via an interfacial exchange interaction, as shown in Fig.~\ref{FIG:SSE-Antiferromagnets}.
This formalism can be applied to SSEs~\cite{Adachi2011,Geprags2016,Hirobe2017d,Hirobe2019,Chen2021} and spin pumpings~\cite{Ohnuma2014,Kato2019,Fyhn2021} in a broad class of magnets including various ordered magnets and quantum spin liquids.
Note that we do not assume any magnetic order in the derivation of the spin current.
For instance, if we consider an SSE in a ferromagnetic insulator, two types of Green's functions, $G_{+-}^{(\textrm{A})\textrm{R}}$ and $G_{+-}^{(\textrm{B})\textrm{R}}$, are replaced with a single Green's function $G_{+-}^{\textrm{R}}$ since there is no sublattice structure in the ferromagnet.
On the other hand, if we consider a magnet with an $M$-sublattice structure, they are replaced with the sum of multiple Green's functions on $M$ different sublattice sites.
\subsection{SSE in the N\'eel phase}
\label{SubSec:SpinCurrent_Neel}
In the N\'eel ordered phase, spin correlators can be computed using the spin-wave approximation Eqs.~(\ref{MagnonRepresentation-AsublatticeNeelSpin}) and (\ref{MagnonRepresentation-BsublatticeNeelSpin}).
Spin correlators can be represented as
\begin{align}
  & G_{+-}^{(\textrm{X})\textrm{A}}\qty(\wn,\omega)= 
 \mathcal{G}_{+-}^{(\textrm{X})\textrm{A}}\qty(\wn,\omega),
 \label{AdvancedGreenFunction_Neel}
 \\
  & G_{+-}^{(\textrm{X})<}\qty(\wn,\omega)=          
 \mathcal{G}_{+-}^{(\textrm{X})<}\qty(\wn,\omega),
 \label{LesserGreenFunction_Neel}
\end{align}
where $\mathcal{G}_{+-}^{(\textrm{X})\textrm{A}[<]}\qty(\wn,\omega)$ denotes the advanced [lesser] part of the magnon propagator (X=A, B).
The detailed forms of $\mathcal{G}_{+-}^{(\textrm{X})\textrm{A}[<]}$ will be given in the next section.

In our perturbation framework, AFM and NM are, respectively, in non-equilibrium steady (almost equilibrium) states with temperatures $T_{\textrm{s}}$ and $T_{\textrm{m}}$.
For such quasi-equilibrium states, we obtain~\cite{HaugJauhoSpringer2008,StefanucciCambridge2013}
\begin{align}
  & \chi_{+-}^{<}\qty(\bm p,\omega)        =           
 2i\textrm{Im}
 \chi_{+-}^{\textrm{R}}\qty(\bm{p},\omega)
 n_{\textrm{B}}\qty(\omega+\delta\mu_{\textrm{s}}/\hbar,T_{\textrm{m}}),
 \label{LesserComponent-Metal}
 \\
  & \mathcal{G}_{+-}^{(\textrm{X})<}\qty(\wn,\omega) = 
 2i\textrm{Im}
 \mathcal{G}_{+-}^{(\textrm{X})\textrm{R}}\qty(\wn,\omega)
 n_{\textrm{B}}\qty(\omega,T_{\textrm{s}}),
 \label{LesserComponent-Antiferromagnet}
\end{align}
through the Lehmann representation of these correlators.
Here, $n_{\textrm{B}}\qty(\omega,T)=\qty(e^{\hbar\omega/k_{\textrm{B}}T}-1)^{-1}$
is the Bose distribution function.
Substituting Eqs.~(\ref{AdvancedGreenFunction_Neel})-(\ref{LesserComponent-Antiferromagnet}) into Eq.~(\ref{TunnelingDCSinCurrent-Expectation-PerturbationCalculation}),
we finally arrive at
\begin{widetext}
 \begin{align}
  \!\!\!\!\!\!\!\!
  \langle I_{\textrm{S}}\rangle
   & =                                         
  -\qty(\frac{J_{\textrm{sd}}}{\hbar})^{2}
  \frac{N_{\inter}/2}{N_{\textrm{m}}\qty(N/2)}
  \sum_{\bm{p},\wn}
  \int_{-\infty}^{\infty}\frac{d\omega}{2\pi}
  \textrm{Im}\chi_{+-}^{\textrm{R}}\qty(\bm p,\omega)
  \qty\Big{
  \textrm{Im}\mathcal{G}_{+-}^{(\textrm{A})\textrm{R}}\qty(\wn,\omega)
  +\textrm{Im}\mathcal{G}_{+-}^{(\textrm{B})\textrm{R}}\qty(\wn,\omega)
  }
  \qty\Big{
  n_{\textrm{B}}\qty(\omega,T_{\textrm{s}})
  -n_{\textrm{B}}\qty(\omega,T_{\textrm{m}})
  }
  \nonumber                                    \\
   & =-\qty(\frac{J_{\textrm{sd}}}{\hbar})^{2} 
  \frac{N_{\inter}}{2}
  \int_{-\infty}^{\infty}\frac{d\omega}{2\pi}
  \textrm{Im}\chi_{+-}^{\textrm{R}}\qty(\omega)
  \qty\Big{
  \textrm{Im}\mathcal{G}_{+-}^{(\textrm{A})\textrm{R}}\qty(\omega)
  +\textrm{Im}\mathcal{G}_{+-}^{(\textrm{B})\textrm{R}}\qty(\omega)
  }
  \qty\Big{
  n_{\textrm{B}}\qty(\omega,T_{\textrm{s}})
  -n_{\textrm{B}}\qty(\omega,T_{\textrm{m}})
  },
  \label{TunnelSinCurrent_finmal1}
 \end{align}
 where $\textrm{Im}\chi_{+-}^{\textrm{R}}\qty(\omega)=\frac{1}{N_{\textrm{m}}}\sum_{\bm p}\textrm{Im}\chi_{+-}^{\textrm{R}}\qty(\bm p,\omega)$ and $\textrm{Im}\mathcal{G}_{+-}^{\textrm{(X)R}}\qty(\omega)=\frac{1}{N/2}\sum_{\bm k}\textrm{Im}\mathcal{G}_{+-}^{\textrm{(X)R}}\qty(\bm k,\omega)$~(X=A, B).
 We have omitted the $\delta \mu_{\rm s}$ dependence of the current $\langle I_{\textrm{S}}\rangle$ since $\delta \mu_{\rm s}$ is very small compared with the typical energy scale of the magnet, i.e., ${\cal O}(J)$.
\end{widetext}
\subsection{SSE in the canted phase}
\label{SubSec:SpinCurrent_Cant}
\begin{figure}[t]
 \includegraphics[width=\linewidth]{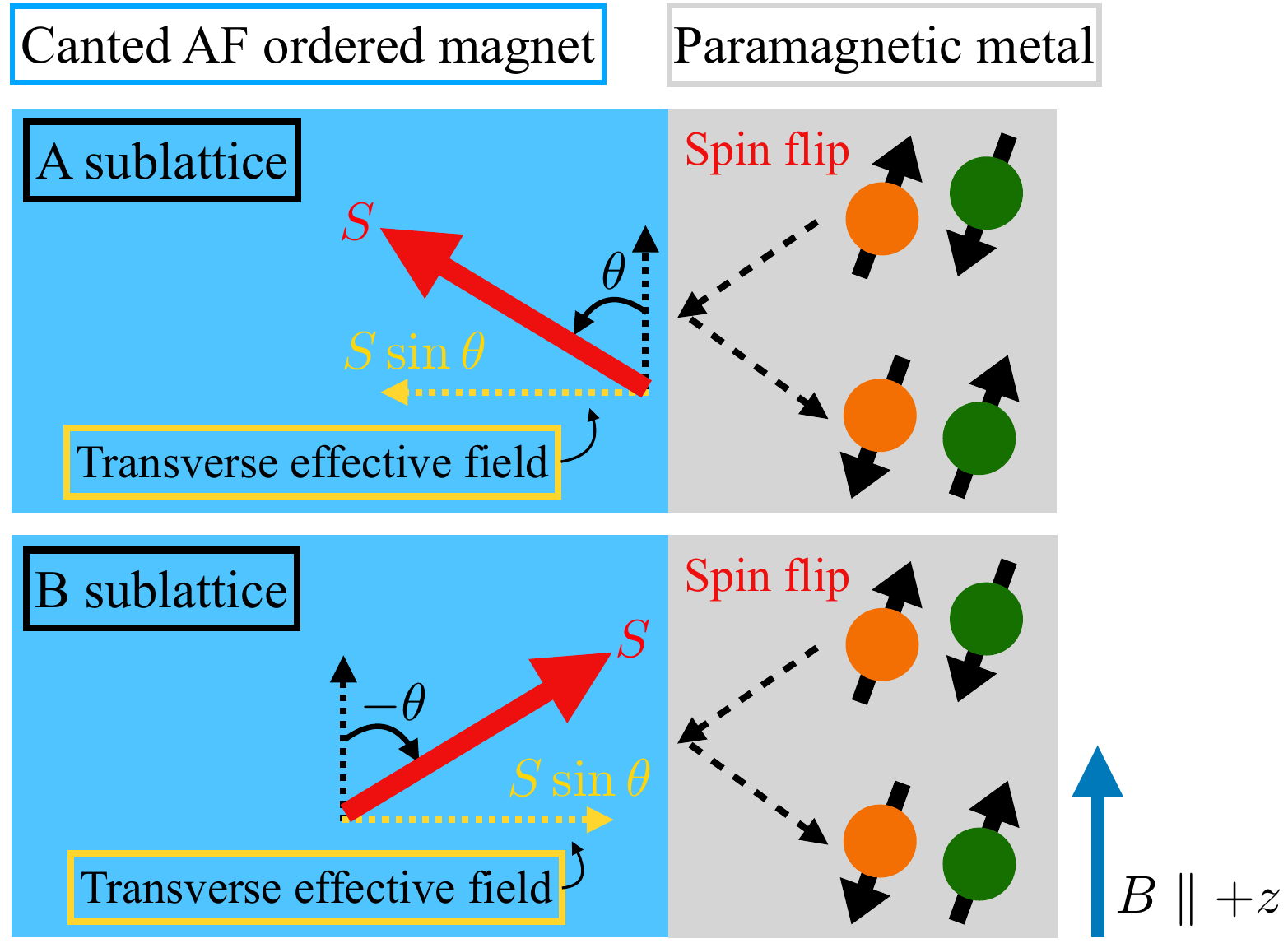}
 \caption{(Color online)
  Schematic view of spin-flip process of conducting electrons at an interface between the antiferromagnet and the metal when the magnet is in the canted phase.
  The effective transverse magnetic fields generated by localized spins cause the spin-flip scattering in the metal.
 }
 \label{FIG:SpinFlip}
\end{figure}
Unlike the N\'eel phase, in the canted AF phase, there is an additional contribution to the spin current $\langle I_{\textrm{S}}\rangle$.
The magnon representations of spins, Eqs.~(\ref{MagnonRepresentation-AsublatticeCantedAFSpin}) and (\ref{MagnonRepresentation-BsublatticeCantedAFSpin}), decompose the spin correlator into two parts. Namely, the advanced and retarded components of the spin correlator are given by the magnon correlation functions, whereas the lesser component is given by the sum of the magnon correlator and a constant term:
\begin{align}
  & G_{+-}^{(\textrm{X})\textrm{A}}\qty(\wn,\omega)= 
 \mathcal{G}_{+-}^{(\textrm{X})\textrm{A}}\qty(\wn,\omega),
 \label{AdvancedGreenFunction_Cant}
 \\
  & G_{+-}^{(\textrm{X})<}\qty(\wn,\omega)=          
 (-i)2\pi
 S^{2}\sin^{2}\theta\frac{N}{2}\delta_{\bm k,\bm 0}\delta\qty(\omega)
 +\mathcal{G}_{+-}^{(\textrm{X})<}\qty(\wn,\omega),
 \label{LesserGreenFunction_Cant}
\end{align}
where X=A, B. The first term of Eq.~(\ref{LesserGreenFunction_Cant}) comes from the transverse magnetization $S\sin\theta$, as shown in Fig.~\ref{FIG:SpinFlip}.
Using the Lehmann representation of the magnon propagator, we again obtain
\begin{equation}
 \mathcal{G}_{+-}^{(\textrm{X})<}\qty(\wn,\omega)=
 2i\textrm{Im}\mathcal{G}_{+-}^{(\textrm{X})\textrm{R}}\qty(\wn,\omega)
 n_{\textrm{B}}\qty(\omega,T_{\textrm{s}}).
 \label{LesserComponent-Cant}
\end{equation}
Substituting Eq.~(\ref{LesserComponent-Metal}) and Eqs.~(\ref{AdvancedGreenFunction_Cant})-(\ref{LesserComponent-Cant}) into Eq.~(\ref{TunnelingDCSinCurrent-Expectation-PerturbationCalculation}), we finally arrive at
\begin{widetext}
 \begin{multline}
  \langle I_{\textrm{S}}\rangle=
  \qty(\frac{J_{\textrm{sd}}}{\hbar})^{2}
  S^{2}\sin^{2}\theta
  \frac{N_{\inter}}{2}
  \textrm{Im}
  \chi_{+-}^{\textrm{R}}\qty(\omega=0)
  \\
  -\qty(\frac{J_{\textrm{sd}}}{\hbar})^{2}
  \frac{N_{\inter}}{2}
  \int_{-\infty}^{\infty}\frac{d\omega}{2\pi}
  \textrm{Im}\chi_{+-}^{\textrm{R}}\qty(\omega)
  \qty\Big{
  \textrm{Im}\mathcal{G}_{+-}^{(\textrm{A})\textrm{R}}\qty(\omega)
  +\textrm{Im}\mathcal{G}_{+-}^{(\textrm{B})\textrm{R}}\qty(\omega)}
  \qty\Big{
  n_{\textrm{B}}\qty(\omega,T_{\textrm{s}})
  -n_{\textrm{B}}\qty(\omega,T_{\textrm{m}})
  },
  \label{TunnelSinCurrent_finmal1_Cant}
 \end{multline}
 where $\textrm{Im}\chi_{+-}^{\textrm{R}}\qty(\omega)=\frac{1}{N_{\textrm{m}}}\sum_{\bm p}\textrm{Im}\chi_{+-}^{\textrm{R}}\qty(\bm p,\omega)$, $\textrm{Im}\mathcal{G}_{+-}^{\textrm{(X)R}}\qty(\omega)=\frac{1}{N/2}\sum_{\bm k}\textrm{Im}\mathcal{G}_{+-}^{\textrm{(X)R}}\qty(\bm k,\omega)$~(X=A, B), and we have again omitted the small quantity $\delta \mu_{\rm s}$ in the second term of the integral.
\end{widetext}

The tunneling spin current of the SSE in the canted AF phase is composed of two parts.
The first term in Eq.~(\ref{TunnelSinCurrent_finmal1_Cant}) indicates that spin flips of conducting electrons caused by transverse magnetization (see Fig.~\ref{FIG:SpinFlip})
contribute to the spin current.
Similar electron-spin flips due to magnetic moments have already been studied in spin Hall magnetoresistance \cite{Nakayama2013,Sugi2023}.
As will be shown later, introducing a small but finite $\delta \mu_{\rm s}$ is essential to obtain a finite value of the first term because $\textrm{Im}\chi_{+-}^{\textrm{R}}\qty(0)=0$ for $\delta \mu_{\rm s}=0$.
The second term is the contribution from magnon dynamics like the spin current in the N\'eel phase of Eq.~(\ref{TunnelSinCurrent_finmal1}).
We stress that
the first term of Eq.~(\ref{TunnelSinCurrent_finmal1_Cant})
seems to survive even in an equilibrium setup, but a
finite spin-dependent potential $\delta\mu_{\textrm{s}}$ emerges only in the NESS with a temperature gradient.
Namely, the first and second terms both stem from the temperature gradient.
\subsection{Magnetic density of states and neutron scattering spectra}
\label{Sec:DoS}
In this subsection, we discuss the physical meaning of the spin-current formulas~(\ref{TunnelSinCurrent_finmal1}) and (\ref{TunnelSinCurrent_finmal1_Cant}) from the viewpoint of the magnetic density of state (DoS).
\begin{table}[h]
 \caption{\label{tab:ElectronGreen}
  Relationship between the one-particle electron Green's function and the electron density of state.}
 \begin{ruledtabular}
  \begin{tabular}{lll}
   DoS                                         & Green's function            & condition \\
   \hline
   electron DoS $D_{\textrm{e}}(\omega)$       & $-\frac{1}{\pi} \textrm{Im}             
   \sum_{\bm k}G^{\textrm{R}}({\bm k},\omega)$ & $\omega>\mu$                            \\
   hole DoS $D_{\textrm{h}}(\omega)$           & $-\frac{1}{\pi} \textrm{Im}             
   \sum_{\bm k}G^{\textrm{R}}({\bm k},\omega)$ & $\omega<\mu$                            
   \\
  \end{tabular}
 \end{ruledtabular}
\end{table}
\begin{table}[h]
 \caption{\label{tab:SpinGreen}
  Relationship between spin Green's functions and spin-up and spin-down density of states.}
 \begin{ruledtabular}
  \begin{tabular}{lll}
   DoS                                                        & Green's function & condition \\
   \hline
   spin-up DoS $D_{\uparrow}(\omega)$                         & $- \textrm{Im}               
   \sum_{\bm k}\mathcal{G}_{-+}^{\textrm{R}}({\bm k},\omega)$ & $\omega> 0$                  \\
   spin-down DoS $D_{\downarrow}(\omega)$                     & $- \textrm{Im}               
   \sum_{\bm k}\mathcal{G}_{+-}^{\textrm{R}}({\bm k},\omega)$ & $\omega> 0$                  
   \\
  \end{tabular}
 \end{ruledtabular}
\end{table}

In conducting electron systems, the imaginary part of the one-particle retarded Green's function, $\sum_{\bm k} G^{\textrm{R}}({\bm k}, \omega)$, is proportional to the electron (hole) DoS $D_{\textrm{e(h)}}(\omega)$ \cite{HaugJauhoSpringer2008,StefanucciCambridge2013}:
\begin{align}
 D_{\textrm{e(h)}}(\omega) & = -\frac{1}{\pi} \textrm{Im} 
 \sum_{\bm k}G^{\textrm{R}}({\bm k},\omega)
 \label{eq:eleDos}
\end{align}
when the frequency $\omega$ is higher (lower) than the chemical potential $\mu$ (see Table~\ref{tab:ElectronGreen}).
Similarly, in magnetic insulators, we can define the magnetic DoS for spin-down [spin-up] excitations from the imaginary part of  $\mathcal{G}_{+-}^{\textrm{R}}({\bm k}, \omega)$ [$\mathcal{G}_{-+}^{\textrm{R}}({\bm k},\omega)$] as
\begin{align}
 D_{\downarrow(\uparrow)}(\omega) & = -\textrm{Im} 
 \sum_{\bm k}\mathcal{G}^{\textrm{R}}_{+-(-+)}({\bm k},\omega).
 \label{eq:magDoS}
\end{align}
Note that the DoS of Eq.~(\ref{eq:magDoS}) is physically relevant only for $\omega> 0$ since the excitation energy is always higher than the ground-state energy in magnetic systems.
In magnon systems with a magnetization along the $S^z$ axis, $\mathcal{G}^{\textrm{R}}_{+-(-+)}({\bm k},\omega)$ can be regarded as the spin-down (spin-up) magnon Green's function because ${\hat S}^{-(+)}$ is almost equivalent to the creation operator of a spin-down (spin-up) magnon.
Taking the complex conjugate of $\mathcal{G}^{\textrm{R}}_{+-(-+)}({\bm k},\omega)$, one can easily find the relationship between the two Green's functions $\mathcal{G}^{\textrm{R}}_{+-}({\bm k},\omega)$ and $\mathcal{G}^{\textrm{R}}_{-+}({\bm k},\omega)$:
\begin{align}
 \textrm{Im}\mathcal{G}^{\textrm{R}}_{+-}({\bm k},\omega)
 =-\textrm{Im}\mathcal{G}^{\textrm{R}}_{-+}({\bm k},-\omega).
 \label{eq:up_down}
\end{align}
This holds in generic spin systems.

Using Eqs.~(\ref{eq:magDoS}) and (\ref{eq:up_down}), we can provide an important interpretation for the tunneling spin current of Eqs.~(\ref{TunnelSinCurrent_finmal1}) and (\ref{TunnelSinCurrent_finmal1_Cant}) as follows.
A key point of the Green's functions in Eq.~(\ref{TunnelSinCurrent_finmal1}) and the second term of Eq.~(\ref{TunnelSinCurrent_finmal1_Cant}) is their $\omega$ dependence.
As we will soon discuss in more detail, the Green's function of a metal generally satisfies $\textrm{Im}\chi_{+-}^{\textrm{R}}\qty(\omega)\propto \omega$ in the low-energy regime, where $\omega$ is much lower than the Fermi energy (chemical potential).
Namely, $\textrm{Im}\chi_{+-}^{\textrm{R}}\qty(\omega)$ is odd w.r.t. the frequency $\omega$.
The temperature factor of
Eqs.~(\ref{TunnelSinCurrent_finmal1}) and (\ref{TunnelSinCurrent_finmal1_Cant}),
$n_{\textrm{B}}\qty(\omega,T_{\textrm{s}})-n_{\textrm{B}}\qty(\omega,T_{\textrm{m}})$, is odd as well.
Therefore, we find that
$\textrm{Im}\mathcal{G}^{\textrm{R}}_{+-}(\omega)$ {\it has to possess a finite $\omega$-even component} if the spin current $\langle I_{\text{S}}\rangle$ becomes a finite value.

Second, the viewpoint of magnetic DoSs enables us to obtain a deep understanding of Eqs.~(\ref{TunnelSinCurrent_finmal1}) and (\ref{TunnelSinCurrent_finmal1_Cant}).
In the integration of Eqs.~(\ref{TunnelSinCurrent_finmal1}) and (\ref{TunnelSinCurrent_finmal1_Cant}),
$\textrm{Im}\mathcal{G}^{\textrm{R}}_{+-}(\omega)$ for $\omega>0$ can be regarded as the spin-down DoS $D_{\downarrow}(\omega)$, whereas that for $\omega<0$ is equivalent to the spin-up DoS $D_{\uparrow}(\omega)$.
Thus, we conclude that the sign of the spin current $\langle I_{\text{S}}\rangle$ is determined by whether the spin-up or spin-down carrier is more dominant.
In other words, the sign of the spin current in the SSE tells us the main magnetic carrier of the target material.
This result is very reminiscent of the standard electronic Seebeck effect, in which the sign of the Seebeck coefficient indicates the type of dominant carrier, i.e., electrons or holes.
This picture based on the magnetic DoS is useful for studying the SSE in antiferromagnets since both spin-up and spin-down excitations occur in antiferromagnetic phases (On the other hand, only down-spin magnons appear in ferromagnetic phases).

The imaginary part of the retarded two-spin Green's function $\textrm{Im}\mathcal{G}^{\textrm{R}}_{+-}(\omega)$ is proportional to spin dynamical structure factor, which can be measured via polarized-neutron scattering experiment~\cite{Nambu2020,Mattis2006}.
Therefore, both experiments on the SSE and neutron scattering enable us to check how quantitatively our formula of the spin current can capture the measured value of SSE voltage.
We note that inelastic-neutron scattering can observe the dynamical structure factor in the $\wn$- and $\omega$-resolved fashion, whereas [as shown in Eqs.~(\ref{TunnelSinCurrent_finmal1}) and (\ref{TunnelSinCurrent_finmal1_Cant})] the SSE spin current includes it in the $(\wn, \omega)$-space integral form.
\section{Spin Seebeck Effect in Antiferromagnetic Insulators}
\label{Sec:SSE}
In this section, based on Eqs. (\ref{TunnelSinCurrent_finmal1}) and (\ref{TunnelSinCurrent_finmal1_Cant}), we estimate the tunneling DC spin currents generated by the SSE in antiferromagnets.
We focus on their magnetic-field and temperature dependences.
\subsection{Normalized spin current}
\label{subSec:Normalized}
In this subsection, substituting some information about the Green's functions into Eqs.~(\ref{TunnelSinCurrent_finmal1}) and (\ref{TunnelSinCurrent_finmal1_Cant}), we simplify the spin-current expression.

For the model described by Eqs.(\ref{Dynamics_Metal}) and (\ref{ThermalAverage_Metal}), the dynamical susceptibility of the conducting electrons can be approximated by (for more detail, App. \ref{App:SpinSusceptibility_NormalMetal})
\begin{equation}
 \textrm{Im}\chi_{+-}^{\textrm{R}}\qty(\omega)\simeq
 -\pi\qty{\mathcal{D}\qty(0)}^{2}\hbar\qty(\hbar\omega+\delta\mu_{\textrm{s}}),
 \label{eq:Fermi}
\end{equation}
where $\textrm{Im}\chi_{+-}^{\textrm{R}}\qty(\omega)=\frac{1}{N_{\textrm{m}}}\sum_{\bm p}\textrm{Im}\chi_{+-}^{\textrm{R}}\qty(\bm p,\omega)$, and $\mathcal{D}\qty(0)$ denotes the electron DoS at the Fermi energy~\cite{To,Shastry1994,Kato2019,WhiteRobert,Kato2020}.
This expression is valid when $\hbar\omega$ is sufficiently lower than the Fermi energy (or band width).
In the SSE setup, a Landau-level structure seems to emerge in the attached metal due to the applied field.
However, (as we already mentioned) a disordered metal with impurities and polycrystalline structure is attached in standard SSE bilayers, and typical energy scales of the metal (the Fermi energy, the impurity potential, etc.) are much larger than those of the magnetic field and the magnetic exchanges.
Thus, the Landau level is expected to be irrelevant, and we can instead use the expression of Eq.~(\ref{eq:Fermi}).

When the temperature difference $\Delta T=T_{\textrm{s}}-T_{\textrm{m}}\ (>0)$ is sufficiently small, the $T$-dependent factor $n_{\textrm{B}}\qty(\omega,T_{\textrm{s}})-n_{\textrm{B}}\qty(\omega,T_{\textrm{m}})$ can be approximated by $\frac{\hbar\omega}{k_{\textrm{B}}T^{2}}\frac{\Delta T}{4\sinh^{2}[\hbar\omega/2k_{\textrm{B}}T]}$, where $T=\qty(T_{\textrm{s}}+T_{\textrm{m}})/2$ is the averaged temperature. Here, the small factor $\delta \mu_{\rm s}$ in the Bose distribution has been again neglected under the assumption that $k_{\rm B}\Delta T$ is sufficiently smaller than the exchange coupling, but larger enough than $\delta \mu_{\rm s}$, namely, $J\gg k_{\rm B}\Delta T\gg \delta \mu_{\rm s}$.

Substituting these relations for the susceptibility of the metal and the $T$-factor into Eq.~(\ref{TunnelSinCurrent_finmal1}), we can define
the normalized spin current $\bar{\langle I_{\textrm{S}}\rangle}$ as $\langle I_{\textrm{S}}\rangle=-\pi\qty{\mathcal{D}\qty(0)}^{2}\qty(J_{\textrm{sd}})^{2}\qty(N_{\inter}/2)\qty(k_{\textrm{B}}\Delta T/\hbar)\bar{\langle I_{\textrm{S}}\rangle}$.
The current in the N\'eel phase is given by
\begin{widetext}
 \begin{align}
  \bar{\langle I_{\textrm{S}}\rangle}
   & \simeq 
  -\frac{1}{4}
  \frac{1}{\qty(k_{\textrm{B}}T)^2}
  \frac{1}{N/2}
  \sum_{\wn}
  \int_{-\infty}^{\infty}\frac{d\omega}{2\pi}
  \qty\Big{
  \textrm{Im}\mathcal{G}_{+-}^{(\textrm{A})\textrm{R}}\qty(\wn,\omega)
  +\textrm{Im}\mathcal{G}_{+-}^{(\textrm{B})\textrm{R}}\qty(\wn,\omega)
  }
  \frac{\qty(\hbar\omega)^{2}}{\sinh^{2}[\hbar\omega/2k_{\textrm{B}}T]}.
  \label{TunnelSinCurrent_finmal2}
 \end{align}
 Similarly, the normalized spin current in the canted AF phase is given by
 \begin{align}
  \bar{\langle I_{\textrm{S}}\rangle}
  \simeq
   & S^{2}\sin^{2}\theta 
  \frac{\delta\mu_{\textrm{s}}}{k_{\textrm{B}}\Delta T}
  \nonumber              \\
   & -\frac{1}{4}        
  \frac{1}{\qty(k_{\textrm{B}}T)^2}
  \frac{1}{N/2}
  \sum_{\wn}
  \int_{-\infty}^{\infty}\frac{d\omega}{2\pi}
  \qty\Big{
  \textrm{Im}\mathcal{G}_{+-}^{(\textrm{A})\textrm{R}}\qty(\wn,\omega)
  +\textrm{Im}\mathcal{G}_{+-}^{(\textrm{B})\textrm{R}}\qty(\wn,\omega)
  }
  \frac{\qty(\hbar\omega)^{2}}{\sinh^{2}[\hbar\omega/2k_{\textrm{B}}T]}.
  \label{TunnelSinCurrent_finmal2_Cant}
 \end{align}
\end{widetext}
The first term in Eq.~(\ref{TunnelSinCurrent_finmal2_Cant}) can be viewed as the contribution from the spin flips of conducting electrons caused by the effective transverse fields in Fig.~\ref{FIG:SpinFlip}.
Its expression clearly indicates that it is necessary to consider the NESS with a finite $\delta\mu_{\textrm{s}}$ to include the effect of the spin flip.
Hereafter, we will discuss the magnetic-field and temperature dependences of the spin current or SSE voltage using these normalized spin currents.
\subsection{Density of states in antiferromagnets}
\label{subSec:DoS}
Equations~(\ref{TunnelSinCurrent_finmal2}) and (\ref{TunnelSinCurrent_finmal2_Cant}) show that
the remaining task is to compute the magnon propagators of antiferromagnets, $\textrm{Im}\mathcal{G}_{+-}^{(\textrm{A})\textrm{R}}\qty(\wn,\omega)$ and $\textrm{Im}\mathcal{G}_{+-}^{(\textrm{B})\textrm{R}}\qty(\wn,\omega)$.
In the N\'eel ordered phase, we use the spin-wave Hamiltonian~(\ref{SpinWave_Neel_Diagonalization}) and the magnon propagators are given by
\begin{widetext}
 \begin{align}
   & \mathcal{G}_{+-}^{(\textrm{A})\textrm{R}}\qty(\wn,\omega)=                                       
  \frac{2S\cosh^{2}\phi_{\wn}}{\omega-\varepsilon_{\alpha}^{(\textrm{N})}(\wn)/\hbar+i\eta}
  -\frac{2S\sinh^{2}\phi_{\wn}}{\omega+\varepsilon_{\beta}^{(\textrm{N})}(\wn)/\hbar+i\eta},\nonumber \\
  \nonumber                                                                                           \\
   & \mathcal{G}_{+-}^{(\textrm{B})\textrm{R}}\qty(\wn,\omega)=                                       
  \frac{2S\sinh^{2}\phi_{\wn}}{\omega-\varepsilon_{\alpha}^{(\textrm{N})}(\wn)/\hbar+i\eta}
  -\frac{2S\cosh^{2}\phi_{\wn}}{\omega+\varepsilon_{\beta}^{(\textrm{N})}(\wn)/\hbar+i\eta},
  \label{SpinCorrelation_Neel}
 \end{align}
 where $1/\eta$ denotes the relaxation time of the magnon.
 Within the phenomenological theory, one may set $\eta\simeq \alpha \omega$ with $\alpha$ being the dimensionless Gilbert damping constant~\cite{AGD,Kirilyuk2010,Oshikawa1999,Oshikawa2002,Furuya2015,Beaurepaire1996,Koopmans2000,Mashkovich2019,Tzschaschel2019,Lenz2006,Vittoria2010}.
 The explicit form of the tunneling DC spin current in the N\'eel ordered phase is therefore given by
 \begin{equation}
  \bar{\langle I_{\textrm{S}}\rangle}=
  \frac{S}{4\pi}
  \frac{1}{\qty(k_{\textrm{B}}T)^{2}}\frac{1}{N/2}
  \sum_{\bm k}
  \int_{-\infty}^{\infty}d\omega
  \frac{\qty(\hbar\omega)^{2}}{\sinh^{2}[\hbar\omega/2k_{\textrm{B}}T]}
  \frac{1}{\sqrt{1-\tanh^{2}(2\phi_{\wn})}}
  \qty[
  \frac{\eta}{\big\{\omega-\varepsilon_{\alpha}^{(\textrm{N})}(\wn)/\hbar\big\}^2+\eta^{2}}
  -\frac{\eta}{\big\{\omega+\varepsilon_{\beta}^{(\textrm{N})}(\wn)/\hbar\big\}^2+\eta^{2}}
  ].
  \label{SSE-TunnelingSpinCurrent-NeelPhase}
 \end{equation}
 Similarly, let us write the explicit form of $\bar{\langle I_{\textrm{S}}\rangle}$ in the canted phase.
 From the spin-wave approximation based on Eq.~(\ref{SpinWaveHamiltonian-CantedPhase}), the magnon Green's functions on A and B sublattices are calculated as
 \begin{equation}
  \mathcal{G}_{+-}^{(\textrm{X})\textrm{R}}\qty(\wn,\omega)=
  \frac{S}{4}
  \frac{A_{\wn}^{\alpha}}{\omega-\varepsilon_{\alpha}^{(\textrm{C})}(\wn)/\hbar+i\eta}
  -\frac{S}{4}
  \frac{B_{\wn}^{\alpha}}{\omega+\varepsilon_{\alpha}^{(\textrm{C})}(\wn)/\hbar+i\eta}
  -\frac{S}{4}
  \frac{B_{\wn}^{\beta}}{\omega+\varepsilon_{\beta}^{(\textrm{C})}(\wn)/\hbar+i\eta}
  +\frac{S}{4}
  \frac{A_{\wn}^{\beta}}{\omega-\varepsilon_{\beta}^{(\textrm{C})}(\wn)/\hbar+i\eta},
  \label{SpinCorrelation_Cant}
 \end{equation}
 where X (=A, B) is the sublattice index, $A_{\wn}^{x}= \qty(1+\cos^{2}\theta)\qty(\cosh^{2}\varphi_{\wn}^{x}+\sinh^{2}\varphi_{\wn}^{x})+2\cos\theta$, $B_{\wn}^{x}= \qty(1+\cos^{2}\theta)\qty(\cosh^{2}\varphi_{\wn}^{x}+\sinh^{2}\varphi_{\wn}^{x}) -2\cos\theta$, and $\cosh^{2}\varphi_{\wn}^{x}+\sinh^{2}\varphi_{\wn}^{x}=1/\sqrt{1-\tanh^{2}(2\varphi_{\wn}^{x})}$ [$x$ ($=\alpha,\ \beta$) is the magnon-band index].
 The spin current in the canted phase is hence given by
 \begin{equation}
  \bar{\langle I_{\textrm{S}}\rangle}=
  \bar{\langle I_{\textrm{S}}\rangle}_{1}
  +\bar{\langle I_{\textrm{S}}\rangle}_{2},
  \label{SSE-TunnelingSpinCurrent-CantedPhase}
 \end{equation}
 where
 \begin{align}
  \bar{\langle I_{\textrm{S}}\rangle}_{1}=
   & S^{2}\sin^{2}\theta     
  \frac{\delta\mu_{\textrm{s}}}{k_{\textrm{B}}\Delta T},
  \label{StaticPart_CantSSE} \\
  \bar{\langle I_{\textrm{S}}\rangle}_{2}=
   & \frac{S}{16\pi}         
  \frac{1}{\qty(k_{\textrm{B}}T)^{2}}
  \frac{1}{N/2}
  \sum_{\bm k}
  \int_{-\infty}^{\infty}d\omega
  \frac{\qty(\hbar\omega)^{2}}{\sinh^{2}[\hbar\omega/2k_{\textrm{B}}T]}
  \nonumber                  \\
   & \times                  
  \qty[
  \frac{\eta A_{\wn}^{\alpha}}{\big\{\omega-\varepsilon_{\alpha}^{(\textrm{C})}(\wn)/\hbar\big\}^2+\eta^{2}}
  -\frac{\eta B_{\wn}^{\alpha}}{\big\{\omega+\varepsilon_{\alpha}^{(\textrm{C})}(\wn)/\hbar\big\}^2+\eta^{2}}
  -\frac{\eta B_{\wn}^{\beta}}{\big\{\omega+\varepsilon_{\beta}^{(\textrm{C})}(\wn)/\hbar\big\}^2+\eta^{2}}
  +\frac{\eta A_{\wn}^{\beta}}{\big\{\omega-\varepsilon_{\beta}^{(\textrm{C})}(\wn)/\hbar\big\}^2+\eta^{2}}
  ].
  \label{DynamicalPart_CantSSE}
 \end{align}
\end{widetext}
The first term $\bar{\langle I_{\textrm{S}}\rangle}_{1}$ is the contribution from the transverse magnetization in Fig.~\ref{FIG:SpinFlip}, and the second, $\bar{\langle I_{\textrm{S}}\rangle}_{2}$, is that from the magnon dynamics such as in Eq.~(\ref{SSE-TunnelingSpinCurrent-NeelPhase}).
Hereafter we will set $\delta\mu_{\textrm{s}}/k_{\textrm{B}}\Delta T$ to be $0.01$. This is a reasonable value since in usual SSE setups, $\delta\mu_{\textrm{s}}$ and $\Delta T$ are estimated as $\delta\mu_{\textrm{s}}\sim\mathcal{O}(10^{1})$ \textmu\textrm{V} $\sim\mathcal{O}(10^{-2})\ \textrm{meV}$~\cite{Cornelissen2016} and $\Delta T\sim10\ \textrm{K}\sim\mathcal{O}(10^{0})\ \textrm{meV}$~\cite{Seki2015}.
\begin{figure}[t]
 \includegraphics[width=\linewidth]{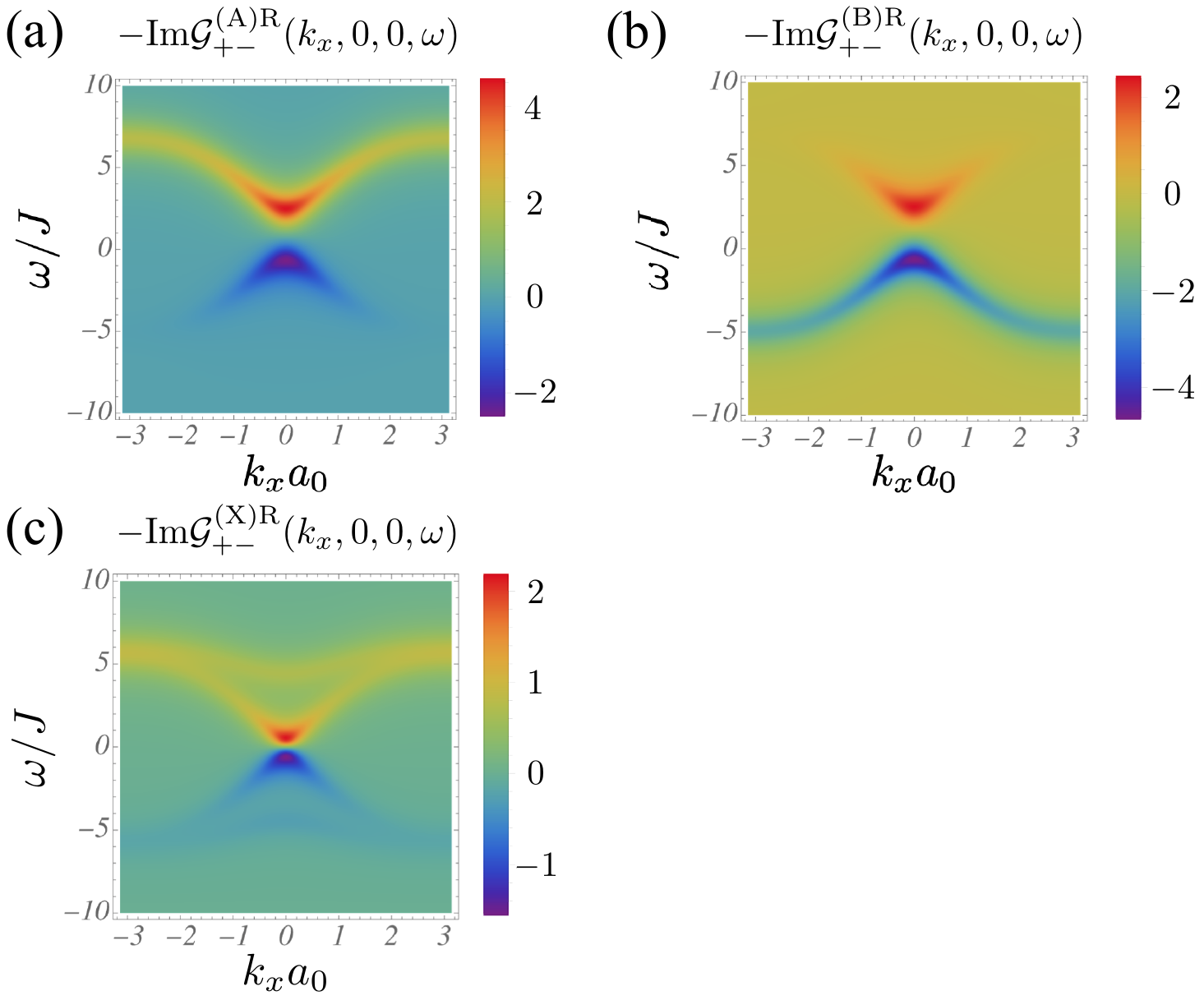}
 \caption{(Color online)
 Magnon's density of states around the $\Gamma$ point, $\wn=(0,0,0)$.
 We set $S=1$, $a_{0}=1$, $J=1$, and $|D|/J=0.1$.
 To enhance the visibility, we take a very high relaxation rate $\eta=0.01|\omega|+1.0$.
 Panels (a) and (b) are, respectively, $-\textrm{Im}\mathcal{G}^{\textrm{R}}_{+-}$ (i.e., the magnon DoS) on the A sublattice and the B one in the N\'eel phase at $B/B_{\textrm{f}}=0.6$.
 Panel (c) shows $-\textrm{Im}\mathcal{G}^{\textrm{R}}_{+-}$ in the canted phase at $B/B_{\textrm{f}}=3.0$.}
 \label{FIG:DoS}
\end{figure}

In the following subsections, we estimate the magnetic-field and temperature dependences of the spin current with Eqs.~(\ref{SSE-TunnelingSpinCurrent-NeelPhase}) and (\ref{SSE-TunnelingSpinCurrent-CantedPhase}).
However, even without these formulas, the Green's functions of Eqs.~(\ref{SpinCorrelation_Neel}) and (\ref{SpinCorrelation_Cant}) are enough to predict the sign of the spin current.
Figure~\ref{FIG:DoS} (a) and (b) respectively show the imaginary parts of $\mathcal{G}^{\textrm{R}}_{+-}$ on the A sublattice and the B one in the N\'eel phase.
On the A sublattice, the spin-down DoS (i.e., $-\textrm{Im}\mathcal{G}^{\textrm{R}}_{+-}$) is higher than the spin-up one, whereas the spin-up one is higher on the B sublattice.
This is reasonable because the spin moment is positive and negative in A and B sublattices, respectively [see Fig.~\ref{FIG:AntiferromagneticPhase}(a)].
Therefore, there is competition between spin-up and spin-down carriers in the N\'eel phase.
By quantitatively comparing Figs.~\ref{FIG:DoS} (a) and (b), we can see that the spin-up carrier is more dominant than the spin-down one in the N\'eel phase, and the sign of spin current is predicted to be opposite to that of ferromagnets.
This conclusion comes from the external magnetic field, and we will discuss this point in more detail in the following subsections.
Figure~\ref{FIG:DoS}(c) shows $-\textrm{Im}\mathcal{G}^{\textrm{R}}_{+-}$ in the canted phase.
The weights of $-\textrm{Im}\mathcal{G}^{\textrm{R}}_{+-}$ on A and B sublattices are equivalent in this phase, and hence we plot $-\textrm{Im}\mathcal{G}^{\textrm{R}}_{+-}$ only on a single sublattice.
Panel (c) shows that the spin-down carrier is more dominant in the canted phase, and the static term $\bar{\langle I_{\textrm{S}}\rangle}_{1}$ is positive.
Therefore, the spin current in the canted phase is predicted to be positive.
This is reasonable because the canted phase has a uniform magnetization along the applied magnetic field, and the dominant carrier should be the same as that of ferromagnets.

In Secs.~\ref{subSec:IntermediateTemp}-\ref{subSec:Low-Temperatures_Neel}, we will quantify the field and temperature dependences of the spin current
using Eqs.~(\ref{SSE-TunnelingSpinCurrent-NeelPhase}) and (\ref{SSE-TunnelingSpinCurrent-CantedPhase}).
As we mentioned in the Introduction, SSEs of the antiferromagnets $\textrm{Cr}_{2}\textrm{O}_3$ \cite{Seki2015,Li2020a} and $\textrm{MnF}_{2}$ \cite{Wu2016} have been experimentally investigated.
Based on the mean-field approximation, we can estimate the exchange and anisotropy interactions for $\textrm{Cr}_{2}\textrm{O}_3$ as $J\sim40\ \textrm{K}$ and $|D|/J\sim\mathcal{O}(10^{-3})$, and those for $\textrm{MnF}_{2}$ as $J\sim4\ \textrm{K}$ and $|D|/J\sim\mathcal{O}(10^{-1})$.
In the remaining parts of this section, we will often refer to these values to quantitatively discuss the properties of the spin current.
\subsection{Field dependence in the intermediate temperature range}
\label{subSec:IntermediateTemp}
\begin{figure}[t]
 \includegraphics[width=\linewidth]{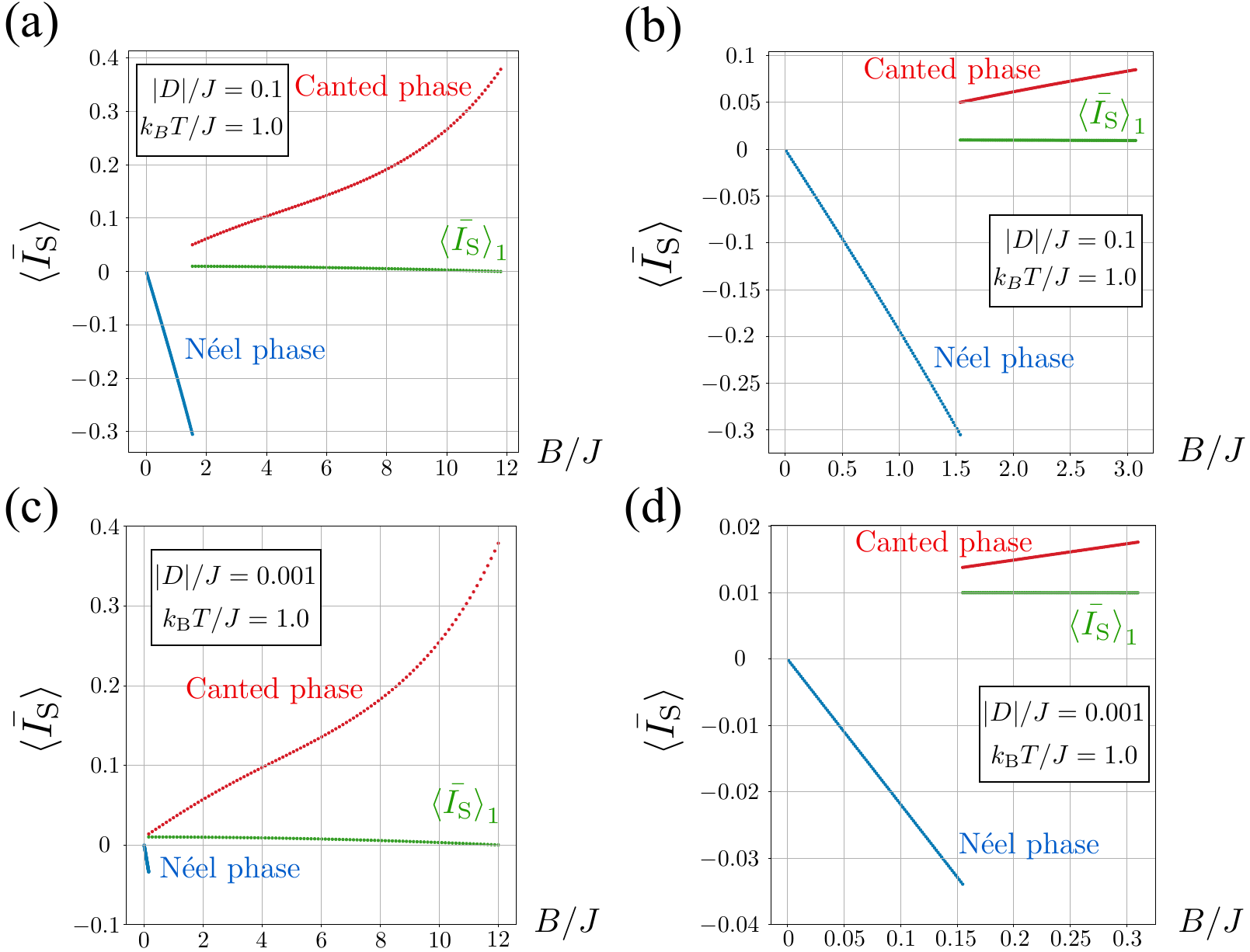}
 \caption{(Color online)
 Magnetic-field dependences of the tunneling spin current in the N\'eel [Eq.~(\ref{SSE-TunnelingSpinCurrent-NeelPhase})] and canted [Eq.~(\ref{SSE-TunnelingSpinCurrent-CantedPhase})] phases at $k_{\textrm{B}}T/J=1.0$.
 Parameters are set to be $S=1$, $a_{0}=1$, $J=1$, and $\hbar=1$.
 The magnon lifetime is set to be $\eta=\alpha|\omega|+\eta_0$ with $\alpha=0.01$ and $\eta_0=0.001$.
 The small constant $\eta_0$ is introduced to enhance the stability of the numerical integral around $\omega\to 0$.
 Since $\eta_{0}$ is set to sufficiently small, our numerical result is almost independent of $\eta_{0}$.
 The blue and red dotted lines respectively correspond to the spin currents in the N\'eel and canted phases.
 The green dotted line corresponds to the static part of the spin current in the canted phase [Eq.~(\ref{StaticPart_CantSSE})], in which the small parameter $\delta\mu_{\textrm{s}}/k_{\textrm{B}}\Delta T$ is set to be $0.01$ in $\bar{\langle I_{\textrm{S}}\rangle}_{1}$.
 Panels (a) and (b) are the field dependences of the spin current for a relatively large anisotropy $|D|/J=0.1$: The former is in a wide range up to the saturation field $B_{\textrm{c}}$, whereas the latter is in a low-field regime $B<2B_{\textrm{f}}$.
 Similarly, panels (c) and (d) are the spin currents for a small anisotropy $|D|/J=0.001$.
 }
 \label{FIG:FieldDepSpinCurrent-Intermediate}
\end{figure}

Let us investigate the magnetic-field and temperature dependences of the tunneling spin current in the N\'eel and canted phases.
This subsection is devoted to the field dependence in the intermediate-temperature regime of $k_{\rm B}T\sim J$ and $k_{\rm B}T\ll k_{\rm B}T_{\rm N}$ [see Fig.~\ref{FIG:AntiferromagneticPhase}(c)].
The magnetic-field dependence of Eqs.~(\ref{SSE-TunnelingSpinCurrent-NeelPhase}) and (\ref{SSE-TunnelingSpinCurrent-CantedPhase}) is shown in Fig.~\ref{FIG:FieldDepSpinCurrent-Intermediate}.
The spin-flop transition field $B_{\rm f}$ is known to be almost independent of temperature, as shown in Fig.~\ref{FIG:AntiferromagneticPhase}(c), and therefore we simply take the approximation $B_{\rm f}= B_{\rm f}^{\rm MF}\equiv 2S\sqrt{6J|D|}$ irrespective of temperature $T$.
Namely, we use Eq.~(\ref{SSE-TunnelingSpinCurrent-NeelPhase}) for $B<B_{\rm f}^{\rm MF}$ and Eq.~(\ref{SSE-TunnelingSpinCurrent-CantedPhase}) for $B_{\rm f}^{\rm MF}<B<B_{\rm c}$.
Hereafter, we will draw the value of the spin current under this approximation.

As expected from the argument based on the magnon DoS in Sec.~\ref{Sec:DoS}, we can confirm that the sign of the tunneling DC spin current reverses at the spin-flop transition in Fig.~\ref{FIG:FieldDepSpinCurrent-Intermediate}.
In the N\'eel phase, the value of the spin current is negative, i.e., the sign is opposite to that of ferromagnets, whereas it becomes positive in the canted phase.
This can be understood from the DoSs for spin-up and spin-down magnons shown in Fig.~\ref{FIG:DoS}.
This field dependence of the spin current (i.e., SSE voltage) agrees well with the experimental result for $\textrm{Cr}_{2}\textrm{O}_3$~\cite{Li2020a} on semi-quantitative level, but not with the two experimental results in Refs.~\cite{Seki2015,Wu2016}, where no sign change is observed.
Figure~\ref{FIG:FieldDepSpinCurrent-Intermediate} also shows that both $\bar{\langle I_{\textrm{S}}\rangle}_{1}$ and $\bar{\langle I_{\textrm{S}}\rangle}_{2}$ can be the main contributions to the spin current in the canted phase, especially in the low-field range [Figs.~\ref{FIG:FieldDepSpinCurrent-Intermediate}(b) and (d)].
\begin{figure}[t]
 \includegraphics[width=\linewidth]{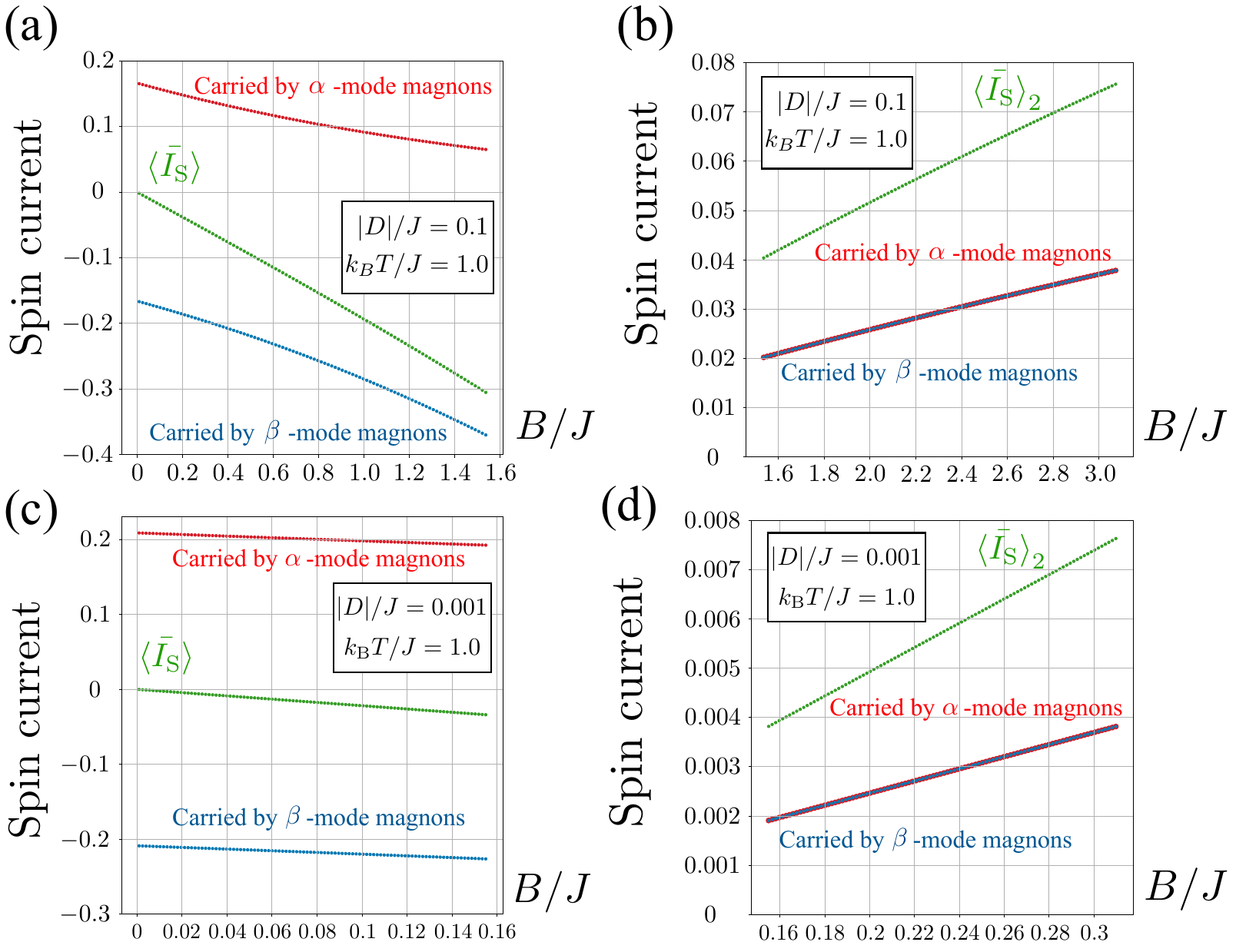}
 \caption{(Color online)
  Magnetic-field dependences of spin currents carried by $\alpha$-mode magnons (red dotted line) and $\beta$-mode magnons (blue dotted line).
  The green dotted line in panels (a) and (c) represents the tunneling spin current in the N\'eel phase [Eq.~(\ref{SSE-TunnelingSpinCurrent-NeelPhase})], whereas that in panels (b) and (d) indicates the dynamical part of the spin current $\bar{\langle I_{\textrm{S}}\rangle}_{2}$ in the canted phase [Eq.~(\ref{DynamicalPart_CantSSE})].
  Parameters $S$, $a_{0}$, $J$, $\eta$, and $\hbar$ are set to the same values as those of Fig.~\ref{FIG:FieldDepSpinCurrent-Intermediate}.
  Panels (a) and (b) are the results for $|D|/J=0.1$ and $k_{\textrm{B}}T/J=1.0$.
  Panels (c) and (d) are for $|D|/J=0.001$ and $k_{\textrm{B}}T/J=1.0$.
 }
 \label{FIG:SeparationSpinCurrent}
\end{figure}

Figure~\ref{FIG:SeparationSpinCurrent} separately depicts the contributions to the spin current from $\alpha$ and $\beta$ modes in both the N\'eel [panels (a) and (c)] and canted [panels (b) and (d)] phases.
From this figure, we can more deeply understand some features of the spin current as follows.
Low-energy excitations in the N\'eel phase are described by two species of magnons with different spin polarizations: The $\alpha$ mode magnons are down-polarized and the $\beta$ mode ones are up-polarized.
Competition between $\alpha$ and $\beta$ magnons determines the sign of the spin current.
The energy gap of the $\alpha$ magnon becomes larger than that of the $\beta$ magnon when we apply a static magnetic field $B$ owing to the Zeeman splitting.
Therefore, the number of up magnons becomes dominant over down ones at a finite $T$ under a magnetic field $B$.
The spin current carried by the up magnons is hence dominant, and we obtain a negative spin current as in Fig.~\ref{FIG:SeparationSpinCurrent}(a) and (c).
This is consistent with the argument based on the magnon DoS in Fig.~\ref{FIG:DoS}.

Similarly, Fig.~\ref{FIG:SeparationSpinCurrent}(b) and (d) also tell us why the spin current takes a positive value in the canted phase.
This phase has two magnon modes (where we again call $\alpha$ and $\beta$ modes), like the N\'eel phase.
As we already mentioned [see Fig.~\ref{FIG:DoS}(c)], both modes have a spin-down polarization on average.
Therefore, as shown in Fig.~\ref{FIG:SeparationSpinCurrent}(b) and (d), both $\alpha$ and $\beta$ modes contribute to a positive spin current and cooperatively carry the spin current (without competition).
This positive value can be roughly understood because, like ferromagnets, the canted phase has a finite uniform magnetization along the magnetic field $B$.
In addition to these two modes, the static part $\bar{\langle I_{\textrm{S}}\rangle}_{1}$ has a positive value as well. The spin current in the canted phase is thus positive.
\subsection{Non-monotonic behavior in the canted phase at low temperature}
\label{subSec:Low-Temperatures}
\begin{figure}[t]
 \includegraphics[width=\linewidth]{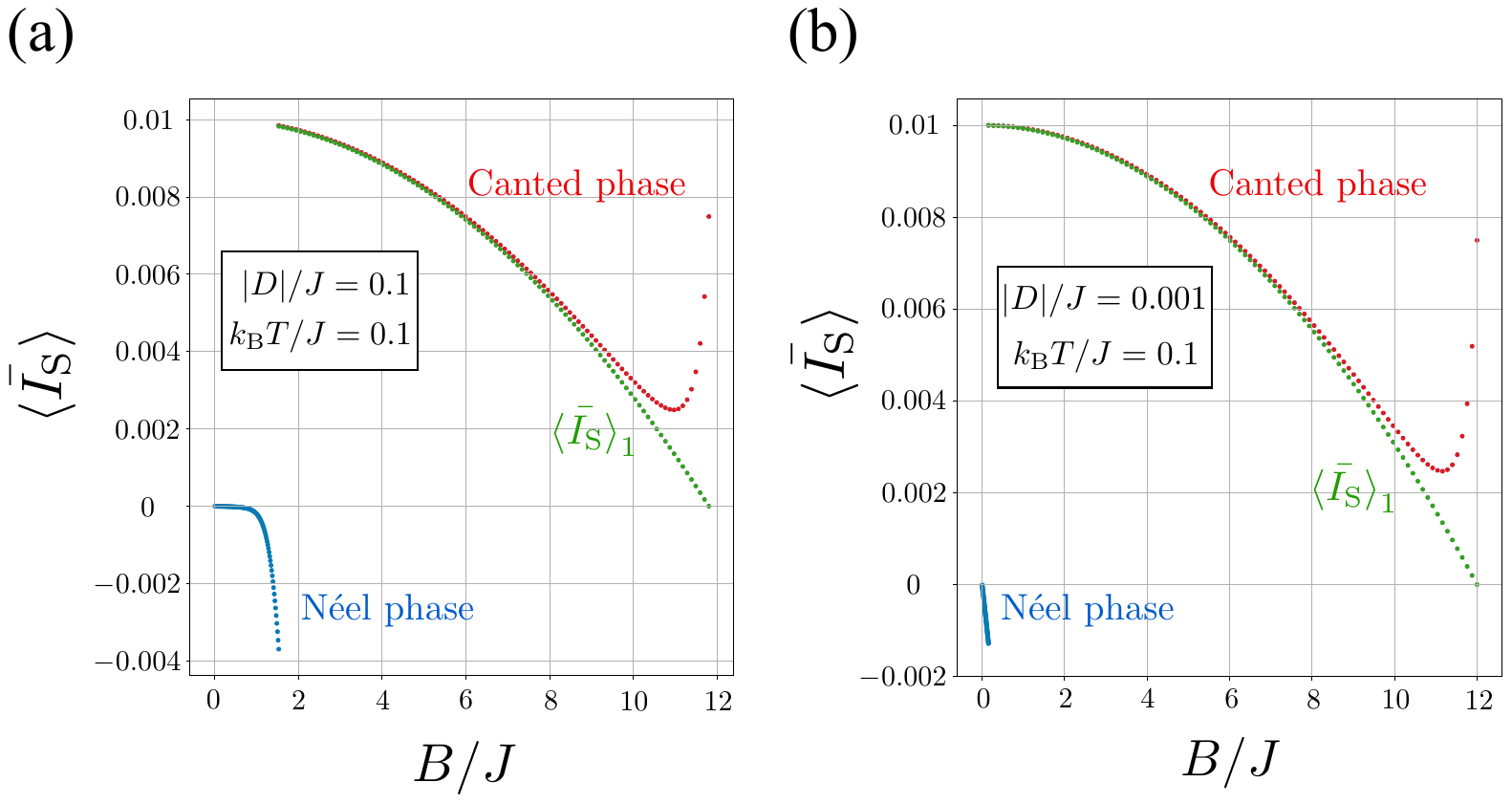}
 \caption{(Color online)
 Magnetic-field dependences of the tunneling spin current in the N\'eel [Eq.~(\ref{SSE-TunnelingSpinCurrent-NeelPhase})] and canted [Eq.~(\ref{SSE-TunnelingSpinCurrent-CantedPhase})] phases at a low temperature of $k_{\textrm{B}}T/J=0.1$.
 Parameters are set to $S=1$, $a_{0}=1$, $J=1$, $\eta=0.01|\omega|+0.001$, $\hbar=1$, and $\delta\mu_{\textrm{s}}/k_{\textrm{B}}\Delta T=0.01$.
 Panel (a) is the result for $|D|/J=0.1$, whereas (b) is for $|D|/J=0.001$.
 Blue and red dotted lines represent the spin currents in the N\'eel and canted phases, respectively.
 The green dotted line corresponds to the static part of the spin current, $\bar{\langle I_{\textrm{S}}\rangle}_{1}$, in the canted phase [Eq.~(\ref{StaticPart_CantSSE})].
 }
 \label{FIG:FieldDepSpinCurrent-Low}
\end{figure}

If we focus on a sufficiently low-temperature regime for the canted phase, our microscopic theory can predict a new property of the SSE spin current.
Figure~\ref{FIG:FieldDepSpinCurrent-Low} shows the field dependences of Eqs.~(\ref{SSE-TunnelingSpinCurrent-NeelPhase}) and (\ref{SSE-TunnelingSpinCurrent-CantedPhase}) at low temperature $k_{\textrm{B}}T< J$.
Even in this low-temperature regime, some properties of the spin current still survive: The sign reversal at the spin-flop transition occurs, and the spin current takes a negative (positive) value in the N\'eel (canted) phase.
However, from Fig.~\ref{FIG:FieldDepSpinCurrent-Low}, one sees that the spin current in the canted phase changes non-monotonically with respect to the magnetic field.
This behavior can be understood as follows.
At low temperature, the magnon density is low and it means a small amount of spin-current carriers.
As a result, the static part $\bar{\langle I_{\textrm{S}}\rangle}_{1}$ of the tunneling spin current [Eq.~(\ref{StaticPart_CantSSE})] is generally dominant over the dynamical part $\bar{\langle I_{\textrm{S}}\rangle}_{2}$ [Eq.~(\ref{DynamicalPart_CantSSE})] in the sufficiently low-temperature region.
Since the canted angle $\theta$ is a monotonically decreasing function of the magnetic field $B$ [see Fig.~\ref{FIG:AntiferromagneticPhase}(b)], the static part decreases as the field increases.
When the field $B$ becomes close to the saturation value $B_{\textrm{c}}$, the static part $\bar{\langle I_{\textrm{S}}\rangle}_{1}$ approaches zero, and hence the dynamical part $\bar{\langle I_{\textrm{S}}\rangle}_{2}$ is again dominant.
Therefore, we can observe the non-monotonic field dependence of the spin current, as shown in Fig.~\ref{FIG:FieldDepSpinCurrent-Low}.

The non-monotonic behavior of the spin current has never been observed experimentally.
Our theory predicts that such behavior can be observed in the canted phase of antiferromagnets if the temperature is set sufficiently low.
\subsection{Shrinkage of the magnetic moment in the N\'eel phase at low temperature}
\label{subSec:Low-Temperatures_Neel}
\begin{figure}[t]
 \includegraphics[width=\linewidth]{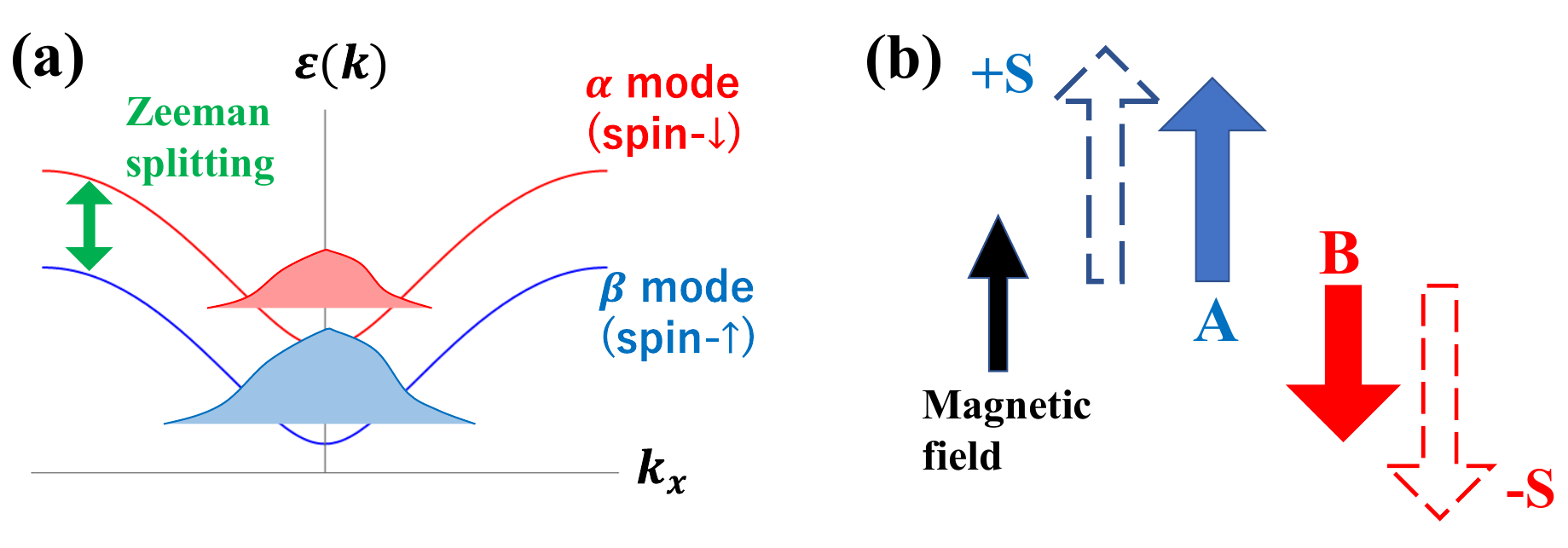}
 \caption{(Color online)
  (a) Image of $\alpha$ and $\beta$ magnon densities in the N\'eel phase with a finite magnetic field $B$.
  (b) Shrinkage of magnetic moments on A and B sublattices in the N\'eel phase.
  Blue and red arrows respectively indicate the spin expectation values on A and B sublattices.
  Dotted arrows show their classical values $\pm S$.
  Due to a finite magnon density, the spin moment generally becomes smaller than the classical value $S$.
  The moment on A sublattice is larger than that on B sublattice because $\beta$ magnons mainly reside on B sublattices, and their high density decreases the moment on B sublattice.
 }
 \label{FIG:MagnonDensity}
\end{figure}
\begin{figure}[t]
 \includegraphics[width=\linewidth]{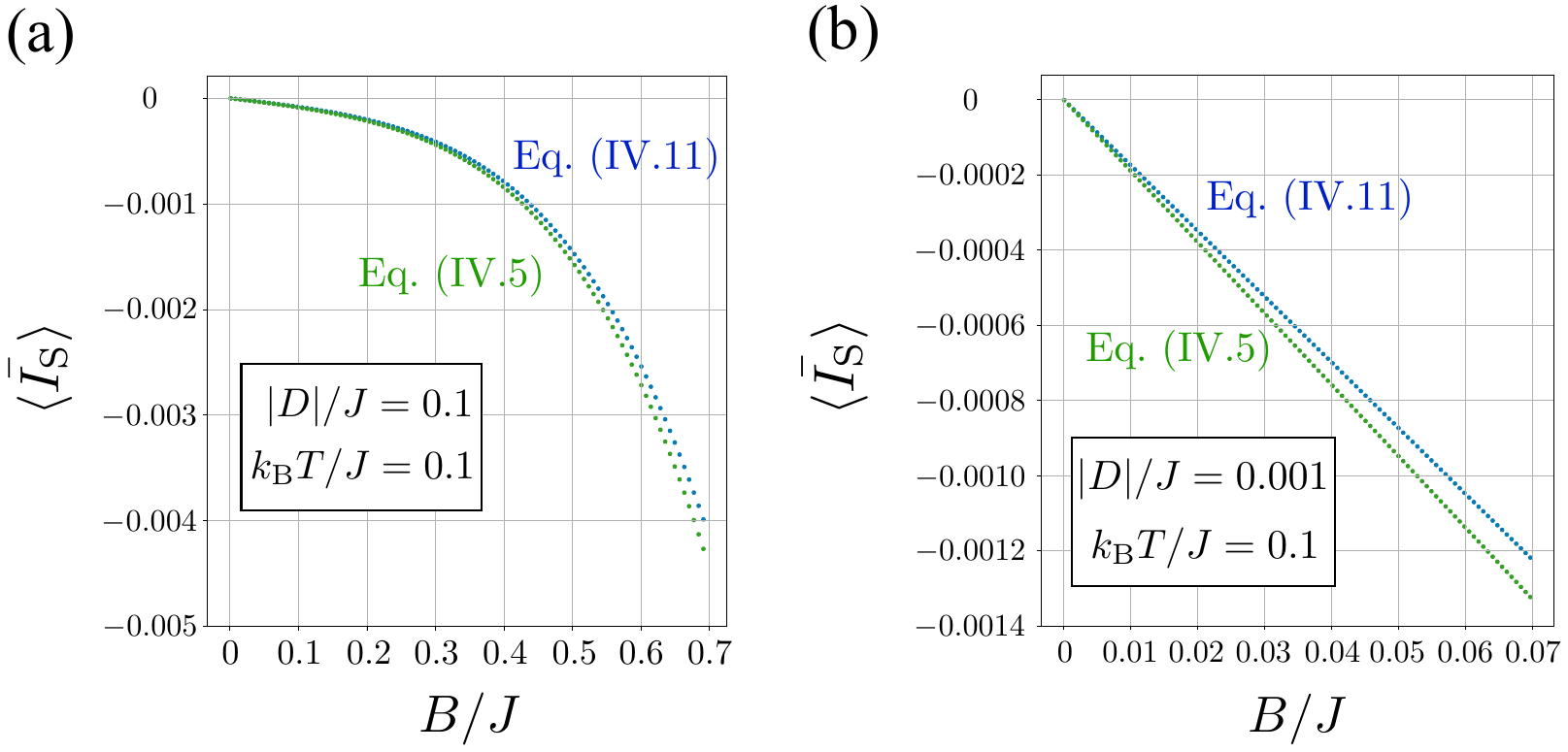}
 \caption{(Color online)
  Magnetic-field dependences of the tunneling spin current in the N\'eel phase at $k_{\textrm{B}}T/J=0.1$.
  Panels (a) and (b) are for $|D|/J=0.1$ and $|D|/J=0.001$, respectively.
  Blue and green dotted lines respectively represent the modified spin current [Eq.~(\ref{TempEffect-SpinCurrent-Neel})] and the simple linear-spin-wave result [Eq.~(\ref{SSE-TunnelingSpinCurrent-NeelPhase})].
  We set $S=1/2$, $a_{0}=1$, $J=1$, $\eta=0.01|\omega|+0.001$, and $\hbar=1$.
 }
 \label{FIG:TempEffect_ModifiedSpinCurrent}
\end{figure}
As we already explained, the reason why the spin current in the N\'eel phase takes a negative value is that the spin-up magnon ($\beta$ mode) density is higher than the spin-down magnon ($\alpha$ mode) density owing to the Zeeman splitting [see Fig.~\ref{FIG:MagnonDensity}(a)].
However, in order to more accurately compute the spin current beyond the linear spin-wave approximation, we have to consider the shrinkage of magnetic moments on A and B sublattices.
The increase in magnon densities generally means a reduction in magnetic moments from the classical spin configuration.
Therefore, in the N\'eel phase, the more the spin-up magnon density increases, the more the spin moment on B sublattice decreases since the spin-up magnons ($\beta$ mode) mainly reside on B sublattice [see Fig.~\ref{FIG:MagnonDensity}(b)].
Let us take this quantum-fluctuation effect into account in calculating the tunneling spin current.
Within the linear spin-wave theory, we have approximated the transverse spin as
$S_{\siteA}^{+}=\sqrt{2S}a_{\siteA}$ and $S_{\siteB}^{+}=\sqrt{2S}b_{\siteB}^\dagger$ in Eq.~(\ref{LSWA-Neel}), but we should replace the approximations with the original expressions $S_{\siteA}^{+}=(2S-a_{\siteA}^\dagger a_{\siteA})^{1/2}a_{\siteA}$ and
$S_{\siteB}^{+}=b_{\siteB}^\dagger(2S-b_{\siteB}^\dagger b_{\siteB})^{1/2}$ if we more quantitatively include the effects of magnon densities.
It is not easy to correctly treat the square root term, but we may take the simple approximation
\begin{align*}
  &                                              
 2S-a_{\siteA}^\dagger a_{\siteA}\to
 2S-\langle a_{\siteA}^\dagger a_{\siteA}\rangle \\
  &                                              
 2S-b_{\siteB}^\dagger b_{\siteB}\to
 2S-\langle b_{\siteB}^\dagger b_{\siteB}\rangle
\end{align*}
for a sufficiently low temperature, in which the magnon densities $\langle a_{\siteA}^\dagger a_{\siteA}\rangle$ and $\langle b_{\siteB}^\dagger b_{\siteB}\rangle$ are low enough.
Within the linear spin-wave approximation, the magnon densities are easily estimated as
\begin{widetext}
 \begin{align}
  \langle a_{\siteA}^{\dagger}a_{\siteA}\rangle=
  \frac{1}{N/2}  \sum_{\wn}
  \frac{\cosh^{2}\phi_{\wn}}{e^{\varepsilon^{(\textrm{N})}_{\alpha}(\wn)/k_{\textrm{B}}T}-1}
  +\frac{1}{N/2}\sum_{\wn} \frac{\sinh^{2}\phi_{\wn}}{e^{\varepsilon^{(\textrm{N})}_{\beta}(\wn)/k_{\textrm{B}}T}-1}
  + \frac{1}{N/2}\sum_{\wn}\sinh^{2}\phi_{\wn},
  \nonumber \\
  \langle b_{\siteB}^{\dagger}b_{\siteB}\rangle=
  \frac{1}{N/2}\sum_{\wn}
  \frac{\cosh^{2}\phi_{\wn}}{e^{\varepsilon^{(\textrm{N})}_{\beta}(\wn)/k_{\textrm{B}}T}-1}
  +\frac{1}{N/2}\sum_{\wn} \frac{\sinh^{2}\phi_{\wn}}{e^{\varepsilon^{(\textrm{N})}_{\alpha}(\wn)/k_{\textrm{B}}T}-1}
  +\frac{1}{N/2}\sum_{\wn}\sinh^{2}\phi_{\wn}.
  \label{MagnonNumber}
 \end{align}
 Using the approximated relations $S_{\siteA}^{+}\simeq\sqrt{2S-\langle a_{\siteA}^\dagger a_{\siteA}\rangle}a_{\siteA}$ and $S_{\siteB}^{+}\simeq b_{\siteB}^\dagger\sqrt{2S-\langle b_{\siteB}^\dagger b_{\siteB}\rangle}$, we can compute the tunneling spin current with the magnon-density effect as
 \begin{multline}
  \bar{\langle I_{\textrm{S}}\rangle}
  =\frac{1}{8\pi}\frac{1}{\qty(k_{\textrm{B}}T)^{2}}\frac{1}{N/2}
  \sum_{\bm k}
  \int_{-\infty}^{\infty}d\omega
  \frac{\qty(\hbar\omega)^{2}}{\sinh^{2}[\hbar\omega/2k_{\textrm{B}}T]}
  \\
  \times
  \left[
  \Big\{\qty(2S-\langle a_{\siteA}^{\dagger}a_{\siteA}\rangle)\cosh^{2}\phi_{\wn}
  +\qty(2S-\langle b_{\siteB}^{\dagger}b_{\siteB}\rangle)\sinh^{2}\phi_{\wn}\Big\}
  \frac{\eta}{\big\{\omega-\varepsilon_{\alpha}^{(\textrm{N})}(\wn)/\hbar\big\}^2+\eta^{2}}
  \right.
  \\
  \left.
  \quad\quad\quad\quad\quad
  -\Big\{\qty(2S-\langle b_{\siteB}^{\dagger}b_{\siteB}\rangle)\cosh^{2}\phi_{\wn}
  +\qty(2S-\langle a_{\siteA}^{\dagger}a_{\siteA}\rangle)\sinh^{2}\phi_{\wn}\Big\}
  \frac{\eta}{\big\{\omega+\varepsilon_{\beta}^{(\textrm{N})}(\wn)/\hbar\big\}^2+\eta^{2}}
  \right].
  \label{TempEffect-SpinCurrent-Neel}
 \end{multline}
\end{widetext}
The magnitude of the spin current is decreased owing to the factors $\langle a_{\siteA}^\dagger a_{\siteA}\rangle$ and $\langle b_{\siteB}^\dagger b_{\siteB}\rangle$ in Eq.~(\ref{TempEffect-SpinCurrent-Neel}). We emphasize that the contribution from B sublattice is more decreased than that from A sublattice because of the inequality $\langle b_{\siteB}^\dagger b_{\siteB}\rangle > \langle a_{\siteA}^\dagger a_{\siteA}\rangle$ in a finite magnetic field $B>0$.

From these arguments, we can say (as shown in Fig.~\ref{FIG:MagnonDensity}) that a magnetic field simultaneously induces (i) a density difference between $\alpha$ and $\beta$ modes and (ii) a reduction in magnetic moments.
The fact (i) favors a negative value of the spin current, whereas the moment reduction (ii) tends to make the spin current positive (i.e., the magnitude of the spin current decreases).
Namely, there is a competition in terms of the sign of the spin current and the possibility that the spin current becomes positive in the N\'eel phase.
Figure~\ref{FIG:TempEffect_ModifiedSpinCurrent} shows that the magnitude of the modified spin current of Eq.~(\ref{TempEffect-SpinCurrent-Neel}) slightly decreases from the linear-spin-wave result, whereas the spin current is still negative.
From this result, we conclude that even at low temperature, the sign of the spin current in the N\'eel phase is always opposite to that in ferromagnets within the formalism of the interface spin current.

In addition, from the above arguments, we can expect that the spin current in the N\'eel phase will be positive if the system possesses very large quantum fluctuations and the moment reduction is larger.
More frustrated antiferromagnets or quasi-two-dimensional ones might be candidates for a positive spin current.

We note that the reduction in the magnetic moment also occurs in the canted phase, but its effect is not so critical because there is no competition in the canted phase.
\section{Comparison with other magnets}
\label{Sec:Comparison}
In this section, we compare the SSE in antiferromagnetic insulators with that in other magnets.
First, we concentrate on ferromagnets whose SSE has been well studied in spintronics.
We quantitatively compare the spin currents in ferromagnets and antiferromagnets in Sec.~\ref{subSec:Comparison_Ferro}.
Second, we simply discuss experimental results for SSEs in various types of magnets in Sec.~\ref{subSec:Comparison}.
\subsection{Comparison with ferromagnetic insulators}
\label{subSec:Comparison_Ferro}
\begin{figure}[t]
 \includegraphics[width=\linewidth]{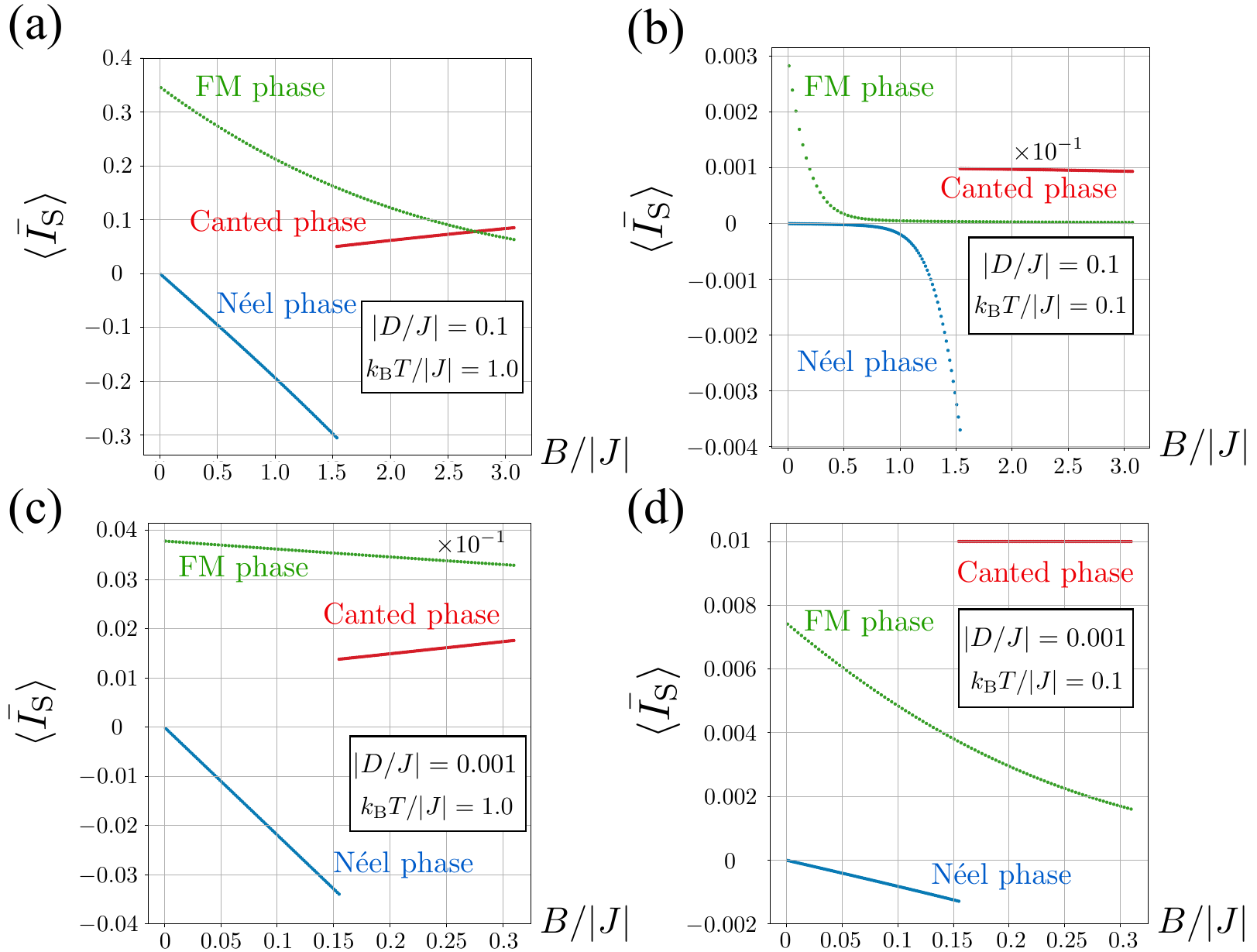}
 \caption{(Color online)
 Magnetic-field dependences of the tunneling spin currents for ferromagnets and antiferromagnets.
 We set $S=1$, $a_{0}=1$, $|J|=1$, $\eta=0.01|\omega|+0.001$, $\hbar=1$, $\delta\mu_{\textrm{s}}/k_{\textrm{B}}\Delta T=0.01$, and $N=N'$.
 Panels (a) and (b) are for $|D/J|=0.1$, and panels (c) and (d) are for $|D/J|=0.001$.
 We assume that the values of $D$, $J_{\textrm{sd}}$, and $N_{\textrm{int}}$ in the antiferromagnet are the same as those in the ferromagnet.
 Blue, red, and green dotted lines respectively correspond to the spin currents for the N\'eel phase, the canted phase in antiferromagnets, and the ferromagnetic ordered phase in ferromagnets.
 }
 \label{FIG:Comparison}
\end{figure}
As we mentioned in the Introduction, the formalism of the tunneling spin current has the following advantages: (i) It can be applied to a broad class of magnetic systems, and (ii) one can compare the value of the spin currents in different magnetic systems without any additional parameters.
In spintronics, the tunneling-current formalism was first applied to the SSE in 3D ordered ferromagnets~\cite{Jauho1994,Adachi2011}.

Here, we briefly review the SSE in ferromagnets in order to compare it with our result for antiferromagnets.
A typical Hamiltonian for ferromagnets is obtained by changing the sign of the exchange coupling $J$ in Eq.~(\ref{ModelHamiltonian-AFM}).
According to the linear spin-wave theory, the spin-wave Hamiltonian for ferromagnets with polarization $\langle S^z\rangle >0$ is approximated by
\begin{equation}
 \mathscr{H}_{\textrm{SW}}^{\textrm{Ferro}}
 =\sum_{\wn}
 \varepsilon_{\textrm{SW}}(\wn)
 a_{\wn}^{\dagger}a_{\wn},
 \label{MagnonDispersion-Ferromagnets}
\end{equation}
where $a_{\wn}^{\dagger}\ (a_{\wn})$ is the creation (annihilation) operator of the magnon with wave vector $\wn$, and $\varepsilon_{\textrm{SW}}(\wn)=2S|J|\qty(3-\gamma_{\wn})+2S|D|+B$ is the magnon dispersion.
In the magnon (spin-wave) picture, the transverse spin correlation function of the ferromagnetic Heisenberg model is given by the magnon Green's function
\begin{equation}
 G_{+-}^{\textrm{R}}\qty(\wn,\omega)
 =\frac{2S}{\omega-\varepsilon_{\textrm{SW}}\qty(\wn)/\hbar+i\eta}.
 \label{SpinCorrelationFunction-FerromagneticOrder}
\end{equation}
Using these tools, we can compute the tunneling spin current for the SSE in ordered ferromagnets.
We can obtain the spin current by replacing $N_{\textrm{int}}/2$ and
$\frac{1}{N/2}\sum_{\wn}\big\{\textrm{Im}\mathcal{G}_{+-}^{(\textrm{A})\textrm{R}}\qty(\wn,\omega)+\textrm{Im}\mathcal{G}_{+-}^{(\textrm{B})\textrm{R}}\qty(\wn,\omega)\}$ with $N_{\textrm{int}}$ and $\frac{1}{N'}\sum_{\wn}\textrm{Im}G_{+-}^{\textrm{R}}\qty(\wn,\omega)$ (where $N'$ is the total number of sites), respectively, in Eq.~(\ref{TunnelSinCurrent_finmal1}).
The explicit form is hence given by
\begin{multline}
 \bar{\langle I_{\textrm{S}}\rangle}=
 \frac{S}{4\pi}\frac{1}{\qty(k_{\textrm{B}}T)^{2}}
 \frac{1}{N'}\sum_{\wn}\int_{-\infty}^{\infty}d\omega
 \frac{\qty(\hbar\omega)^{2}}{\sinh^{2}[\hbar\omega/2k_{\textrm{B}}T]}
 \\
 \times
 \frac{\eta}{\big\{\omega-\varepsilon_{\textrm{SW}}(\wn)/\hbar\big\}^2+\eta^{2}},
 \label{SSE-SpinCurrent-Ferromagnet}
\end{multline}
where we have assumed that the Hamiltonians for the metal and the interface are the same as those of antiferromagnets.

In Fig.~\ref{FIG:Comparison}, we quantitatively compare the value of the tunneling spin currents in an antiferromagnet [Eqs.~(\ref{SSE-TunnelingSpinCurrent-NeelPhase}) and (\ref{SSE-TunnelingSpinCurrent-CantedPhase})] and a ferromagnet [Eq.~(\ref{SSE-SpinCurrent-Ferromagnet})].
We have set $S=1$, $|J|=1$, and $N=N'$ in the ferromagnet and the antiferromagnet.
We have also supposed that values of $D$, $J_{\textrm{sd}}$, and $N_{\textrm{int}}$ in the antiferromagnet are the same as those in the ferromagnet.
Figure~\ref{FIG:Comparison} shows that (i) the magnitude of the spin current in the antiferromagnet approaches the same order as that in the ferromagnet in an intermediate-temperature range $k_{\rm B}T\sim |J|$, and (ii) the former is larger than that of the ferromagnet in a low-temperature range $k_{\rm B}T\ll |J|$.
In several SSE experiments~\cite{Uchida2010,Kikkawa2015,Seki2015,Wu2016,Li2020a} on ferromagnetic and antiferromagnetic insulators, the measured SSE voltages have been usually on the order of micro volts.
Therefore, we can conclude that the computed spin current for antiferromagnets is sufficiently reliable on a semi-quantitative level.
The study in Ref.~\cite{Hirobe2017d} has also performed a similar comparison between spin currents in SSEs for a ferromagnet and an antiferromagnetic spin-$1/2$ chain.
It concludes that the spinon spin current in the spin-$1/2$ chain is $10^{-3}\sim 10^{-4}$ order of magnitude smaller than that in ferromagnets, which agrees well with the experimental result.
These experimental and theoretical works imply that the microscopic formalism for the tunneling current~\cite{Jauho1994} correctly captures some essential features of the spin currents in SSEs (especially their magnetic-field dependence) for different types of magnets.
\subsection{Comparison with various magnets}
\label{subSec:Comparison}
\begin{figure*}[t]
 \includegraphics[width=\linewidth]{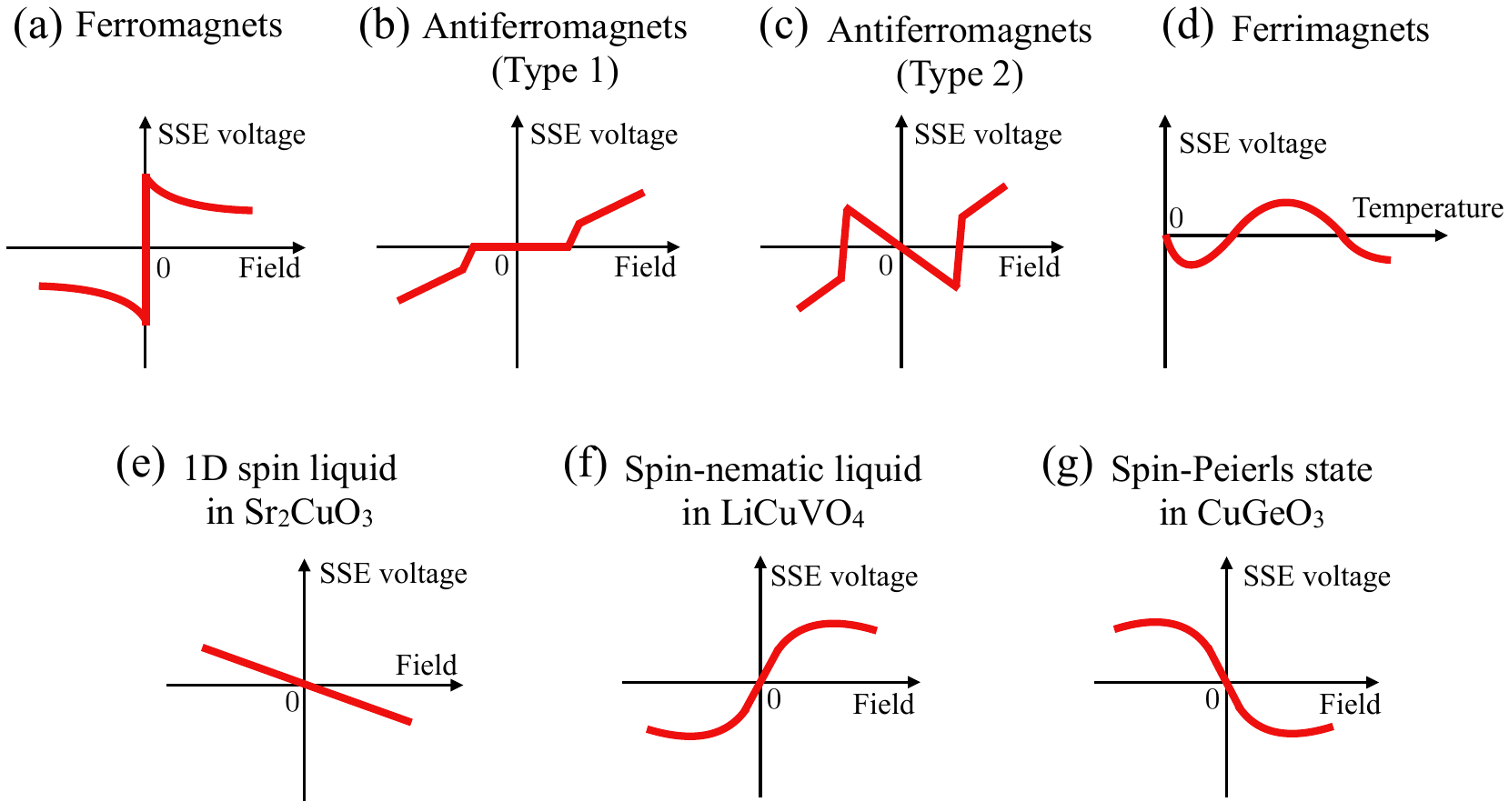}
 \caption{(Color online)
  Schematic views of field or temperature dependences of SSE voltages in magnetically ordered phases [(a)-(d)] and quantum disordered states [(e)-(g)].
  These schematic images all represent the typical behavior observed in experiments.
  Several features of these results have been theoretically explained by using the tunnel current formula.
  Panels (a) is for a typical ferromagnet (see, e.g., Ref.~\cite{Kikkawa2015}).
  Panel (b) is the SSE voltage for the antiferromagnets $\rm MnF_2$~\cite{Wu2016} and $\rm Cr_2O_3$~\cite{Seki2015} without a sign change, and panel (c) is the experimental result accompanying a sign reversal for the antiferromagnet $\rm Cr_2O_3$ observed by another group~\cite{Li2020a}.
  Panel (d) is the temperature dependence of the SSE voltage in a ferrimagnet~\cite{Geprags2016}, which includes two sign reversals.
  Panels (e), (f), and (g) are respectively the magnetic-field dependence of the SSE voltage in a 1D spin liquid with spinon excitations~\cite{Hirobe2017d}, a spin-nematic liquid with both magnons and magnon-pairs~\cite{Hirobe2019}, and a spin-Peierls state with triplons~\cite{Chen2021}.
 }
 \label{FIG:SpinSeebeckEffect}
\end{figure*}

Here, we briefly discuss experimental results for SSEs in different magnets.
As we mentioned in the Introduction, SSEs have recently been investigated both experimentally and theoretically in various magnets in addition to the usual ferro or ferrimagnets.
Figure~\ref{FIG:SpinSeebeckEffect} shows typical magnetic-field or temperature dependences of SSE voltage (proportional to the tunneling spin current) in different magnets.
In panels (e)-(g), magnetic states even without magnetic order are shown to potentially carry the spin current through a thermal gradient.
Theoretical studies~\cite{Adachi2010,Hirobe2017d,Hirobe2019,Chen2021} show that the tunneling spin-current formalism can explain the main features (especially the magnetic-field dependence) of SSEs in different magnets.
This means that the magnetic DoS plays an important role in SSE spin currents in a broad class of magnets.
As we discussed in Sec.~\ref{Sec:DoS}, the magnetic DoS $\propto -{\rm Im}\mathcal{G}^{\rm R}_{\pm\mp}(\omega)$ can be observed via inelastic neutron scattering.
Therefore, experiments on the SSE and neutron scattering cooperatively leads to more deep understanding of various magnets and their excitations.

From Fig.~\ref{FIG:SpinSeebeckEffect}, we see that SSE voltages exhibit rich variety as functions of $k_{\rm B}T$ and $B$, depending on the types of magnetic orders and excitations.
SSE research has far focused on application, and therefore ferromagnets (typically YIG) have been mainly investigated.
However, as we mentioned earlier, Fig.~\ref{FIG:SpinSeebeckEffect} indicates that the SSE is also useful for detecting some characteristic features of various magnets.
In other words, the SSE can be used to probe magnetic properties.
For instance, it is generally difficult to experimentally detect the features of quantum spin liquids and topological states in frustrated magnets~\cite{Balents2010,Savary2016,Han2020}.
The SSE would be useful for studying such "invisible" quantum states in magnets.
\section{Discussion}
\label{Sec:Discussion}
In Secs.~\ref{Sec:SpinCurrent}$-$\ref{Sec:Comparison}, based on the spin-wave theory and the non-equilibrium Green's function method, we analyze the SSE of antiferromagnets.
Our theoretical result nicely explains the sign reversal and the magnetic-field dependence of the spin current observed in a recent experiment for $\rm Cr_2O_3$~\cite{Li2020a}.
However, our theory does not agree with two other experimental results~\cite{Seki2015,Wu2016} where no sign reversal occurred.
In this section, we qualitatively consider important effects and possibilities not taken into account by the tunneling spin-current formalism.

First, we discuss the missing interaction term on the interface in the canted phase.
Our spin-current formula starts from the Heisenberg's equation of motion for total spin in the metal (or antiferromagnet), like Eq.~(\ref{TunnelingDCSpinCurrent-Operator}).
To calculate this spin current, we simply neglect the $z$ component of the interface interaction $S_{\bm r}^z\sigma_{\bm r}^z$.
This treatment is justified when the $z$ component of "local" spins is conserved in the system, in which
$S_{\bm r}^z\sigma_{\bm r}^z$ does not induce any dynamics of the $z$ component of spins.
However, in the canted phase, the local $S_{\bm r}^z$ conservation is broken due to the transverse magnetization (see Fig.~\ref{FIG:SpinFlip}). As a result, $S^z_{\bm r}$ possesses a constant term and one-magnon operators like transverse spins $S^{\pm}_{\bm r}$ [see Eqs.~(\ref{MagnonRepresentation-AsublatticeCantedAFSpin}) and (\ref{MagnonRepresentation-BsublatticeCantedAFSpin})].
Within the spin-wave theory, the static and one-magnon parts of $S^z_{\bm r}$ are
\begin{widetext}
 \begin{multline}
  S_{\siteA}^{z}
  \simeq
  \frac{\sqrt{S}}{2}
  \sin\theta
  \sqrt{\frac{1}{N/2}}\sum_{\wn}e^{i\wn\cdot\siteA}
  \qty(\cosh\varphi_{\wn}^{\alpha}\alpha_{\wn}
  +\sinh\varphi_{\wn}^{\alpha}\alpha_{-\wn}^{\dagger}
  +\sinh\varphi_{\wn}^{\beta}\beta_{\wn}^{\dagger}
  +\cosh\varphi_{\wn}^{\beta}\beta_{-\wn})
  \\
  +\frac{\sqrt{S}}{2}
  \sin\theta
  \sqrt{\frac{1}{N/2}}\sum_{\wn}e^{-i\wn\cdot\siteA}
  \qty(\cosh\varphi_{\wn}^{\alpha}\alpha_{\wn}^{\dagger}
  +\sinh\varphi_{\wn}^{\alpha}\alpha_{-\wn}
  +\sinh\varphi_{\wn}^{\beta}\beta_{\wn}
  +\cosh\varphi_{\wn}^{\beta}\beta_{-\wn}^{\dagger})
  +S\cos\theta,
  \label{SpinZMagnon_AsubCanted}
 \end{multline}
 \begin{multline}
  S_{\siteB}^{z}
  \simeq
  \frac{\sqrt{S}}{2}
  \sin\theta
  \sqrt{\frac{1}{N/2}}\sum_{\wn}e^{-i\wn\cdot\siteB}
  \qty(\sinh\varphi_{\wn}^{\alpha}\alpha_{\wn}^{\dagger}
  +\cosh\varphi_{\wn}^{\alpha}\alpha_{-\wn}
  -\cosh\varphi_{\wn}^{\beta}\beta_{\wn}
  -\sinh\varphi_{\wn}^{\beta}\beta_{-\wn}^{\dagger})
  \\
  +\frac{\sqrt{S}}{2}
  \sin\theta
  \sqrt{\frac{1}{N/2}}\sum_{\wn}e^{i\wn\cdot\siteB}
  \qty(\sinh\varphi_{\wn}^{\alpha}\alpha_{\wn}
  +\cosh\varphi_{\wn}^{\alpha}\alpha_{-\wn}^{\dagger}
  -\cosh\varphi_{\wn}^{\beta}\beta_{\wn}^{\dagger}
  -\sinh\varphi_{\wn}^{\beta}\beta_{-\wn})
  +S\cos\theta.
  \label{SpinZMagnon_BsubCanted}
 \end{multline}
\end{widetext}
Therefore, in addition to the transverse spin interaction on the interface, the Ising interaction $S_{\bm r}^z\sigma_{\bm r}^z$ potentially contributes to the tunneling spin current in the first-order perturbation calculation.
Roughly speaking, the perturbation term is given by a product of two correlators,
$\langle S_{\bm r}^z(t)S_{\bm r}^{\pm}(t')\rangle$ and $\langle \sigma_{\bm r}^z(t)\sigma_{\bm r}^{\pm}(t')\rangle$.
From Eqs.~(\ref{SpinZMagnon_AsubCanted}) and (\ref{SpinZMagnon_BsubCanted}), the former correlator $\langle S_{\bm r}^z(t)S_{\bm r}^{\pm}(t')\rangle$ includes both a static term and magnon Green's functions.
However, the value of the correlation function $\langle \sigma_{\bm r}^z(t)\sigma_{\bm r}^{\pm}(t')\rangle$ strongly depends on the types and strengths of spin-orbit couplings in the metal.
We thus expect that these correlators somewhat change the value of spin current in the canted phase, especially near the spin-flop transition, $B\sim B_{\rm f}$.
Formulating a theory including the effect of these correlators is an important future issue in spintronics.

We have also ignored higher-order magnon interaction terms in calculating spin correlation functions.
However, the effects of such higher-order terms are generally weak if we consider magnetically ordered phases at sufficiently low temperature.

\begin{figure}[t]
 \includegraphics[width=\linewidth]{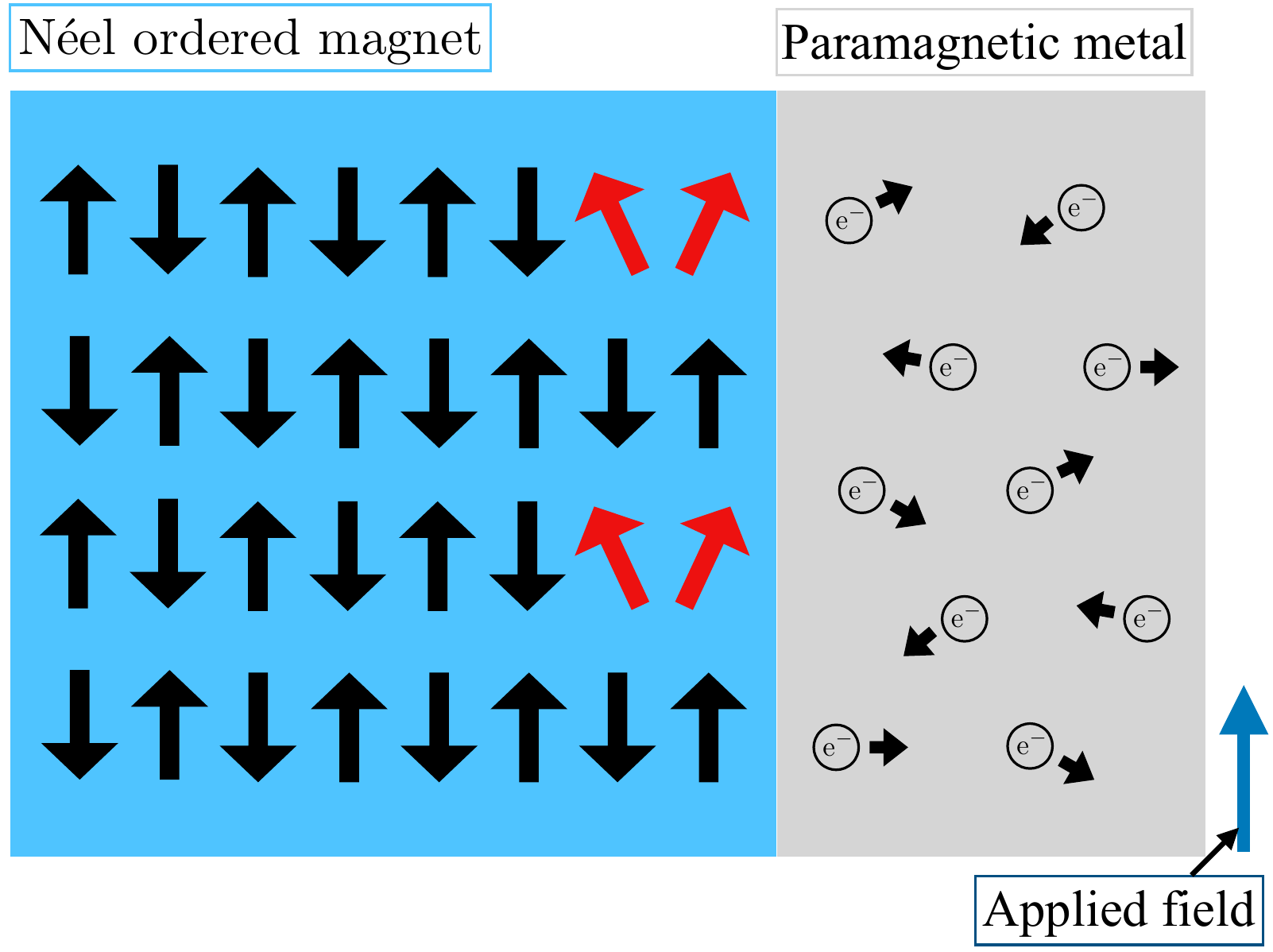}
 \caption{(Color online)
  Schematic view of a magnetic structure near the interface between a N\'eel ordered magnet and a paramagnetic metal.
  It is possible that the magnetic order partially changes in the vicinity of the interface because of interface-driven magnetic anisotropies.
 }
 \label{FIG:Interface}
\end{figure}

Second, we consider the interface properties.
As already mentioned, a sign reversal of the SSE voltage has been observed in an experiment on a $\rm Cr_2O_3$-Pt bilayer system ~\cite{Li2020a}, whereas another experiment on the same bilayer set-up did not detect any sign reversal~\cite{Seki2015}.
As the authors in Ref.~\cite{Li2020a} briefly discussed, these two experimental results indicate that the type of interface and the method of creating it strongly affect the tunneling spin current and the resulting SSE voltage.
The presence of an interface generally reduces the symmetry of a focused system compared with a bulk material.
Inversion and translation symmetries are clearly broken owing to the interface.
Such low-symmetry systems allow the appearance of some types of magnetic anisotropies such as Dzyaloshinskii-Moriya and additional single-ion interactions in the vicinity of the interface.
For instance, we expect that additional magnetic anisotropies near the interface change the spin orientation from that of the bulk, as shown in Fig.~\ref{FIG:Interface}.
If a canted state appears near the interface with a certain probability when the bulk system is in the N\'eel ordered phase, the tunneling spin current might become positive, which is different from the negative value predicted in the N\'eel phase.
This scenario potentially (partially) explains the positive SSE voltage observed in Refs.~\cite{Seki2015,Wu2016,Li2020a}.

Finally, we discuss the spin-current transport in the bulk antiferromagnet.
The microscopic theory in this paper focuses on the tunneling process at the interface, and (as we discussed in the last section) this formalism has successfully explained SSEs in various magnets (see Fig.~\ref{FIG:SpinSeebeckEffect}) including antiferromagnets.
However, in real experiments, a spin current exists not only around the interface but also in the whole bulk region: The spin current flows along the temperature gradient throughout the whole antiferromagnet, and part of it arrives at the interface.
To complete the microscopic theory of the SSE in a bilayer system of a magnet and a metal, we have to combine the tunneling spin-current formalism and a bulk transport theory such as Boltzmann's equation approach, without any phenomenological parameter.
In the N\'eel phase, both spin-up and spin-down magnons flow along the thermal gradient, and hence spin-moment accumulation near the interface is expected to be relatively small because of the cancellation between spin-up and spin-down magnons.
This implies that the bulk transport of spin current in the N\'eel phase is more important than that in the canted phase.
Unifying bulk and boundary transports is another critical future step in understanding spin transport phenomena including the SSE.
\section{Conclusion}
\label{Sec:Conclusion}
In summary, we have developed a microscopic theory for the SSE of antiferromagnetic insulators.
Our theory well explains the sign reversal and the magnetic-field dependence of the SSE voltage ($\propto$ the tunneling spin current) in a semi-quantitative level.
In this last section, we simply summarize the contents of this paper.

In Secs.~\ref{Sec:Model}$-$\ref{Sec:SSE}, based on the non-equilibrium Green's function and the spin-wave theory, we analyzed the tunneling DC spin current generated by a thermal gradient in a bilayer system consisting of an antiferromagnet and a metal (see Fig.~\ref{FIG:SSE-Antiferromagnets}).
Our analysis shows that in antiferromagnets, the dominant carrier of the spin current suddenly changes at the first-order spin-flop transition and then a sign reversal of the SSE voltage occurs (see Fig.~\ref{FIG:FieldDepSpinCurrent-Intermediate}).
From the argument based on the spin-current formula and the magnon DoSs, we confirmed that the spin current due to spin-up magnons is dominant in the N\'eel phase, whereas spin-down magnons are dominant in the canted phase.
We explained that the tunneling spin current is strongly associated with the magnetic DoS, which can be observed via inelastic neutron scattering.
Namely, we emphasized the close relationship between the SSE and neutron-scattering spectra.

In the canted phase, the spin current was shown to consist of two parts, a static term and magnon correlation.
The former can be interpreted as the effect of electron spin-flip caused by the transverse magnetization (see Fig.~\ref{FIG:SpinFlip}).
This is very reminiscent of the spin flip in spin Hall magnetoresistance~\cite{Nakayama2013}.
From the result of two main parts of the spin current, we predicted that in the canted phase at low temperature, the spin current non-monotonically change as a function of the magnetic field $B$ (see Fig.~\ref{FIG:FieldDepSpinCurrent-Low}).
Such non-monotonic behavior has never been observed in the SSE of antiferromagnets, and its detection would be a test to enhance the reliability of our theory.

In Sec.~\ref{Sec:Comparison}, we compared the SSE of antiferromagnets with that of various magnets.
In particular, we quantitatively compared ferromagnets and antiferromagnets, and showed that the magnitude of the SSE voltage in antiferromagnets can approach that of ferromagnets.
This is consistent with observed SSE voltages.
Moreover, Fig.~\ref{FIG:SpinSeebeckEffect} shows that the magnetic-field and temperature dependences of the SSE voltage strongly depend on the types of excitations and orders in magnetic systems, and it indicates that the SSE can also be useful for detecting characteristic features of different magnets~\cite{Uchida2010,Kikkawa2023,Seki2015,Wu2016,Li2020a,Geprags2016,Hirobe2017d,Hirobe2019,Chen2021,Xing2022}.

Finally, in Sec.~\ref{Sec:Discussion}, we discussed important pieces missing from the tunneling spin-current formalism used in this paper.
We pointed out the importance of the Ising interaction $S_{\bm r}^z\sigma_{\bm r}^z$ on the magnet-metal interface in the canted phase, the material dependences of interface properties, and the effects of the spin-current transport in bulk far from the interface.
\begin{acknowledgments}
 We are grateful to Yuichi~Ohnuma, Takeo~Kato, and Minoru~Kanega for fruitful discussions.
 We also thank Masahito~Mochizuki and Hiroyuki~Chudo for useful comments.
 K.~M is supported by JST, the establishment of university fellowships towards the creation of science technology innovation, Grant Number JPMJFS2105.
 M.~S. is supported by JSPS KAKENHI (Grant No. 20H01830 and No. 20H01849) and a Grant-in-Aid for Scientific Research on Innovative Areas~"Quantum Liquid Crystals"~(Grant No. 19H05825) and “Evolution of Chiral Materials Science using Helical Light Fields”~(Grants No. JP22H05131 and No. JP23H04576) from JSPS of Japan.
\end{acknowledgments}
\appendix
\section{Mean-field approximation for antiferromagnets}
\label{App:MFA}
In this appendix, we shortly review the mean-field results for antiferromagnets needed to draw the phase diagram in Fig.~\ref{FIG:AntiferromagneticPhase}(c).
We write down the mean-field free energies for the N\'eel and canted phases.
\subsection{N\'eel phase}
\label{App:MFA_Neel}
First, we consider the N\'eel phase in the antiferromagnetic Heisenberg model~(\ref{ModelHamiltonian-AFM}).
In the mean-field approximation, spin operators are given by
\begin{align}
  & \bm S_{\siteA}=\qty(\delta S_{\siteA}^{x},\delta S_{\siteA}^{y},m_{\textrm{A}}+\delta S_{\siteA}^{z}),  
 \\
  & \bm S_{\siteB}=\qty(\delta S_{\siteB}^{x},\delta S_{\siteB}^{y},-m_{\textrm{B}}+\delta S_{\siteB}^{z}), 
\end{align}
where $\langle S_{\siteA}^{z}\rangle=m_{\textrm{A}}>0$ and $\langle S_{\siteB}^{z}\rangle=-m_{\textrm{B}}>0$ are respectively the spin moment on A and B sublattices, and $\delta S_{\bm r_{\textrm{X}}}^{\alpha}$ is the fluctuation from the mean value of the spin $\alpha=x,y,z$ component on site $\bm r_{\textrm{X}}$.
Substituting these spin operators into Eq.~(\ref{ModelHamiltonian-AFM}) and then
ignoring terms of $\mathcal O\qty(\{\delta S_{\bm r_{\textrm{X}}}^{\alpha}\}^{2})$, we obtain the effective one-body Hamiltonian
\begin{equation}
 \mathscr{H}^{(\textrm{N\'eel})}_{\textrm{MF}}=
 E_{\textrm{cl}}
 -B^{\textrm{A}}_{\textrm{eff}}\sum_{\siteA}S_{\siteA}^{z}
 -B^{\textrm{B}}_{\textrm{eff}}\sum_{\siteB}S_{\siteB}^{z},
 \label{MFModel_Neel}
\end{equation}
where the classical ground-state energy $E_{\textrm{cl}}$ and
the effective fields $B_{\textrm{eff}}^{\textrm{A,B}}$ are defined as
\begin{align}
  & E_{\textrm{cl}}=                                     
 \frac{N}{2}\qty{ 6Jm_{\textrm{A}}m_{\textrm{B}}
 +|D|m_{\textrm{A}}^{2}+|D|m_{\textrm{B}}^{2}},\nonumber \\
  & B_{\textrm{eff}}^{\textrm{A}}=                       
 B+6Jm_{\textrm{B}}+2|D|m_{\textrm{A}},
 \nonumber                                               \\
  & B_{\textrm{eff}}^{\textrm{B}}=                       
 B-6Jm_{\textrm{A}}-2|D|m_{\textrm{B}}.
 \nonumber
\end{align}
The partition function $Z_{\textrm{MF}}^{(\textrm{N\'eel})}$
of the model Eq.~(\ref{MFModel_Neel}) is easily calculated.
The mean-field free energy $F_{\textrm{MF}}^{(\textrm{N\'eel})}=-k_{\rm B}T\ln Z_{\textrm{ MF}}^{(\textrm{N\'eel})}$ is given by
\begin{align}
 F_{\textrm{MF}}^{(\textrm{N\'eel})}=E_{\textrm{cl}}
   & -\frac{N}{2}k_{\textrm{B}}T\log 
 \qty(\frac{\sinh\big[\beta B_{\textrm{eff}}^{\textrm{A}}\left(S+\frac{1}{2}\right)\big]}{\sinh\left(\beta B_{\textrm{eff}}^{\textrm{A}}/2\right)})
 \nonumber                           \\
 - & \frac{N}{2}k_{\textrm{B}}T\log  
 \qty(\frac{\sinh\big[\beta B_{\textrm{eff}}^{\textrm{B}}\left(S+\frac{1}{2}\right)\big]}{\sinh\left(\beta B_{\textrm{eff}}^{\textrm{B}}/2\right)}).
 \label{FreeEnergy_Neel_MF}
\end{align}
The self-consistent condition determines spin expectation values as follows:
\begin{equation}
 \left\{ \,
 \begin{aligned}
   & \langle S_{\siteA}^{z}\rangle                   
  =m_{\textrm{A}}
  =SB_{S}\qty(S\beta B^{\textrm{A}}_{\textrm{eff}}), \\
   & \langle S_{\siteB}^{z}\rangle                   
  =-m_{\textrm{B}}
  =SB_{S}\qty(S\beta B^{\textrm{B}}_{\textrm{eff}}),
 \end{aligned}
 \right.
 \label{Moment_Neel_MF}
\end{equation}
where the Brillouin function $B_{S}(x)$ is defined as
\begin{equation}
 B_{S}(x)=
 \frac{2S+1}{2S}\coth\left(\frac{2S+1}{2S}x\right)
 -\frac{1}{2S}\coth\left(\frac{x}{2S}\right).
 \label{App_BrillouinFunction}
\end{equation}
\subsection{Canted AF phase}
\label{App:MFA_Canted}
Similarly, we can build up the mean-field theory for the canted phase, in which spin operators are represented as
\begin{align}
  & \bm S_{\siteA}=\qty(-m\sin\theta+\delta S_{\siteA}^{x},\delta S_{\siteA}^{y},m\cos\theta+\delta S_{\siteA}^{z}), 
 \label{Spin_A_Canted_MF}                                                                                            \\
  & \bm S_{\siteB}=\qty(m\sin\theta+\delta S_{\siteB}^{x},\delta S_{\siteB}^{y},m\cos\theta+\delta S_{\siteB}^{z}),  
 \label{Spin_B_Canted_MF}
\end{align}
where $\langle S_{\siteA}^{x}\rangle=-m\sin\theta$, $\langle S_{\siteA}^{z}\rangle=m\cos\theta$, $\langle S_{\siteB}^{x}\rangle=m\sin\theta$, and $\langle S_{\siteB}^{z}\rangle=m\cos\theta$ ($m>0$).
We have assumed that spins are in $S^z$-$S^x$ plane, and $\delta S_{\bm r_{\textrm{X}}}^{\alpha}$ is again
the fluctuation from the mean value of $S^\alpha$ on site
$\bm r_{\textrm{X}}$. From these tools, the mean-field Hamiltonian is given by
\begin{align}
 \mathscr{H}_{\textrm{MF}}^{(\textrm{Cant})}
 =E_{\rm{cl}}
 + & B_{\textrm{eff}}^{x}\sum_{\siteA}S_{\siteA}^{x}  
 -B_{\textrm{eff}}^{z}\sum_{\siteA}S_{\siteA}^{z}
 \nonumber                                            \\
   & -B_{\textrm{eff}}^{x}\sum_{\siteB}S_{\siteB}^{x} 
 -B_{\textrm{eff}}^{z}\sum_{\siteB}S_{\siteB}^{z},
 \label{MFHamiltonian_Canted}
\end{align}
where
\begin{align}
  & E_{\rm{cl}}=\frac{N}{2}\qty{6Jm^{2}\left(1-2\cos^{2}\theta\right)+2|D|m^{2}\cos^{2}\theta}, 
 \nonumber                                                                                      \\
  & B_{\textrm{eff}}^{x}=6Jm\sin\theta,                                                         
 \nonumber                                                                                      \\
  & B_{\textrm{eff}}^{z}=B-2\qty(3J-|D|)m\cos\theta.                                            
 \nonumber
\end{align}
In order to simplify the mean-field Hamiltonian, we define new coordinates so that the $\tilde{z}$-direction is aligned to the direction of the effective magnetic field:
\begin{align}
 \mqty(
 S_{\siteA}^{\tilde{x}}          \\
 S_{\siteA}^{\tilde{y}}          \\
 S_{\siteA}^{\tilde{z}}
 )=
 \mqty(
 \cos\varphi  & 0 & \sin\varphi  \\
 0            & 1 & 0            \\
 -\sin\varphi & 0 & \cos\varphi  
 )
 \mqty(
 S_{\siteA}^{x}                  \\
 S_{\siteA}^{y}                  \\
 S_{\siteA}^{z}
 ) ,
 \label{Coordinate_A_MF}         \\
 \mqty(
 S_{\siteB}^{\tilde{x}}          \\
 S_{\siteB}^{\tilde{y}}          \\
 S_{\siteB}^{\tilde{z}}
 )=
 \mqty(
 \cos\varphi  & 0 & -\sin\varphi \\
 0            & 1 & 0            \\
 \sin\varphi  & 0 & \cos\varphi  
 )
 \mqty(
 S_{\siteB}^{x}                  \\
 S_{\siteB}^{y}                  \\
 S_{\siteB}^{z}
 ),
 \label{Coordinate_B_MF}
\end{align}
where the angle $\varphi$ is defined by $\tan\varphi=B_{\textrm{eff}}^{x}/B_{\textrm{eff}}^{z}$.
Using these spin operators on the new coordinates, the effective Hamiltonian is transformed to
\begin{equation}
 \mathscr{H}_{\textrm{MF}}^{(\textrm{Cant})}
 =E_{\rm{cl}}
 -\tilde{B}_{\textrm{eff}}\sum_{\siteA}S_{\siteA}^{\tilde{z}}
 -\tilde{B}_{\textrm{eff}}\sum_{\siteB}S_{\siteB}^{\tilde{z}},
 \label{MFHamiltonian_Canted_2}
\end{equation}
where the effective magnetic field is given by
\begin{equation}
 \tilde{B}_{\textrm{eff}}=\sqrt{\qty(B_{\textrm{eff}}^{x})^{2}+\qty(B_{\textrm{eff}}^{z})^{2}}.
 \label{EffectiveField-Canted}
\end{equation}
The mean-field free energy are estimated as
\begin{equation}
 F_{\textrm{MF}}^{(\textrm{Cant})}=
 E_{\textrm{cl}}
 -Nk_{\textrm{B}}T
 \log
 \qty(\frac{\sinh\big[\beta \tilde{B}_{\textrm{eff}}\left(S+\frac{1}{2}\right)\big]}{\sinh\left(\beta \tilde{B}_{\textrm{eff}}/2\right)}).
 \label{FreeEnergy_Canted_MF}
\end{equation}
The spin expectation values on A and B sublattices are
\begin{equation}
 \left\{ \,
 \begin{aligned}
   & \langle S_{\siteA}^{x}\rangle                          
  =-m\sin\theta
  =-S\sin\varphi B_{S}\qty(S\beta\tilde{B}_{\textrm{eff}}), \\
   & \langle S_{\siteA}^{z}\rangle                          
  =m\cos\theta
  =S\cos\varphi B_{S}\qty(S\beta\tilde{B}_{\textrm{eff}}),
 \end{aligned}
 \right.
 \label{Spin_A_Canted}
\end{equation}
\begin{equation}
 \left\{ \,
 \begin{aligned}
   & \langle S_{\siteB}^{x}\rangle                         
  =m\sin\theta
  =S\sin\varphi B_{S}\qty(S\beta\tilde{B}_{\textrm{eff}}), \\
   & \langle S_{\siteB}^{z}\rangle                         
  =m\cos\theta
  =S\cos\varphi B_{S}\qty(S\beta\tilde{B}_{\textrm{eff}}).
 \end{aligned}
 \right.
 \label{Spin_B_Canted}
\end{equation}
If we numerically compute the free energy and spin moments (order parameters) in both N\'eel and canted phases, we can determine the magnetic phase diagram for the antiferromagnetic Heisenberg model~(\ref{ModelHamiltonian-AFM}).
\section{Bogoliubov transformation for the N\'eel phase}
\label{App:BGT_Neel}
Here, we shortly explain the Bogoliubov transformation to obtain the effective spin-wave Hamiltonian in the N\'eel state of Eq.~(\ref{ModelHamiltonian-AFM}).
After the HP transformation and Fourier one for magnon operators, the Hamiltonian is approximated as
\begin{align}
 \mathscr{H}_{\textrm{SW}}^{(\textrm{N\'eel})} & =            \qty(6JS+2|D|S+B)\sum_{\wn}a_{\wn}^{\dagger}a_{\wn} \nonumber \\
                                               &                                                                            
 \quad\quad+\qty(6JS+2|D|S-B)\sum_{\wn}b_{\wn}^{\dagger}b_{\wn}\nonumber                                                    \\
                                               & \quad\quad\quad+2JS\sum_{\wn}\gamma_{\wn}\qty(a_{\wn}b_{\wn}+              
 a_{\wn}^{\dagger}b_{\wn}^{\dagger})+(\textrm{const.}).
 \label{App_SpinWaveHamiltonian-NeelPhase-WaveNumberSpace}
\end{align}
Hereafter, we neglect the final constant term of $\mathscr{H}_{\textrm{SW}}^{(\textrm{N\'eel})}$.
In order to diagonalize Eq.~(\ref{App_SpinWaveHamiltonian-NeelPhase-WaveNumberSpace}), we apply the following Bogoliubov transformation
\begin{equation}
 \mqty(\alpha_{\wn}\\\beta_{\wn}^{\dagger})=
 \mqty(\cosh\phi_{\wn}&-\sinh\phi_{\wn}\\-\sinh\phi_{\wn}&\cosh\phi_{\wn})\mqty(a_{\wn}\\b_{\wn}^{\dagger}),
 \label{App_BogoliubovTransformation-NeelPhase}
\end{equation}
where $\alpha_{\wn}$ and $\alpha_{\wn}^{\dagger}$, and $\beta_{\wn}$ and $\beta_{\wn}^{\dagger}$ are satisfy the Bose-Einstein statistics, respectively.
If we tune the value of the angle $\phi_{\wn}$ as
\begin{equation}
 \tanh\qty(2\phi_{\wn})=\frac{-J\gamma_{\wn}}{3J+|D|},
 \nonumber
\end{equation}
we obtain the diagonalized Hamiltonian of Eq.~(\ref{SpinWave_Neel_Diagonalization}).
\section{Bogoliubov transformation for the canted AF phase}
\label{App:BGT_Canted}
In this appendix, we explain the Bogoliubov transformation to obtain the effective spin-wave Hamiltonian in the canted AF state of Eq.~(\ref{ModelHamiltonian-AFM}).
In order to obtain that Hamiltonian, we introduce a local spin coordinates $(S^{\zeta},S^{\eta},S^{\xi})$ where the $S^{\xi}$ direction is aligned to the spin-polarization direction.
We define the relationship between original spins and $(S^{\zeta},S^{\eta},S^{\xi})$ as follows:
\begin{align}
 \mqty(
 S_{\siteA}^{\zeta}                      \\
 S_{\siteA}^{\eta}                       \\
 S_{\siteA}^{\xi}
 )=
 \mqty(
 \cos\theta  & 0 & \sin\theta            \\
 0           & 1 & 0                     \\
 -\sin\theta & 0 & \cos\theta            
 )
 \mqty(
 S_{\siteA}^{x}                          \\
 S_{\siteA}^{y}                          \\
 S_{\siteA}^{z}
 ) ,
 \label{App_LocalCoordinate-Asublattice} \\
 \mqty(
 S_{\siteB}^{\zeta}                      \\
 S_{\siteB}^{\eta}                       \\
 S_{\siteB}^{\xi}
 )=
 \mqty(
 \cos\theta  & 0 & -\sin\theta           \\
 0           & 1 & 0                     \\
 \sin\theta  & 0 & \cos\theta            
 )
 \mqty(
 S_{\siteB}^{x}                          \\
 S_{\siteB}^{y}                          \\
 S_{\siteB}^{z}
 ) ,
 \label{App_LocalCoordinate-Bsublattice}
\end{align}
where $\bm S_{\siteA}$ ($\bm S_{\siteB}$) is the spin-$S$ operators on an A-sublattice site $\siteA$ (B-sublattice site $\siteB$).
In the local coordinate, the canted state in the $S^{z}$-$S^{x}$ plane is regarded as a ferromagnetic state along the $S^{\xi}$ axis.
Therefore, we may perform a standard HP transformation for the spins $\qty(S^{\zeta},S^{\eta},S^{\xi})$:
\begin{align}
  & \!\!\! S_{\siteA}^{\xi}\!\!=\!S\!-a_{\siteA}^{\dagger}\!a_{\siteA}, \,\,\,\, 
 S_{\siteA}^{\zeta}\!\!\!+\!iS_{\siteA}^{\eta}\!\!\simeq\!\sqrt{2S}a_{\siteA},\,\,\,\,
 S_{\siteA}^{\zeta}\!\!\!-\!iS_{\siteA}^{\eta}\!\!\simeq\!\sqrt{2S}a_{\siteA}^{\dagger},
 \nonumber                                                                       \\
  & \!\!\! S_{\siteB}^{\xi}\!\!=\!S\!-b_{\siteB}^{\dagger}\!b_{\siteB}, \,\,\,\, 
 S_{\siteB}^{\zeta}\!\!\!+\!iS_{\siteB}^{\eta}\!\!\simeq\!\sqrt{2S}b_{\siteB},\,\,\,\,
 S_{\siteB}^{\zeta}\!\!\!-\!iS_{\siteB}^{\eta}\!\!\simeq\!\sqrt{2S}b_{\siteB}^{\dagger},
 \label{App_LSWA-Cant}
\end{align}
where $a_{\siteA}^{\dagger}$ and $a_{\siteA}$~($b_{\siteB}^{\dagger}$ and $b_{\siteB}$) are creation and annihilation operators of magnon on A-sublattice site $\siteA$ (B-sublattice site $\siteB$), respectively.
Substituting this magnon representation into the Hamiltonian~(\ref{ModelHamiltonian-AFM}) and neglecting the magnon interaction, we have an effective Hamiltonian with quadratic and linear terms of magnon operators.
The single magnon terms all vanish by imposing the condition of minimizing the classical ground-state energy of the canted state, i.e., Eq.~(\ref{CantedAngle}).

Through the Fourier transformation for magnon operators, the wavenumber representation of the Hamiltonian with $a_{\wn}$ and $b_{\wn}$ is simply derived.
If we perform the following two Bogoliubov transformations, the Hamiltonian is diagonalized.
The first simple Bogoliubov transformation is as follows:
\begin{equation}
 \mqty(c_{\wn}\\d_{-\wn})=
 \frac{1}{\sqrt{2}}\mqty(1&-1\\1&1)
 \mqty(a_{\wn}\\b_{-\wn}).
 \label{App_BogoliubovTransformation-Cant-First}
\end{equation}
The Hamiltonian is then represented as
\begin{widetext}
 \begin{align}
  \mathscr{H}_{\textrm{SW}}^{(\textrm{Cant})}=
  \sum_{\wn} & \qty\Big{\qty\big(\textrm{C}_{1}-\textrm{C}_{3}(\wn))c_{\wn}^{\dagger}c_{\wn}            
  +\qty\big(\textrm{C}_{4}-\frac{\textrm{C}_{2}(\wn)}{2})
  \qty(c_{\wn}c_{-\wn}+c_{\wn}^{\dagger}c_{-\wn}^{\dagger})}
  \nonumber                                                                                             \\
             & +\sum_{\wn}\qty\Big{\qty\big(\textrm{C}_{1}+\textrm{C}_{3}(\wn))d_{\wn}^{\dagger}d_{\wn} 
  +\qty\big(\textrm{C}_{4}+\frac{\textrm{C}_{2}(\wn)}{2})
  \qty(d_{\wn}d_{-\wn}+d_{\wn}^{\dagger}d_{-\wn}^{\dagger})}
  +(\textrm{const.}),
  \label{App_SpinWaveHamiltonian-CantPhase-WaveNumberSpace}
 \end{align}
\end{widetext}
where parameters $\textrm{C}_{1-4}$ are given by
\begin{align}
  & \textrm{C}_{1}=6JS\qty(1-2\cos^{2}\theta)-|D|S\qty(1-3\cos^{2}\theta)+B\cos\theta, 
 \nonumber                                                                             \\
  & \textrm{C}_{2}(\wn)=-2JS\qty(1-\cos^{2}\theta)\gamma_{\wn},                        
 \nonumber                                                                             \\
  & \textrm{C}_{3}(\wn)=2JS\cos^{2}\theta\gamma_{\wn},                                 
 \nonumber                                                                             \\
  & \textrm{C}_{4}=-\frac{|D|S}{2}\qty(1-\cos^{2}\theta).                              
 \label{App_ParametersC1_4}
\end{align}
Secondly, if we apply the Bogoliubov transformation
\begin{align}
                                   & \mqty(\alpha_{\wn}           \\
 \alpha_{-\wn}^{\dagger})=
 \mqty(\cosh\varphi_{\wn}^{\alpha} & -\sinh\varphi_{\wn}^{\alpha} \\
 -\sinh\varphi_{\wn}^{\alpha}      & \cosh\varphi_{\wn}^{\alpha}) 
 \mqty(c_{\wn}                                                    \\ c_{-\wn}^{\dagger}),
 \nonumber                                                        \\
                                   & \mqty(\beta_{\wn}            \\
 \beta_{-\wn}^{\dagger})=
 \mqty(\cosh\varphi_{\wn}^{\beta}  & -\sinh\varphi_{\wn}^{\beta}  \\
 -\sinh\varphi_{\wn}^{\beta}       & \cosh\varphi_{\wn}^{\beta})  
 \mqty(d_{\wn}                                                    \\   d_{-\wn}^{\dagger}),
 \label{App_BogoliubovTransformation-Cant-Second}
\end{align}
we finally arrive at the diagonalized spin-wave Hamiltonian of Eq.~(\ref{SpinWaveHamiltonian-CantedPhase}).
\begin{widetext}
 \section{Derivation of Eq.~(\ref{LocalSpinCurrentA})}
 \label{App:PerturbativeCalculation}
 In this appendix, we explain the derivation of Eq.~(\ref{LocalSpinCurrentA}).
 First, we expand the exponential factor in the statistical average of $F_{+-}\qty(\siteA,t;\siteA,t')=-i\big\langle T_{\textrm{C}}\sigma_{\siteA}^{+}(t)S_{\siteA}^{-}(t')\big\rangle$ w.r.t. $\mathscr{H}_{\inter}$:
 \begin{align}
  F_{+-}\qty(\siteA,t;\siteA,t')
   & =      
  -i
  \sum_{n=0}^{\infty}
  \qty(\frac{-i}{\hbar})^{n}
  \frac{1}{n!}
  \int_{\textrm{C}}dt_{1}
  \cdots
  \int_{\textrm{C}}dt_{n}
  \Big\langle
  T_{\textrm{C}}
  \tilde{\sigma}_{\siteA}^{+}(t)
  \tilde{S}_{\siteA}^{-}(t')
  \tilde{\mathscr{H}}_{\inter}\qty(t_{1})
  \cdots
  \tilde{\mathscr{H}}_{\inter}\qty(t_{n})
  \Big\rangle_{0}
  \nonumber \\
   & =      
  -i
  \qty(\frac{-i}{\hbar})
  \int_{\textrm{C}}dt_{1}
  \Big\langle
  T_{\textrm{C}}
  \tilde{\sigma}_{\siteA}^{+}(t)
  \tilde{S}_{\siteA}^{-}(t')
  \tilde{\mathscr{H}}_{\inter}\qty(t_{1})
  \Big\rangle_{0}
  +\cdots,
 \end{align}
 where we have written only the leading term.
 Since the perturbed Hamiltonian is given by
 \begin{equation}
  \tilde{\mathscr{H}}_{\inter}\qty(t_{1})
  =
  \sum_{\siteAd\in\interA}
  J_{\textrm{sd}}\qty(\siteAd)
  \tilde{\bm S}_{\siteAd}\qty(t_{1})
  \cdot
  \tilde{\bm\sigma}_{\siteAd}\qty(t_{1})
  +
  \sum_{\siteBd\in\interB}
  J_{\textrm{sd}}\qty(\siteBd)
  \tilde{\bm S}_{\siteBd}\qty(t_{1})
  \cdot
  \tilde{\bm\sigma}_{\siteBd}\qty(t_{1}),
 \end{equation}
 we obtain
 \begin{align}
  F_{+-}\qty(\siteA,t;\siteA,t')
   & =      
  (-i)^{2}
  \sum_{\siteAd\in\interA}
  \frac{J_{\textrm{sd}}\qty(\siteAd)}{2\hbar}
  \int_{\textrm{C}}dt_{1}
  \Big\langle
  T_{\textrm{C}}
  \tilde{\sigma}_{\siteA}^{+}(t)
  \tilde{\sigma}^{-}_{\siteAd}\qty(t_{1})
  \Big\rangle_{0}
  \Big\langle
  T_{\textrm{C}}
  \tilde{S}^{+}_{\siteAd}\qty(t_{1})
  \tilde{S}_{\siteA}^{-}(t')
  \Big\rangle_{0}
  \nonumber \\
   & =      
  \sum_{\siteAd\in\interA}
  \frac{J_{\textrm{sd}}\qty(\siteAd)}{2\hbar}
  \int_{\textrm{C}}dt_{1}
  \chi_{+-}\qty(\siteA,t;\siteAd,t_{1})
  G_{+-}^{(\textrm{A})}\qty(\siteAd,t_{1};\siteA,t').
  \label{App_FirstOrderPerturbation}
 \end{align}
 In the last line, we have used Eqs.~(\ref{SpinCorrelationFunction_Metal}) and (\ref{SpinCorrelationFunction_Magnet}).
 Using Langreth rule~\cite{HaugJauhoSpringer2008,StefanucciCambridge2013}, we arrive at the following expression of $F_{+-}^{<}\qty(\siteA,t;\siteA,t')$:
 \begin{equation}
  F_{+-}^{<}\qty(\siteA,t;\siteA,t')
  =
  \sum_{\siteAd\in\interA}
  \frac{J_{\textrm{sd}}\qty(\siteAd)}{2\hbar}
  \int_{-\infty}^{\infty}dt_{1}
  \qty\Big{
  \chi_{+-}^{\textrm{R}}\qty(\siteA,t;\siteAd,t_{1})
  G_{+-}^{(\textrm{A})<}\qty(\siteAd,t_{1};\siteA,t')
  +
  \chi_{+-}^{<}\qty(\siteA,t;\siteAd,t_{1})
  G_{+-}^{(\textrm{A})\textrm{A}}\qty(\siteAd,t_{1};\siteA,t')
  },
  \label{App_LangrethRule}
 \end{equation}
 where $\chi_{+-}^{\textrm{R}[<]}\qty(\siteA,t;\siteAd,t_{1})$ is the retarded [lesser] part of $\chi_{+-}\qty(\siteA,t;\siteAd,t_{1})$, and $G_{+-}^{(\textrm{A})\textrm{A}[<]}\qty(\siteAd,t_{1};\siteA,t')$ is the advanced [lesser] part of $G_{+-}^{(\textrm{A})}\qty(\siteAd,t_{1};\siteA,t')$.
 Finally, applying the Fourier transformations on Eq.~(\ref{App_LangrethRule}), we get
 \begin{align}
  F_{+-}^{<}\qty(\siteA,t;\siteA,t')
   & =      
  \sum_{\siteAd\in\interA}
  \frac{J_{\textrm{sd}}\qty(\siteAd)}{2\hbar}
  \int_{-\infty}^{\infty}\frac{d\omega}{2\pi}
  e^{-i\omega(t-t')}
  \qty\Big{
  \chi_{+-}^{\textrm{R}}\qty(\siteA,\siteAd,\omega)
  G_{+-}^{(\textrm{A})<}\qty(\siteAd,\siteA,\omega)
  +
  \chi_{+-}^{<}\qty(\siteA,\siteAd,\omega)
  G_{+-}^{(\textrm{A})\textrm{A}}\qty(\siteAd,\siteA,\omega)
  }
  \nonumber \\
   & =      
  \frac{J_{\textrm{sd}}\qty(\siteA)}{2\hbar}
  \frac{1}{N_{\textrm{m}}\qty(N/2)}
  \sum_{\bm{p},\wn}
  \int_{-\infty}^{\infty}\frac{d\omega}{2\pi}
  e^{-i\omega(t-t')}
  \qty\Big{
  \chi_{+-}^{\textrm{R}}\qty(\bm p,\omega)
  G_{+-}^{(\textrm{A})<}\qty(\bm k,\omega)
  +
  \chi_{+-}^{<}\qty(\bm p,\omega)
  G_{+-}^{(\textrm{A})\textrm{A}}\qty(\bm k,\omega)
  }.
  \label{App_LesserComponent}
 \end{align}
 In the first line, we have performed the integral calculations for $t_{1}$ and then for $\omega'$.
 In the last line, we have neglected the correlation between different interfacial sites $\siteA$ and $\siteAd$ ($\siteA\neq\siteAd$) (see Fig.~\ref{Fig:Interface}).
 Substituting Eq.~(\ref{App_LesserComponent}) into the local spin current Eq.~(\ref{LocalSpinCurrent-Asublattice}) and taking the limit of $\delta\to+0$, we obtain Eq.~(\ref{LocalSpinCurrentA}).
 \section{Dynamical susceptibility of the normal metal}
 \label{App:SpinSusceptibility_NormalMetal}
 In this appendix, using the Hamiltonian Eqs.(\ref{Dynamics_Metal}) and (\ref{ThermalAverage_Metal}), we calculate the imaginary part of the local spin susceptibility in the NESS of the metal.
 The retarded part of $\chi_{+-}\qty(\bm r,t;\bm r',t')$ is defined by Eq.~(\ref{RetardedFunction_Metal}):
 \begin{equation}
  \chi_{+-}^{\textrm{R}}\qty(\bm r,t;\bm r',t')=
  -i\theta\qty(t-t')
  \Big\langle
  \qty[\tilde{\sigma}_{\bm r}^{+}(t),\tilde{\sigma}_{\bm r'}^{-}(t')]
  \Big\rangle_{0},
  \nonumber
 \end{equation}
 where $\sigma^{+}_{\bm r}(t)=f^{\dagger}_{\bm r,\uparrow}(t)f_{\bm r,\downarrow}(t)$, $\sigma^{-}_{\bm r}(t)=\big\{\sigma^{+}_{\bm r}(t)\big\}^{\dagger}$, and $\theta\qty(t-t')$ is the step function.
 The Fourier transformations of fermion operators are defined as ($\alpha=\uparrow,\downarrow$)
 \begin{align}
   & f_{\bm q,\alpha}           
  =\frac{1}{\sqrt{N_{\textrm{m}}}}\sum_{\bm r}e^{-i\bm q\cdot\bm r}f_{\bm r,\alpha},
  \nonumber                     \\
   & f_{\bm q,\alpha}^{\dagger} 
  =\frac{1}{\sqrt{N_{\textrm{m}}}}\sum_{\bm r}e^{i\bm q\cdot\bm r}f_{\bm r,\alpha}^{\dagger}.
  \label{Fourier_Fermion}
 \end{align}
 Substituting Eq.~(\ref{Fourier_Fermion}) into Eq.~(\ref{RetardedFunction_Metal}), we obtain
 \begin{equation}
  \chi_{+-}^{\textrm{R}}\qty(\bm r-\bm r',t-t')=
  -i\theta\qty(t-t')\frac{1}{\qty(N_{\textrm{m}})^2}
  \sum_{\bm q,\bm q'}
  e^{-i\qty(\bm q-\bm q')\cdot\qty(\bm r-\bm r')}
  e^{i\qty(\epsilon_{\bm q}-\epsilon_{\bm q'})\qty(t-t')/\hbar}
  \Big\{
  n_{\textrm{F}}\qty(\xi_{\bm q}-\delta\mu_{\textrm{s}}/2)
  -n_{\textrm{F}}\qty(\xi_{\bm q'}+\delta\mu_{\textrm{s}}/2)
  \Big\},
 \end{equation}
 where $n_{\textrm{F}}\qty(\xi_{\bm q})=\qty(e^{\beta\xi_{\bm q}}+1)^{-1}$ is the Fermi distribution function.
 Using the Fourier transformations, we get
 \begin{equation}
  \chi_{+-}^{\textrm{R}}\qty(\bm p,\omega)=
  \frac{-\hbar}{N_{\textrm{m}}}\sum_{\bm q}
  \frac{n_{\textrm{F}}\qty(\xi_{\bm p+\bm q}+\delta\mu_{\textrm{s}}/2)-n_{\textrm{F}}\qty(\xi_{\bm q}-\delta\mu_{\textrm{s}}/2)}{\hbar\omega+\xi_{\bm q}-\xi_{\bm p+\bm q}+i\delta},
 \end{equation}
 where $\delta=+0$.
 Hence, the local spin susceptibility, $\chi_{+-}^{\textrm{R}}\qty(\omega)=\frac{1}{N_{\textrm{m}}}\sum_{\bm p}\chi_{+-}^{\textrm{R}}\qty(\bm p,\omega)$, is calculated as
 \begin{align}
  \chi_{+-}^{\textrm{R}}\qty(\omega)
   & =\frac{-\hbar}{\qty(N_{\textrm{m}})^2}\sum_{\bm p,\bm q} 
  \frac{n_{\textrm{F}}\qty(\xi_{\bm p+\bm q}+\delta\mu_{\textrm{s}}/2)-n_{\textrm{F}}\qty(\xi_{\bm q}-\delta\mu_{\textrm{s}}/2)}{\hbar\omega+\xi_{\bm q}-\xi_{\bm p+\bm q}+i\delta}
  \nonumber                                                   \\
   & \simeq                                                   
  -\hbar\{D(0)\}^{2}\int_{-\infty}^{\infty}d\xi~d\xi'
  \frac{n_{\textrm{F}}\qty(\xi+\delta\mu_{\textrm{s}}/2)-n_{\textrm{F}}\qty(\xi'-\delta\mu_{\textrm{s}}/2)}{\hbar\omega+\xi'-\xi+i\delta}.
 \end{align}
 In the last line, we have replaced the wavenumber summation according to $1/N_{\textrm{m}}\sum_{\bm~q}\simeq~D(0)\int_{-\infty}^{\infty}d\xi$, where $D(\xi)$ is the density of states per spin and per unit cell.
 Taking the limit $\delta=+0$, we obtain
 \begin{equation}
  \textrm{Im}\chi_{+-}^{\textrm{R}}\qty(\omega)
  \simeq
  -\pi\{D(0)\}^{2}\hbar\qty(\hbar\omega+\delta\mu_{\textrm{s}}).
 \end{equation}
\end{widetext}
%
\end{document}